\documentclass[sigconf,nonacm]{acmart}

\AtBeginDocument{%
  }

\usepackage{xspace}
\usepackage{bm}
\usepackage{booktabs}
\usepackage{multirow}
\usepackage{subcaption}
\usepackage{enumitem}
\usepackage[linesnumbered, ruled, vlined]{algorithm2e}
\usepackage{wasysym}

\makeatletter
\DeclareRobustCommand\onedot{\futurelet\@let@token\@onedot}
\def\@onedot{\ifx\@let@token.\else.\null\fi\xspace}

\def\eg{\emph{e.g}\onedot}

\makeatother

\usepackage{bbm}

{\begin{list}               %
    {$\bullet$ \hfill}{
        \setlength{\leftmargin}{\parindent}
        \setlength{\parsep}{0.04\baselineskip}
        \setlength{\itemsep}{0.5\parsep}
        \setlength{\labelwidth}{\leftmargin}
        \setlength{\labelsep}{0em}}
    }
{\end{list}}

\providecommand{\cref}[1]{Chapter~\ref{#1}}

\newcommand{\proposed}{\textsc{AdvScene}\xspace}

\begin{document}

\title{
AdvScene: Rethinking Adversarial Patch Evaluation Through Scene Robustness
} 
\author{Xiaoyong (Brian) Yuan}
\affiliation{%
  \institution{Clemson University}
  \city{Clemson}
  \state{SC}
  \country{USA}
}
\email{xiaoyon@clesmon.edu}

\author{Lan (Emily) Zhang}
\affiliation{%
  \institution{Clemson University}
  \city{Clemson}
  \state{SC}
  \country{USA}
}
\email{lan7@clemson.edu}

\renewcommand{\shortauthors}{AdvScene}

\begin{abstract} 
Adversarial patches are physical patterns attached to real objects to mislead AI vision systems. Their real-world risk is not determined by a single successful prediction, but by whether they remain effective after deployment under changing viewpoints, distances, and scene conditions. We refer to this property as \emph{scene robustness}, the effectiveness of a deployed patch across conditions in a real environment. Yet existing evaluations do not measure scene robustness well: real image benchmarks are realistic but fixed, while simulators are controllable but not grounded in a specific real scene.

We present AdvScene, a scene-grounded framework for measuring the scene robustness of adversarial patches in reconstructed real environments. AdvScene reframes evaluation as operational measurement: given a fixed deployed patch, it characterizes the patch's operational envelope - where and when the attack succeeds - as a function of viewpoint, distance, and scene context. A key challenge is that the attack is typically defined only in a single anchor view, while evaluation requires a representation that remains faithful under viewpoint changes. We formalize this as a constrained lifting problem and introduce Adversarial Patch-to-Scene Embedding (APSE), which resolves cross-view ambiguity while preserving attack-critical appearance and enforcing locality, target-surface attachment, and cross-view consistency. Applied to reconstructed scenes with controllable rendering and calibrated using matched real captures, APSE enables systematic and behaviorally meaningful evaluation of attack effectiveness. We validate AdvScene using real-world physical data and conduct a comprehensive evaluation of existing adversarial patches across image classification and object detection tasks, including targeted, untargeted, and disappearing settings, on the CO3D and Waymo datasets. Our results show that AdvScene reveals substantial scene-dependent variation in attack effectiveness that is not captured by existing image-centric or simulator-based evaluations.
\end{abstract}

\maketitle
\section{Introduction}
\textbf{From physical attacks to scene-level risk.}
Adversarial attacks~\cite{szegedy2013intriguing,goodfellow2014explaining,carlini2017towards,athalye2018synthesizing,madry2018towards,yuan2019adversarial} have exposed significant vulnerabilities in modern vision systems. Early work primarily studied digital perturbations, where an attacker modifies image pixels directly to induce model failure. More recent work has pushed these attacks toward physical deployment, asking whether adversarial effects can survive printing, capture, viewpoint change, and environmental variation. Among these attacks, adversarial patches are especially concerning because they can be attached to real objects and repeatedly observed by downstream vision systems after physical capture \cite{brown2017adversarial,liu2018dpatch}. Prior work has demonstrated patch-based attacks against both image classification and object detection systems under a range of real-world or physically motivated conditions~\cite{kurakin2018adversarial,eykholt2018robust,chen2018shapeshifter,hu2022adversarial,hu2023physically,zhang2023capatch,zhu2023tpatch,cheng2024fda}. This shift from digital perturbations to deployable artifacts changes the security question: the central concern is no longer only whether a patch succeeds once, but \emph{where, when, and under what scene conditions} it remains effective after deployment.

We refer to this property as \emph{scene robustness}: the effectiveness of a fixed deployed patch across the operational conditions induced by a real scene. Scene robustness is different from standard image-level robustness. It is shaped by the interaction between the patch, the target object, the camera trajectory, the viewing geometry, the projected patch footprint, and the surrounding visual context. A patch that fools a model from one carefully chosen view may pose limited practical risk, while a patch that remains effective across a broad range of naturally occurring views represents a substantially stronger threat. Thus, physical patch evaluation should characterize not only whether an attack succeeds, but also the operational region in which it succeeds.

\vspace{.3em}
\noindent \textbf{A gap between realism and controllability.}
Existing evaluation methods do not measure scene robustness well. Image-plane transformation-based evaluation, including EOT-style transformation sweeps~\cite{athalye2018synthesizing}, is scalable because a patch can be rotated, scaled, perspective warped, or brightness adjusted before being pasted into an image. However, this treats the patch as a two-dimensional overlay rather than as a deployed physical artifact, and can therefore preserve a front-facing patch appearance even when a real patch would become oblique, partially invisible, or too small after projection.

Real-capture benchmarks such as APRICOT~\cite{braunegg2020apricot} and REAP~\cite{hingun2023reap} provide valuable physical evidence, but each capture records only one realized pose, distance, illumination condition, and background. Densely repeating the same deployed-patch experiment over many camera trajectories is labor-intensive, difficult to reproduce, and often unsafe in settings such as driving perception. Simulator-based platforms, including CARLA-based adversarial evaluation frameworks~\cite{dosovitskiy2017carla,nesti2022ss,nesti2024carlagear,lan2024carlaa3}, provide controlled sweeps over camera pose, weather, lighting, and traffic conditions, but their scenes are simulator-native and may not correspond to a specific real deployment site.

These settings leave a missing evaluation regime: real-scene grounded, controllable, and faithful to the behavior of the same fixed deployed attack. We refer to the last requirement as \emph{attack-behavior fidelity}: the evaluation should preserve the success and failure behavior of the deployed patch under scene-induced changes, rather than introducing artifacts from image overlays or simulator-specific assumptions. Table~\ref{tab:evaluation_settings} summarizes this gap.

We therefore formulate adversarial patch evaluation as an \emph{operational measurement} problem. Given a fixed deployed patch and a real scene, the goal is to characterize the patch's \emph{operational envelope}: the region of scene-induced observation conditions under which the attack succeeds. Scene robustness is the property we seek to evaluate, and the operational envelope is the measured region where that property holds.

\begin{table}[!t]
\centering
\caption{Comparison of physical patch evaluation settings. \(\CIRCLE\), \(\LEFTcircle\), and \(\Circle\) denote strong, partial, and limited support for the corresponding evaluation goal.}
\small
\label{tab:evaluation_settings}
\resizebox{\linewidth}{!}{%
\begin{tabular}{lccc}
\toprule
\textbf{Setting} &
\begin{tabular}{c}real-scene\\grounded\end{tabular} &
\begin{tabular}{c}controlled\\view changes\end{tabular} &
\begin{tabular}{c}attack-behavior\\fidelity\end{tabular} \\
\midrule
Real-capture benchmark~\cite{braunegg2020apricot,hingun2023reap} 
& \(\CIRCLE\) & \(\Circle\) & \(\CIRCLE\) \\
Image-plane transformation~\cite{athalye2018synthesizing} 
& \(\LEFTcircle\) & \(\CIRCLE\) & \(\Circle\) \\
Synthetic simulator~\cite{huang2020universal,nesti2022ss,nesti2024carlagear,lan2024carlaa3} 
& \(\Circle\) & \(\CIRCLE\) & \(\LEFTcircle\) \\
\proposed 
& \(\CIRCLE\) & \(\CIRCLE\) & \(\CIRCLE\) \\
\bottomrule
\end{tabular}
}
\vspace{-0.5em}
\end{table}

\vspace{.3em}
\noindent\textbf{Reconstructed scenes as an evaluation substrate.}
A natural way to enable this measurement is to reconstruct the real scene and evaluate the deployed patch inside that reconstruction. Neural rendering and multiview reconstruction have made it increasingly practical to build controllable scene representations from real captures~\cite{mildenhall2021nerf,muller2022instant}. Recent advances in 3D Gaussian Splatting (3DGS)~\cite{kerbl20233d,Huang2DGS2024} further make this direction plausible by supporting high-fidelity reconstruction and efficient novel-view rendering from multiview captures. In principle, a reconstructed scene provides a scene-specific digital twin that combines the grounding of real capture with the controllability of simulation. However, adversarial patch evaluation introduces a challenge that is not addressed by standard 3DGS editing: given only a 2D attacked anchor view, how can we insert the patch into a clean reconstructed scene so that the resulting adversarial scene remains \emph{attack faithful}, \emph{localized on the target surface}, and \emph{stable across viewpoints}? Existing 3DGS editing methods are designed primarily for photorealistic or semantic modification, often relying on diffusion or generative priors~\cite{chen2024gaussianeditor,cao20263dot,xiao2025localizedgs,galerne2025sgsst}. Such methods may produce visually plausible edits, but adversarial evaluation requires preserving fine-grained image structures that determine model behavior.

\begin{figure}[!t]
\centering
\includegraphics[width=0.9\linewidth]{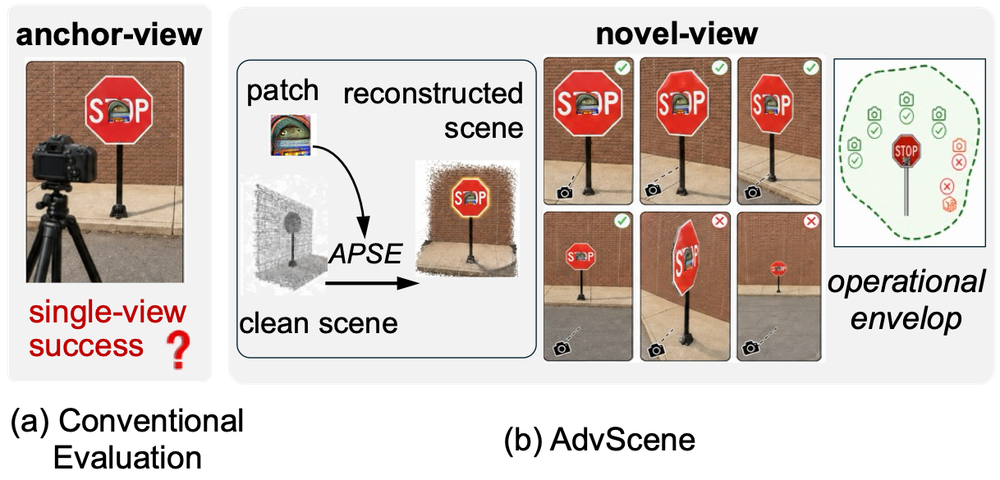}
\caption{
\textbf{From single-view success to scene-level risk.}
(a) Conventional evaluation reports success from a single anchor view, but this does not capture deployment risk.
(b) \proposed reconstructs the scene, embeds the adversarial patch, and evaluates it across viewpoints and distances.
The resulting operational envelope summarizes where the deployed patch remains effective and where scene-induced changes, such as distance and viewing angle, cause the attack to fail.
}
\vspace{-0.5em}
\label{fig:intro_teaser}
\end{figure}

\vspace{.3em}
\noindent\textbf{Our approach.}
To fill this gap, we present \proposed, a scene-grounded framework for measuring the operational envelope of deployed adversarial patches in reconstructed real environments. \proposed starts from a clean multiview capture, generates a fixed adversarial patch from an anchor view, embeds that patch into the reconstructed 3DGS scene, and renders the resulting adversarial scene under controlled changes in viewpoint, distance, camera pose, and imaging conditions. The core technical component is \emph{Adversarial Patch-to-Scene Embedding} (APSE), a hybrid Gaussian embedding method that reuses target-surface Gaussians for structural grounding and introduces new patch Gaussians for attack-critical appearance. 
APSE enforces anchor-view fidelity while using auxiliary-view constraints to suppress leakage into the clean scene, maintain surface attachment, and improve stability under viewpoint changes.
This distinction is important: \proposed is not a stronger attack generator, but a scene-grounded measurement framework for fixed attacks produced by existing pipelines.

\proposed enables a different kind of adversarial evaluation. In our physical evaluation, \proposed achieves 99.30\% ASR consistency with physical captures, compared with 93.54\% for an image-plane transformation baseline. More importantly, it reveals a systematic failure mode of image-plane evaluation: in close oblique views, the baseline can predict high attack success even when the physical patch almost never succeeds. Across CO3D and Waymo, we find that anchor-view success often collapses outside a narrow operational envelope, driving scenes can induce directional and model-dependent envelopes, and physical attack techniques play different roles. Geometry-aware EOT expands scene coverage, while non-printability score (NPS) and total variation (TV) mainly affect printability or smoothness and $\ell_2$ can constrain patch magnitude without substantially shrinking the evaluated envelope.

\vspace{.5em}
\noindent\textbf{Contributions.}
This paper makes the following contributions.
\begin{itemize}
    \item We identify \emph{scene robustness} as a missing dimension in physical adversarial patch evaluation and formulate fixed-patch evaluation as operational-envelope measurement in reconstructed real scenes.
    
    \item We present \proposed, a scene-grounded evaluation framework that combines real-scene reconstruction, fixed-patch embedding, and controlled novel-view rendering to measure attack behavior under changes in viewpoint, distance, projected scale, camera pose, and scene context.
    
    \item We introduce APSE, a constrained patch-to-scene embedding method for 3D Gaussian Splatting. APSE preserves attack-critical anchor-view appearance while enforcing locality, clean-scene preservation, target-surface attachment, and cross-view stability.
    
    \item We validate \proposed with matched physical captures across $7$ real-world scenes. \proposed achieves $99.30\%$ aggregate ASR consistency with physical captures, compared with $93.54\%$ for an image-plane transformation baseline, and avoids the baseline's severe overestimation under close oblique viewing.
    
    \item We evaluate adversarial patches across $59$ CO3D scenes and $28$ Waymo scenes, covering targeted classification, untargeted classification, and detection disappearance. We evaluate over $2{,}400$ images across different views per scene. The results show that anchor-view success can misrepresent deployment risk, that driving scenes induce directional and model-dependent envelopes, and that attack-design choices play different roles: geometry-aware EOT expands scene coverage, while NPS, TV, and $\ell_2$ mainly affect printability, smoothness, or magnitude.
\end{itemize}

\section{Background and Threat Model}
\label{sec:background_problem}

\subsection{Physical Patch Attacks and Evaluation Challenge}
\label{subsec:bg_physical_patch}

Adversarial patches are localized perturbations designed to induce model failure after being placed on or near a target object~\cite{brown2017adversarial,liu2018dpatch,lee2019physical,lovisotto2021slap,nassi2020phantom}. Unlike digital perturbations, physical patches must survive printing, camera capture, viewpoint change, distance variation, illumination change, and sensing noise. Prior work has demonstrated physical patch attacks against traffic signs and object detectors~\cite{eykholt2018robust,chen2018shapeshifter}, person detectors through clothing or texture manipulation~\cite{hu2022adversarial,hu2023physically}, and other security-critical patch settings~\cite{zhang2023capatch,zhu2023tpatch,cheng2024fda}.

To improve physical robustness, patch-generation methods commonly use Expectation over Transformation (EOT)~\cite{athalye2018synthesizing}, which optimizes the patch over sampled image transformations such as rotation, scaling, translation, perspective changes, and brightness changes. Physical regularizers such as object masks, total variation, and the Non-Printability Score (NPS) further constrain the patch to remain localized, smooth, and printable~\cite{eykholt2018robust}. These mechanisms improve how patches are generated, but they do not by themselves solve the evaluation problem. After deployment, the patch is fixed, and its effectiveness depends on the scene-induced observations produced by camera trajectory, target pose, distance, projected scale, and visual context.

The evaluation gap summarized in Table~\ref{tab:evaluation_settings} motivates our setting: rather than optimizing a new attack for each rendered view or relying on a finite set of physical captures, we evaluate how one fixed deployed patch behaves across controlled observation changes in the same reconstructed real scene.

\begin{figure*}[!t]
\centering
\includegraphics[width=0.8\textwidth]{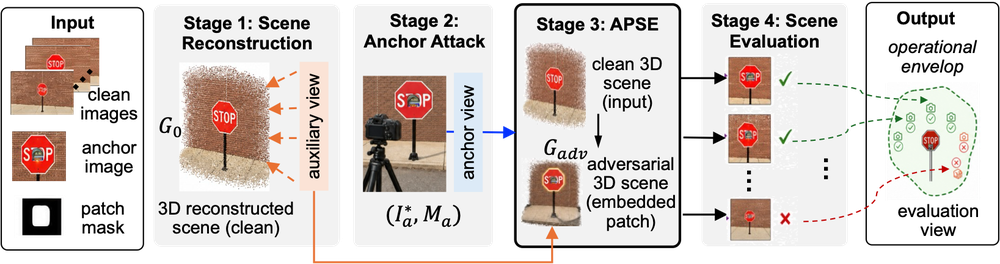}
\vspace{-0.5em}
\caption{
\textbf{Overview of AdvScene.}
Given clean multi-view images, an anchor attack, and a patch mask, AdvScene reconstructs a 3DGS scene \(G_0\), instantiates a fixed anchor-view attack \((I_a^\star, M_a)\), embeds the patch using APSE to obtain \(G_{\mathrm{adv}}\), and evaluates it across diverse viewpoints and distances. 
Distinct view roles are used for supervision, embedding, calibration, and final measurement. 
The output summarizes scene-level attack behavior as an operational envelope.
}
\label{fig:overview}
\vspace{-0.5em}
\end{figure*}

\subsection{Scene-Grounded Evaluation with 3DGS}
\label{subsec:bg_3dgs}

A scene-grounded evaluation substrate should be reconstructed from real captures while still allowing controlled rendering under new viewpoints, distances, and camera conditions. Classical structure-from-motion and multiview stereo estimate camera poses and scene structure from image collections~\cite{schonberger2016structure,schonberger2016pixelwise}. Neural rendering methods such as NeRF and its efficient variants turn multiview captures into novel-view renderers~\cite{mildenhall2021nerf,muller2022instant,barron2022mip,haque2023instruct}. We use 3D Gaussian Splatting (3DGS)~\cite{kerbl20233d,Huang2DGS2024} because it provides high-fidelity reconstruction and efficient rendering.

A 3DGS scene represents the scene as a set of anisotropic Gaussian primitives, 
$G = \{g_i\}_{i=1}^{N},$
where each Gaussian $g_i$ is parameterized by its center $\mu_i$, covariance or scale-rotation parameters, opacity $\alpha_i$, and appearance coefficients $c_i$. Given a camera pose $\bm{\theta}$, a differentiable renderer $\mathcal{R}$ produces an image 
$I(\bm{\theta}) = \mathcal{R}(G, \bm{\theta}).$
Because 3DGS can be reconstructed from multiview real captures and rendered from novel viewpoints, it can, in principle, combine the grounding of real scenes with the controllability needed for operational measurement.

Standard 3DGS editing is not sufficient for adversarial patch evaluation. Existing editing methods primarily target photorealistic or semantic modification, such as object insertion, texture editing, or localized appearance transfer~\cite{chen2024gaussianeditor,cao20263dot,xiao2025localizedgs,galerne2025sgsst}. In contrast, adversarial evaluation requires preserving attack-critical image details while keeping the modification localized and surface-attached. Small changes in color, opacity, boundary alignment, or cross-view placement can alter model behavior even when the edit appears visually plausible to humans.

\subsection{Threat Model and Evaluator Access}
\label{subsec:problem_formulation}

We study the security risk posed by a fixed adversarial patch after deployment in a real scene. The central question is: \textit{How effective is a deployed adversarial patch under the operational conditions induced by a realistic scene?} Equivalently, from the defender's perspective, we ask how robust a vision model is to the deployed patch as viewpoint, distance, camera pose, and scene context change. This setting treats the patch as fixed after deployment and focuses on measuring its behavior under scene-induced observation changes.

\textbf{Adversary goal and capabilities.}
The adversary aims to induce task-specific failure of a victim vision model after real capture. We consider three commonly investigated tasks: targeted attack against classification, untargeted attack against classification, and disappearance attack against object detection. The adversary deploys a fixed adversarial patch on a target object in a real scene. The patch is generated by an existing attack pipeline and may incorporate standard physical robustness mechanisms, such as EOT, NPS, and TV regularization. Once deployed, the patch is fixed: the adversary does not adapt it online, alter it across views, or jointly optimize it with the 3D scene representation used by our framework.

\textbf{Victim and deployment setting.}
The victim is a vision model $f(\cdot)$ operating on captured RGB images. We consider both image classification and object detection settings. After deployment, the patch is observed under realistic operational variation, including changes in distance, azimuth, elevation, camera pose, and scene context. These factors are not fully controlled by the adversary at test time, yet they determine whether the deployed patch remains visible, geometrically aligned, and effective in practice.

\textbf{Evaluator access.}
We assume the evaluator has access to a clean multiview capture of the target scene or object, from which a clean scene representation can be reconstructed. The evaluator also has an attacked anchor-view image $I_a^\star$ generated by the external attack pipeline and a binary patch mask $M_a$ identifying the patch region in that view. 

\textbf{Scope and assumptions.}
We focus on fixed, printable, surface-attached 2D adversarial patches in reconstructed, mostly static scenes. This scope matches a common physical-patch deployment model: the attacker optimizes a patch once, attaches it to a target object, and the deployed patch is then observed from different viewpoints and distances. We do not consider full 3D adversarial objects, adversarial shape changes, volumetric perturbations, time-varying displays, dynamic scene editing, non-rigid deformation, or re-optimizing the attack directly in 3D.
This is a deliberate scope choice. Our goal is to evaluate the deployment robustness of fixed patches produced by existing 2D physical attack pipelines, not to generate a new scene-optimized 3D attack. Three-dimensional attacks require different assumptions about fabrication, geometry, material properties, object replacement, and scene-specific optimization. 
We leave this to future work.

We focus on viewpoint, physical distance, camera pose, projected scale, and scene context because they are first-order deployment variables for 2D patches: they directly affect the patch's pixel resolution, viewing geometry, partial self-occlusion, local appearance distortion, and placement within the image. These choices reflect our goal of measuring deployed patch behavior under dominant operational variables, rather than building a universal simulator for all physical-world adversarial conditions.

\section{\proposed: Operational Measurement Framework}
\label{sec:advscene_overview}
Given the above threat model, AdvScene turns a fixed anchor-view attack into a scene-grounded measurement pipeline. Its input is a clean multiview capture, an attacked anchor image $I_a^\star$, and a patch mask $M_a$; its output is an adversarial scene $G_{\mathrm{adv}}$ and rendered observations used to estimate the operational envelope.

\subsection{End-to-End Pipeline}
\label{subsec:pipeline}

\proposed realizes these requirements through the four-stage pipeline shown in Figure~\ref{fig:overview}. \\
\textbf{Stage 1: Scene reconstruction} reconstructs a clean 3DGS scene \(G_0\) from multiview images of a clean real scene, providing the scene-grounded foundation for controllable rendering. \\
\textbf{Stage 2: Anchor attack instantiation} selects an anchor view \(a\) and applies an external adversarial patch generation pipeline to obtain the attacked image \(I_a^\star\) and patch mask \(M_a\). \\
\textbf{Stage 3: Adversarial Patch-to-Scene Embedding} (APSE) lifts the attacked anchor-view appearance into the reconstructed scene, producing an adversarial scene \(G_{\mathrm{adv}}\) that preserves attack-critical appearance while keeping the patch localized on the target surface. \\
\textbf{Stage 4: Scene-level robustness evaluation} renders \(G_{\mathrm{adv}}\) under controlled observation conditions and evaluates the victim model on the resulting images. This produces a scene-grounded measurement of where an adversarial patch succeeds and where it fails. 

\subsection{Operational Envelope}
\label{subsec:operational_envelope}

\begin{figure}[!tb]
\centering
\includegraphics[width=0.9\linewidth]{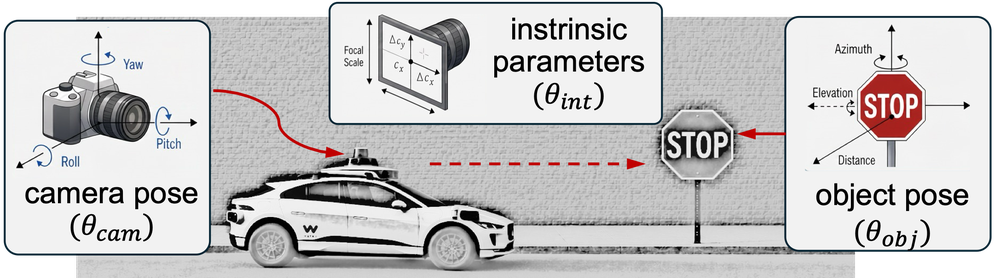}
\caption{Illustration of scene-induced observation condition.}
\label{fig:eval_illustration}
\vspace{-0.5em}
\end{figure}

The output of \proposed is not a single attack success rate, but an operational envelope that characterizes where the deployed patch remains effective. Let $G_0$ denote a clean reconstructed scene, and let $P$ denote a fixed adversarial patch optimized externally in 2D. Starting from the attacked anchor-view image $I_a^\star$ and mask $M_a$, \proposed constructs an adversarial scene $G_{\mathrm{adv}}$ that preserves the patch's attack-critical appearance while keeping the modification localized and geometrically consistent.

Let $\bm{\theta} \in \Theta$ denote a scene-induced observation condition. We parameterize each condition by
\begin{equation}
\bm{\theta} = (\bm{\theta}_{\mathrm{cam}}, \bm{\theta}_{\mathrm{obj}}, \bm{\theta}_{\mathrm{int}}),
\end{equation}
where
\(\bm{\theta}_{\mathrm{cam}} = (\mathrm{yaw}, \mathrm{pitch}, \mathrm{roll})\)
controls camera orientation,
\(\bm{\theta}_{\mathrm{obj}} = (\mathrm{azimuth}, \mathrm{elevation}, \mathrm{distance})\)
controls the camera position relative to the target object, and
\(\bm{\theta}_{\mathrm{int}} = (\mathrm{focal\_scale}, \Delta c_x, \Delta c_y)\)
controls camera intrinsics. Figure~\ref{fig:eval_illustration} illustrates these control groups.
Rendering the adversarial scene under condition $\bm{\theta}$ gives
$I_{\mathrm{adv}}(\bm{\theta})
=
\mathcal{R}(G_{\mathrm{adv}}, \bm{\theta}).$
Let $\mathcal{S}(\bm{\theta}) \in \{0,1\}$ denote whether the attack succeeds under condition $\bm{\theta}$. We define the patch's \emph{operational envelope} as
\begin{equation}
\mathcal{E}
=
\{
\bm{\theta} \in \Theta
\mid
\mathcal{S}(\bm{\theta}) = 1
\}.
\end{equation}
This envelope is the scene-level measurement target of \proposed. It captures not only whether the patch succeeds at the anchor view, but also the region of observation conditions under which it remains effective after deployment. In practice, we estimate this envelope by sampling observation conditions and reporting attack behavior over the corresponding condition bins. 

\section{APSE: Faithful Patch-to-Scene Embedding}
\label{sec:apse}

APSE realizes Stage~3 in Section~\ref{subsec:pipeline}. Given the attacked anchor view \(I_a^\star\), the patch mask \(M_a\), and the clean scene \(G_0\), APSE constructs an adversarial scene \(G_{\mathrm{adv}}\) for scene-level evaluation.

APSE must satisfy three embedding goals. First, it should preserve the \emph{adversarial patch}: the rendered patch should retain the attack-critical appearance of the anchor-view patch. Second, it should preserve the \emph{clean scene}: regions outside the patch should remain unchanged, so downstream failures can be attributed to the deployed patch rather than unrelated scene edits. Third, it should preserve the \emph{patch-surface geometry}: the patch should remain attached to the target surface, rather than becoming a floating layer or view-specific artifact.

This problem is fundamentally \textbf{underconstrained}: only the anchor view provides attacked-image supervision, while the resulting adversarial scene must be evaluated under novel viewpoints. This constraint is practical and intentional. Existing physical patch pipelines typically optimize the patch from one selected view, and our goal is to evaluate the fixed attack produced by such pipelines. Although one could physically attach the optimized patch and collect attacked images from many views, doing so would require additional physical experiments for every patch, scene, and attack variant, which is infeasible for systematic scene-level analysis. APSE therefore uses only the attacked anchor view as direct supervision, while using clean auxiliary views to constrain locality, scene preservation, and surface attachment.

APSE is organized around four components. Section~\ref{subsec:hybrid_gaussian} introduces a hybrid Gaussian representation that combines reused scene Gaussians with newly spawned patch Gaussians. Section~\ref{subsec:anchor_fidelity} transfers the adversarial appearance from the anchor view while preserving the clean scene outside the patch. Section~\ref{subsec:aux_containment} uses auxiliary views to suppress leakage and view-specific artifacts. Section~\ref{subsec:thin_object} initializes and constrains spawned Gaussians so that the patch remains attached to the target surface, including thin objects such as traffic signs. Together, these components turn a single-view attack into a localized, surface-attached adversarial scene for operational-envelope measurement.

\subsection{Hybrid Gaussian Representation}
\label{subsec:hybrid_gaussian}

Let \(G_0\) denote the clean reconstructed 3DGS scene, \(I_a^\star\) the attacked anchor-view image, and \(M_a\) the patch mask. APSE constructs \(G_{\mathrm{adv}}\) by editing selected Gaussians from \(G_0\) and appending new Gaussians in the patch region.

A naive design fails in two ways. Reusing only existing scene Gaussians preserves surface geometry, but these Gaussians were optimized to reconstruct the clean object appearance. Their spatial density, scale, and opacity may not provide enough local capacity to reproduce high-frequency adversarial patterns in the patch. Conversely, using only newly spawned Gaussians provides appearance capacity, but the result is weakly constrained by a single attacked anchor view. New Gaussians can then explain the anchor image through a detached foreground layer, leakage outside the target surface, or a view-specific 3D shortcut that does not correspond to a physical patch.

APSE therefore adopts a hybrid representation with two Gaussian subsets: \(O\) for reused target-surface Gaussians and \(N\) for newly spawned patch Gaussians. APSE selects \(O\) by projecting clean scene Gaussians into the attacked anchor view and retaining Gaussians whose centers project inside \(M_a\), are sufficiently opaque, and have depth consistent with the rendered anchor-view depth. Thus, \(O\) preserves the reconstructed surface support of the object. APSE spawns \(N\) densely from the patch mask to provide additional local capacity. Each spawned Gaussian is initialized from the attacked anchor view: its color is initialized from \(I_a^\star\), its opacity is initialized to a fixed value, its scale is initialized from a target projected radius, and its center is initialized on an estimated patch surface in 3D.

APSE optimizes the two subsets asymmetrically. For reused Gaussians \(g\in O\), only appearance parameters are updated, including color \(c_g\) and opacity \(\alpha_g\), while geometry is fixed. This preserves the reconstructed object surface and prevents the clean scene geometry from being distorted to match the attacked image. For newly spawned Gaussians \(g\in N\), APSE optimizes color, opacity, center \(x_g\), and scale \(s_g\), while keeping orientation fixed. This gives the patch additional local appearance capacity without introducing another weakly constrained geometric degree of freedom under single-view attacked supervision.

The subsequent movement of newly spawned Gaussians is constrained by the surface-attachment mechanism described in Section~\ref{subsec:thin_object}. Thus, this section defines the editable Gaussian representation, while Section~\ref{subsec:thin_object} specifies how the spawned Gaussians are kept attached to the target surface during optimization.

\subsection{Anchor-View Fidelity and Scene Preservation}
\label{subsec:anchor_fidelity}

The hybrid representation defines the local degrees of freedom APSE can modify. The anchor view \(a\) is the only view with attacked-image supervision, so it must both transfer the adversarial appearance inside the patch and prevent unrelated scene changes outside the patch.

Let \(I_a^0=\mathcal{R}(G_0,a)\) denote the clean render at the anchor view, and let \(I_a=\mathcal{R}(G_{\mathrm{adv}},a)\) denote the current adversarial render. Inside the patch mask, APSE matches the attacked anchor image:
\begin{equation}\label{eq:patch}
\mathcal{L}_{\mathrm{patch}} =
\frac{\sum_p M_a(p)\|I_a(p)-I_a^\star(p)\|_1}
{\sum_p M_a(p)} .
\end{equation}
Outside the patch mask, APSE preserves the clean reconstruction:
\begin{equation}\label{eq:keep}
\mathcal{L}_{\mathrm{keep}} =
\frac{\sum_p (1-M_a(p))\|I_a(p)-I_a^0(p)\|_1}
{\sum_p (1-M_a(p))} .
\end{equation}

These two losses form a constrained appearance transfer: \(\mathcal{L}_{\mathrm{patch}}\) injects the attack-critical signal only where the patch is allowed to appear, while \(\mathcal{L}_{\mathrm{keep}}\) prevents global color shifts, boundary leakage, or unrelated scene edits. The remaining ambiguity is cross-view behavior: many 3D embeddings can match the same anchor view but behave differently from novel viewpoints. APSE addresses this ambiguity using auxiliary-view containment, which constrains how the embedded patch behaves outside the anchor view.

\subsection{Auxiliary-View Containment}
\label{subsec:aux_containment}

Auxiliary views \(V_{\mathrm{aux}}\) constrain the embedded patch under viewpoint changes without introducing additional attacked-image supervision. These views have no target attacked images. Instead, they enforce a containment principle: the patch should appear only where its trainable Gaussian support projects, and the rest of the scene should remain unchanged.

For an auxiliary view \(v\in V_{\mathrm{aux}}\), let \(I_v=\mathcal{R}(G_{\mathrm{adv}},v)\) and \(I_v^0=\mathcal{R}(G_0,v)\) denote the adversarial and clean renders, respectively. Let \(S_v^{\mathrm{proj}}\) denote the projected support of the trainable patch Gaussians in view \(v\). APSE uses two auxiliary constraints outside this support: an RGB preservation term that suppresses unintended appearance changes, and an opacity suppression term that discourages floating patch artifacts.
The RGB preservation term is defined as
\begin{equation}\label{eq:aux_rgb}
\mathcal{L}_{\mathrm{aux\mbox{-}rgb}}^{(v)} =
\frac{\sum_p (1 - S_v^{\mathrm{proj}}(p)) \| I_v(p) - I_v^0(p) \|_1}
{\sum_p (1 - S_v^{\mathrm{proj}}(p))}.
\end{equation}
This term prevents the patch optimization from contaminating unrelated regions of the scene.

The opacity suppression term penalizes spurious foreground opacity outside the projected patch support. Let \(A_v\) denote the rendered opacity map at view \(v\). We define
\begin{equation}\label{eq:aux_alpha}
\mathcal{L}_{\mathrm{aux\mbox{-}alpha}}^{(v)} =
\frac{\sum_p (1 - S_v^{\mathrm{proj}}(p)) A_v(p)^2}
{\sum_p (1 - S_v^{\mathrm{proj}}(p))}.
\end{equation}
This term discourages diffuse or floating patch artifacts that do not align with the target surface.

The auxiliary losses are averaged over all auxiliary views:
\(\mathcal{L}_{\mathrm{aux\mbox{-}rgb}}
=
\frac{1}{|V_{\mathrm{aux}}|}
\sum_{v\in V_{\mathrm{aux}}}
\mathcal{L}_{\mathrm{aux\mbox{-}rgb}}^{(v)},
\quad
\mathcal{L}_{\mathrm{aux\mbox{-}alpha}}
=
\frac{1}{|V_{\mathrm{aux}}|}
\sum_{v\in V_{\mathrm{aux}}}
\mathcal{L}_{\mathrm{aux\mbox{-}alpha}}^{(v)}.\)
Together, these losses restrict the feasible 3D embeddings to those that remain localized to the projected patch support while preserving the clean scene across auxiliary viewpoints.

\subsection{Surface Attachment and Thin-Object Handling}
\label{subsec:thin_object}

The auxiliary-view constraints keep the patch local in rendered views, but they do not fully determine where newly spawned Gaussians should be placed in 3D. This is especially important for thin objects such as traffic signs, where small depth errors near object boundaries or reconstruction noise can place Gaussians in front of the sign, behind it, or on nearby background geometry. Such errors can make the patch appear detached even if the anchor-view render looks correct.

APSE then estimates a foreground surface before spawning new Gaussians. Given \(M_a\) and the clean depth map \(D_a^0\), APSE collects valid depth pixels inside the mask and keeps the nearest depth quantile, which corresponds to the visible front surface. Back-projecting these pixels yields foreground 3D points, from which APSE fits a plane. New Gaussians are initialized by intersecting anchor-view camera rays with this fitted plane and shifting the centers slightly along the plane normal. This produces a coherent foreground surface instead of copying noisy per-pixel depths.

During optimization, APSE constrains each spawned Gaussian around its initial surface anchor \(x_g^0\). For each Gaussian, APSE defines a local surface frame with two tangent directions and one normal direction. After each optimizer update, APSE projects the center displacement \(x_g-x_g^0\) into this local frame, clips the tangent components by \(\rho_g\), and clips the normal component by \(\tau_n\). The tangent bound limits how far the Gaussian can slide within the patch neighborhood, while the normal bound prevents it from floating in front of or behind the surface.

In addition to this hard clamp, APSE applies a soft normal-attachment loss. Let \(x_g\) be the current center of a spawned Gaussian, \(x_g^0\) its initial center, and \(n_g\) the local surface normal. Define \(\Delta_g^n=n_g^\top(x_g-x_g^0)\). APSE uses
\begin{equation}\label{eq:normal}
\mathcal{L}_{\mathrm{normal}}
=
\frac{1}{|N|}
\sum_{g \in N}
\left(
\frac{\min(|\Delta_g^n|,\tau_{\mathrm{attach}})}
{\tau_{\mathrm{attach}}}
\right)^2 .
\end{equation}
Here, \(\tau_{\mathrm{attach}}\) controls the normal-distance scale, and the cap prevents a single outlier from dominating the loss. Together, the foreground-plane initialization, local-frame clamp, and normal-attachment loss keep the patch attached to the target surface instead of forming a floating layer or leaking onto background geometry.

\subsection{Overall Objective}
\label{subsec:apse_objective}
The APSE objective combines anchor-view transfer, scene preservation, auxiliary-view containment, and surface attachment:
\begin{equation}
\begin{aligned}
\mathcal{L}_{\mathrm{APSE}} =\;&
\lambda_{\mathrm{patch}} \mathcal{L}_{\mathrm{patch}}
+ \lambda_{\mathrm{keep}} \mathcal{L}_{\mathrm{keep}} + \lambda_{\mathrm{aux\mbox{-}rgb}} \mathcal{L}_{\mathrm{aux\mbox{-}rgb}}\\
&+ \lambda_{\mathrm{aux\mbox{-}alpha}} \mathcal{L}_{\mathrm{aux\mbox{-}alpha}} + \lambda_{\mathrm{normal}} \mathcal{L}_{\mathrm{normal}} .
\end{aligned}
\end{equation}
The first two terms enforce patch fidelity and scene preservation at the anchor view via Eq.~(\ref{eq:patch}) and Eq.~(\ref{eq:keep}). 
The auxiliary terms preserve locality across views via Eqs.~(\ref{eq:aux_rgb}) and~(\ref{eq:aux_alpha}), and the normal term promotes surface attachment via Eq.~(\ref{eq:normal}).
Only the anchor view provides supervision for attacked images. Auxiliary views provide preservation and containment constraints, and no attacked target image is imposed outside the anchor view. As a result, APSE transfers the adversarial signal locally while avoiding global scene edits and detached geometric artifacts.

\section{Evaluation}

Our evaluation examines whether \proposed provides a faithful and useful measurement of scene-level adversarial patch robustness. We focus on four main questions.

\begin{itemize}[leftmargin=10pt]
    \item \textbf{RQ1: Behavioral fidelity.}
    Does \proposed preserve the visual appearance and attack behavior of a deployed physical patch under matched physical and rendered observations?
    
    \item \textbf{RQ2: Operational-envelope measurement.} 
    How does the effectiveness of a fixed deployed patch change as viewpoint, distance, and scene-induced observation conditions vary? 
    
    \item \textbf{RQ3: Scene factors.}
    Which observation factors, including viewing angle, distance, projected patch scale, and image-plane location, most strongly explain changes in attack success?
 
    \item \textbf{RQ4: Attack design.}
    Which common patch-generation techniques, such as EOT and regularization terms, actually expand the operational envelope after deployment?
\end{itemize}
These questions reflect the central goal of \proposed: evaluating an adversarial patch as a fixed physical object deployed in a real scene, rather than as an isolated image perturbation. Accordingly, our evaluation reports how attack behavior varies across scene-induced conditions and uses the operational envelope to characterize where the deployed patch remains effective.

We organize the evaluation accordingly through four steps. We first validate \proposed using matched physical captures to test whether the reconstructed adversarial scene
preserves both visual fidelity and attack behavior. We then
measure scene robustness at scale across object-centered and driving scenes.
Next, we analyze which scene factors dominate the degradation of attack
success after deployment. Finally, we study whether common physical-attack
design choices expand the operational envelope of a fixed deployed patch.

\subsection{Experimental Protocol and Metrics}
\label{sec:eval:common}

\textbf{Attack objectives.}
We evaluate three attack objectives. For image classification, we consider
targeted attacks, where the patch aims to induce an attacker-specified
label, and untargeted attacks, where the patch aims to suppress the clean
prediction. For object detection, we consider disappearance attacks, where
the patch aims to suppress the detector output for the attacked object.
In all cases, the patch is optimized only from the anchor view and then
kept fixed during scene-level evaluation.

\vspace{.3em}
\noindent\textbf{Default settings.}
Unless otherwise stated, the default patch size is $20\%$ of the target
object. The default physical-attack objective includes EOT,
non-printability score (NPS), total variation (TV), and $\ell_2$
regularization. Detailed settings are reported in
Appendix~\ref{app:settings}. Section~\ref{sec:eval:attack-components} compares these attack-design choices to determine which ones improve scene-level robustness after deployment.

\vspace{.3em}
\noindent\textbf{Evaluation metrics.}
We report attack behavior over view bins rather than as a single scene-averaged number. Let $\mathcal{B}$ denote a set of observations corresponding to a particular condition bin, such as a distance-angle bin in a heatmap or a value range in a one-dimensional sweep. For an observation type $m$, we define attack success rate (ASR) as
\begin{equation}
\mathrm{ASR}_{m}(\mathcal{B})=\frac{1}{|\mathcal{B}|}\sum_{(j,\boldsymbol{\theta})\in \mathcal{B}}s\!\left(x^{m}_{j,\boldsymbol{\theta}}\right),
\label{eq:asr_bin}
\end{equation}
where $j$ indexes the scene, $\boldsymbol{\theta}$ indexes the observation condition, $x^{m}_{j,\boldsymbol{\theta}}$ is the corresponding image, and $s(\cdot)$ is the task-specific attack success indicator. In large-scale rendered evaluation, $m\in\{\mathrm{adv},\mathrm{benign}\}$. In matched physical-rendered evaluation, $m\in\{\mathrm{phy},\mathrm{ren}\}$.
For untargeted classification and disappearance attacks, benign failures can occur under difficult views even without the adversarial patch. We therefore report attack gain,
\begin{equation}
\mathrm{AG}(\mathcal{B})=\mathrm{ASR}_{\mathrm{adv}}(\mathcal{B}) - \mathrm{ASR}_{\mathrm{benign}}(\mathcal{B}).
\label{eq:attack_gain}
\end{equation}
ASR measures raw attack success within a condition bin, while AG isolates the additional failure caused by the deployed patch.

\begin{figure}[!t]
\centering
\begin{subfigure}[t]{0.42\columnwidth}
\centering
\includegraphics[width=\linewidth]{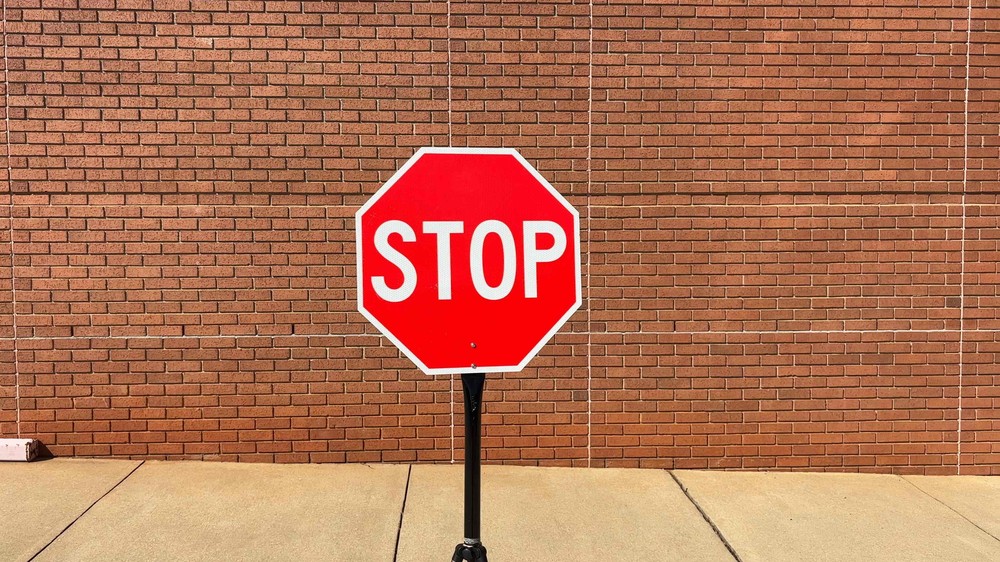}
\caption{Clean capture}
\end{subfigure}
\begin{subfigure}[t]{0.42\columnwidth}
\centering
\includegraphics[width=\linewidth]{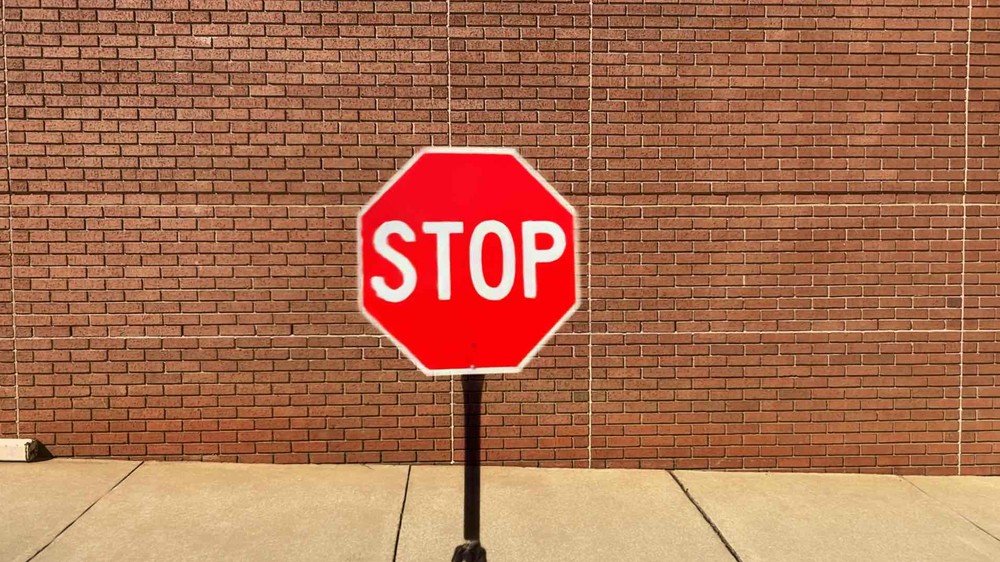}
\caption{Clean render}
\end{subfigure}

\vspace{0.25em}

\begin{subfigure}[t]{0.42\columnwidth}
\centering
\includegraphics[width=\linewidth]{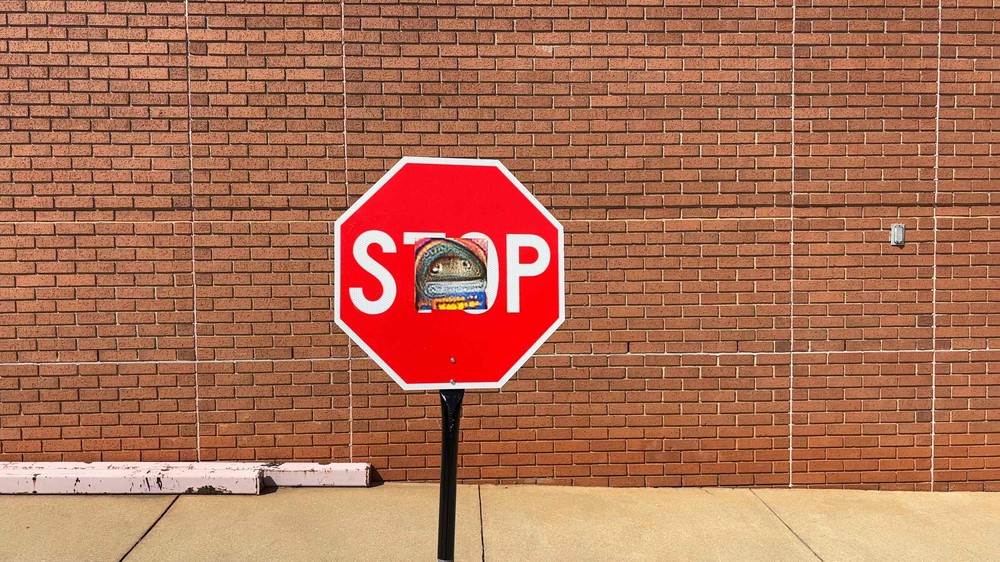}
\caption{Patched capture 1}
\end{subfigure}
\begin{subfigure}[t]{0.42\columnwidth}
\centering
\includegraphics[width=\linewidth]{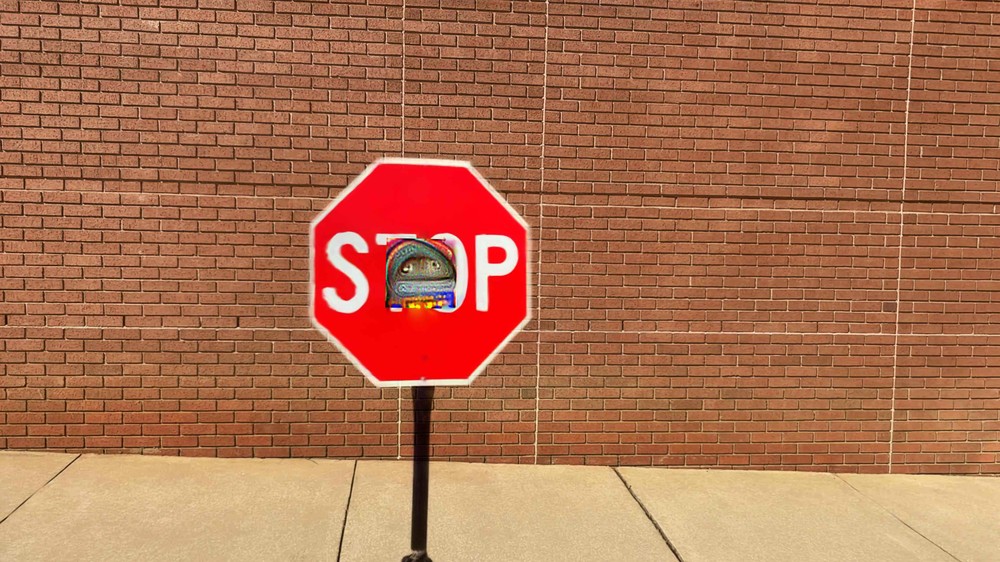}
\caption{Matched render 1}
\end{subfigure}

\begin{subfigure}[t]{0.42\columnwidth}
\centering
\includegraphics[width=\linewidth]{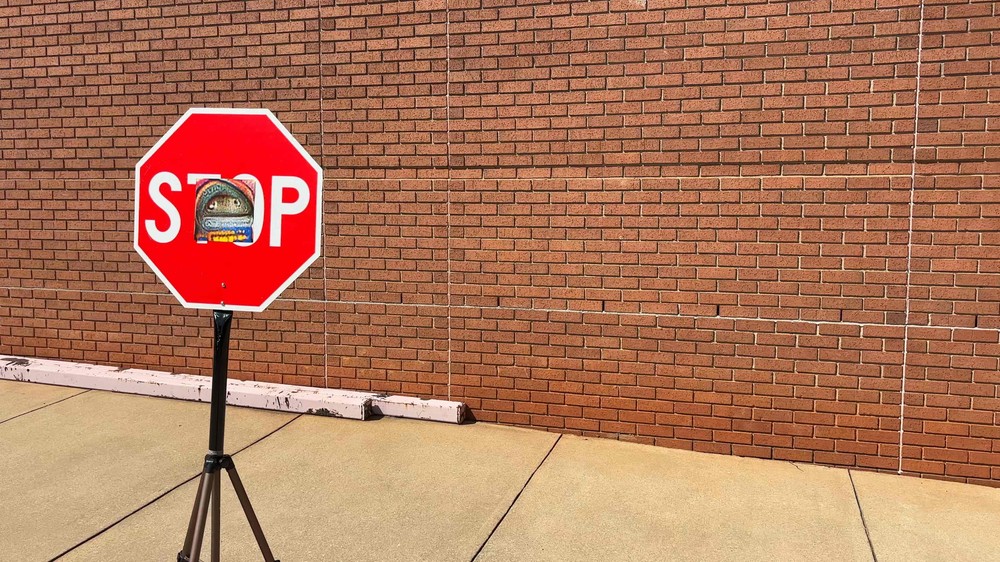}
\caption{Patched capture 2}
\end{subfigure}
\begin{subfigure}[t]{0.42\columnwidth}
\centering
\includegraphics[width=\linewidth]{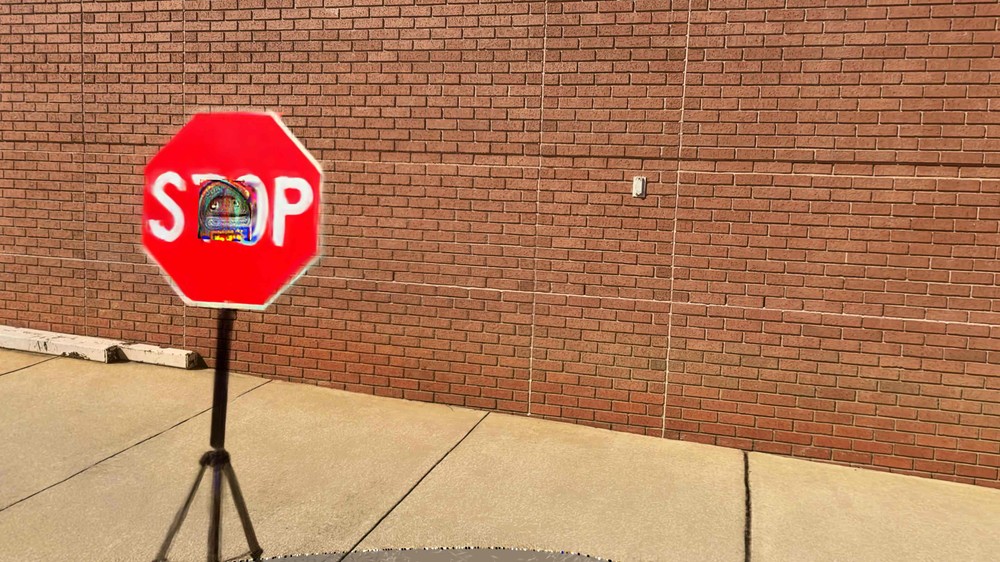}
\caption{Matched render 2}
\end{subfigure}
\caption{
\textbf{Visual fidelity under matched poses.}
(a) and (b) compare a clean physical capture with the corresponding clean 3DGS render at the anchor pose. (c)-(f) compare physical captures after attaching the printed adversarial patch with AdvScene renders from the matched estimated camera poses. The paired layout shows that APSE preserves patch placement, target-surface attachment, and surrounding background appearance under the same observation conditions.
}
\label{fig:physical_examples}
\vspace{-1em}
\end{figure}

\subsection{Fidelity Validation of \proposed with Physical Captures}
\label{sec:eval:fidelity}
We first evaluate whether \proposed preserves both the \emph{appearance} and \emph{attack behavior} of a deployed adversarial patch under matched physical and rendered observations. This validation tests whether the rendered adversarial scene can serve as a reliable proxy for downstream scene-level attack evaluation.

\vspace{.3em}
\noindent\textbf{Evaluation protocol.}
We construct a physical validation set of $7$ scenes, each containing one target object: a stop sign, a slow sign, a speed limit sign, a laptop, a book, an envelope, or a binder, with about $305$ images per scene on average from different views. For each scene, we first capture benign multi-view images and reconstruct a clean 3DGS model. We then select an \emph{anchor view} and generate a targeted adversarial patch. The patch is embedded into the 3DGS using APSE to obtain an adversarial scene.

Independently, we print and physically attach the same patch to the target object and collect images under varying camera poses. For each physical image, we estimate the corresponding camera pose $p_i$ and render the adversarial 3DGS from the same pose, yielding matched observations
\(
\{x^{\mathrm{phy}}(p_i), x^{\mathrm{ren}}(p_i)\}_{i=1}^{N}.
\)
These pairs enable direct comparison between physical and rendered behavior under matched viewpoints. For this matched physical-rendered study, we use a Vision Transformer (ViT)~\cite{dosovitskiy2021image} classifier as the victim model for both patch optimization and matched physical-rendered evaluation. This controlled setting investigates whether the reconstructed adversarial scene preserves the attack behavior observed in physical captures.

\vspace{.3em}
\noindent\textbf{Visual fidelity.}
We first inspect whether APSE preserves the appearance and placement of the deployed patch under matched physical and rendered observations. The goal is to verify that the rendered patch remains attached to the target surface and that the surrounding scene is preserved before evaluating decision-level attack behavior. Figure~\ref{fig:physical_examples} shows three paired comparisons: a clean physical capture and its corresponding clean render, followed by two physical patched captures and their matched adversarial renders from the same estimated camera poses. 
Table~\ref{tab:visual_fidelity} further quantifies this alignment by measuring the distances between raw image, clean render, and adversarial render. APSE preserves the local patch appearance while keeping the surrounding scene largely unchanged. The adversarial render achieves low patch-region error relative to the optimized digital patch, and the background difference between clean and adversarial renders remains small. This supports using the rendered scene for downstream attack-behavior validation, since model failures can be attributed primarily to the deployed patch rather than unrelated scene edits.

\begin{table}[!t]
\centering
\caption{Visual fidelity of rendering. Lower values indicate better alignment. We report $\ell_1$ and $\ell_2$ distances between raw image, clean render, and adversarial render in anchor views. }
\label{tab:visual_fidelity}
\small
\begin{tabular}{lcc}
\toprule
\textbf{Comparison} & $\ell_1$ & $\ell_2$ \\
\midrule
Full frame (clean render vs. raw image) & 0.137 & 0.117 \\
Patch region (clean render vs. raw image) & 0.119 & 0.113 \\
Non-patch region (clean render vs. raw image) & 0.137 & 0.116 \\
\midrule
Patch region (adv render vs. digital patch) & \textbf{0.101} & \textbf{0.080} \\
Background (adv render vs. clean render) & \textbf{0.049} & \textbf{0.068} \\
\bottomrule
\end{tabular}
\end{table}

\vspace{.3em}
\noindent\textbf{Attack-behavior fidelity under matched poses.}
We next evaluate whether rendered and physical observations lead to consistent attack behavior under the same camera poses. For a matched physical-rendered pair at pose $p_i$, let $x^{\mathrm{phy}}(p_i)$ be the physical capture and $x^{\mathrm{ren}}(p_i)$ be the corresponding APSE render. For a condition bin $\mathcal{B}$, we report the physical ASR, the rendered ASR, and their aggregate ASR consistency.
The ASR consistency is defined as
\begin{equation}
\mathrm{cons}_{ren}(\mathcal{B}) = 1 -|\mathrm{ASR}_{ren}(\mathcal{B}) - \mathrm{ASR}_{phy}(\mathcal{B})|.
\label{eq:consistency}
\end{equation}
This metric measures aggregate agreement between physical and rendered attack outcomes within the same condition bin. Higher consistency indicates that the rendered setting better matches the physical attack success rate in that bin.

We compare APSE with a baseline via image-plane transformation, which reflects the common evaluation practice (\eg, EOT) used when only a single attacked view is available. Given an evaluation pose, the baseline geometrically transforms the optimized 2D patch according to that view and composites it onto the corresponding benign 3DGS render. This baseline uses the same patch and the same clean scene render as APSE, but does not construct a scene-embedded adversarial patch. It therefore tests whether a simple image-plane patch transformation is sufficient to predict physical attack behavior.

\begin{table}[!tb]
\centering
\caption{ASR consistency comparison between physical captures, an image-plane transformation baseline, and \proposed across distance-angle bins. The global row reports the average ASR across all matched observations.}
\label{tab:asr_consistency_baseline}
\small
\resizebox{\linewidth}{!}{%
\begin{tabular}{ccccccc}
\toprule
\begin{tabular}{c}distance\\range (m)\end{tabular} &
\begin{tabular}{c}angle\\range ($^\circ$)\end{tabular} &
\begin{tabular}{c}physical\\$\mathrm{ASR}_{phy}$\end{tabular} &
\multicolumn{2}{c}{baseline} &
\multicolumn{2}{c}{\proposed} \\
\cmidrule(lr){4-5}
\cmidrule(lr){6-7}
& & &
$\mathrm{ASR}_{ren}$ &
$\mathrm{cons}_{ren}$ &
$\mathrm{ASR}_{ren}$ &
$\mathrm{cons}_{ren}$ \\
\midrule
$[0, 1)$ & $[0, 15)$
& $62.57$
& $50.02$ & $87.45$
& $58.55$ & $\textbf{95.98}$ \\
$[0, 1)$ & $[15, 45)$
& $30.22$
& $57.66$ & $72.56$
& $31.71$ & $\textbf{98.51}$ \\
$[0, 1)$ & $[45, +\infty)$
& $0.19$
& $59.85$ & $40.34$
& $0.00$ & $\textbf{99.81}$ \\
\midrule
$[1, 3)$ & $[0, 15)$
& $60.71$
& $44.47$ & $83.76$
& $63.74$ & $\textbf{96.97}$ \\
$[1, 3)$ & $[15, 45)$
& $48.96$
& $47.19$ & $98.23$
& $53.40$ & $\textbf{95.56}$ \\
$[1, 3)$ & $[45, +\infty)$
& $66.89$
& $55.60$ & $88.71$
& $61.63$ & $\textbf{94.74}$ \\
\midrule
$[3, +\infty)$ & $[0, 15)$
& $24.70$
& $13.89$ & $89.19$
& $42.17$ & $\textbf{82.53}$ \\
$[3, +\infty)$ & $[15, 45)$
& $20.24$
& $8.33$ & $88.09$
& $34.79$ & $\textbf{85.45}$ \\
\midrule
all & all
& $45.26$
& $51.72$ & $93.54$
& $45.96$ & $\textbf{99.30}$ \\
\bottomrule
\end{tabular}
}
\end{table}

\vspace{.3em}
\noindent\textbf{Finding 1.}
\textit{\proposed matches physical attack behavior better than image-plane overlays.}
Table~\ref{tab:asr_consistency_baseline} shows that the physical ASR is 45.26\%, while \proposed reports 45.96\%, yielding 99.30\% aggregate consistency. The image-plane transformation baseline reaches only 93.54\% consistency and fails most clearly in oblique views. In the close, large-angle bin, the physical ASR is nearly zero, while the image-plane baseline predicts high attack success. This indicates that overlay-based evaluation can hallucinate deployment risk by preserving a fronto-parallel patch appearance that a real surface-attached patch would lose under oblique viewing.

AdvScene is not equally reliable in all regimes. At longer distances, \proposed overestimates physical ASR, likely because the target occupies fewer pixels and reconstruction or pose errors have larger behavioral impact. We therefore treat matched physical validation as a calibration step. In the following experiments, we focus on the high-confidence near and medium-distance regimes, where physical-rendered ASR agreement is strongest.

\subsection{Scene Robustness at Scale}
\label{sec:eval:robustness}
After validating behavioral fidelity with physical captures, we use \proposed to measure adversarial scene robustness at scale.

\vspace{.3em}
\noindent\textbf{Evaluation protocol.}
For each scene, we select an anchor view, generate one fixed adversarial patch from that view, embed the patch into the reconstructed scene using APSE, and render both benign and adversarial observations over controlled scene-induced conditions. This protocol keeps the patch fixed and varies only the observation condition, allowing us to measure the operational envelope of the deployed patch. Consistent with the physical validation in Section~\ref{sec:eval:fidelity}, we interpret the main scene-robustness trends in the high-confidence region where the rendered behavior is well aligned with matched physical captures.

\vspace{.3em}
\noindent\textbf{Datasets, tasks, and victim models.}
For large-scale scene robustness evaluation, we use two complementary real-scene datasets. The CO3D dataset~\cite{reizenstein2021common} provides object-centered multi-view captures, where viewpoint changes can be controlled around a dominant target object. This setting isolates how a fixed patch transfers across object-relative view changes. The Waymo dataset~\cite{Sun_2020_CVPR} provides driving scenes captured from vehicle-mounted cameras, where the target undergoes natural changes in distance, projected scale, viewpoint, background, and partial occlusion. This setting tests the same fixed-patch assumption in a more deployment-realistic environment. We use $59$ CO3D scenes and $28$ Waymo scenes containing traffic signs.
On CO3D, we evaluate targeted classification, untargeted classification, and detection disappearance. Targeted classification measures whether the patch induces an attacker-specified label; untargeted classification measures whether it suppresses the clean label; and disappearance measures whether it suppresses the detector output for the attacked object. On Waymo, we focus on disappearance attacks, which are the most relevant to driving perception. For classification, we evaluate ViT-B/16~\cite{dosovitskiy2021image}, ResNet-50~\cite{he2016deep}, and ConvNeXt~\cite{liu2022convnet}. For detection, we evaluate YOLOv8~\cite{yolov8_ultralytics} and RT-DETR~\cite{lv2024rt}.

\vspace{.3em}
\noindent\textbf{Scene generation.}
For each scene, we render both benign and adversarial views under $9$ controlled dimensions: yaw, pitch, roll, azimuth, elevation, radius, focal scale, horizontal shift, and vertical shift. Each dimension is sampled at $91$ values with $3$ noise variants per value, resulting in $2{,}457$ rendered views per scene. Since object distance can vary substantially across scenes, we report radius as a normalized distance multiplier: radius $=1$ corresponds to the anchor-view distance, while values below or above $1$ move the camera closer to or farther from the target.

\subsubsection{CO3D Results}
\label{sec:eval_co3d}
\begin{figure}[!tb]
\centering
\includegraphics[width=\linewidth]{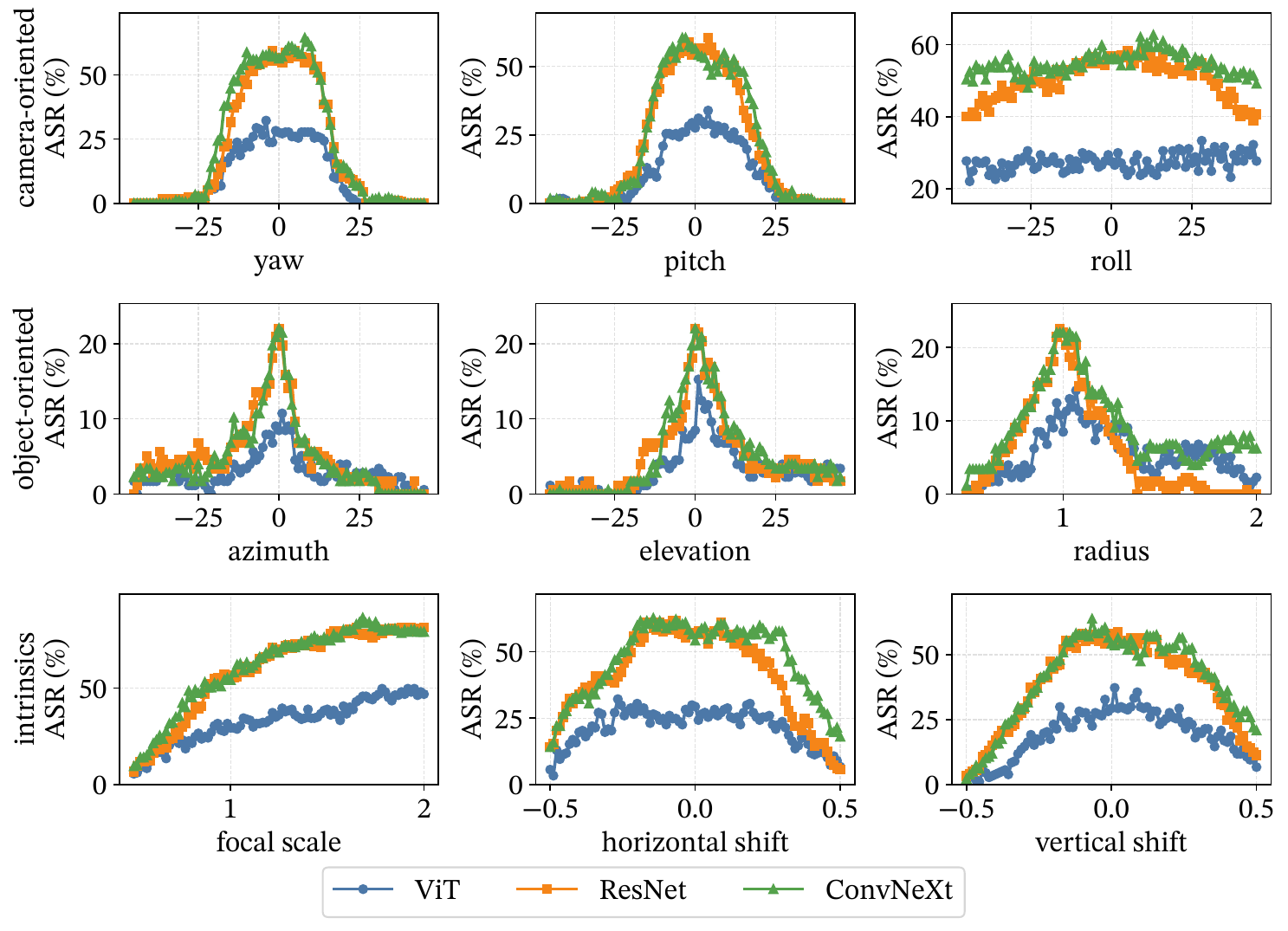}
\caption{Targeted attack ASR across observation dimensions on CO3D.
Anchor-view success is local: ASR is highest near the anchor-centered
condition and drops sharply as geometry changes. }
\label{fig:co3d_targeted_asr}
\vspace{-1em}
\end{figure}

\begin{figure}[!tb]
\centering
\includegraphics[width=\linewidth]{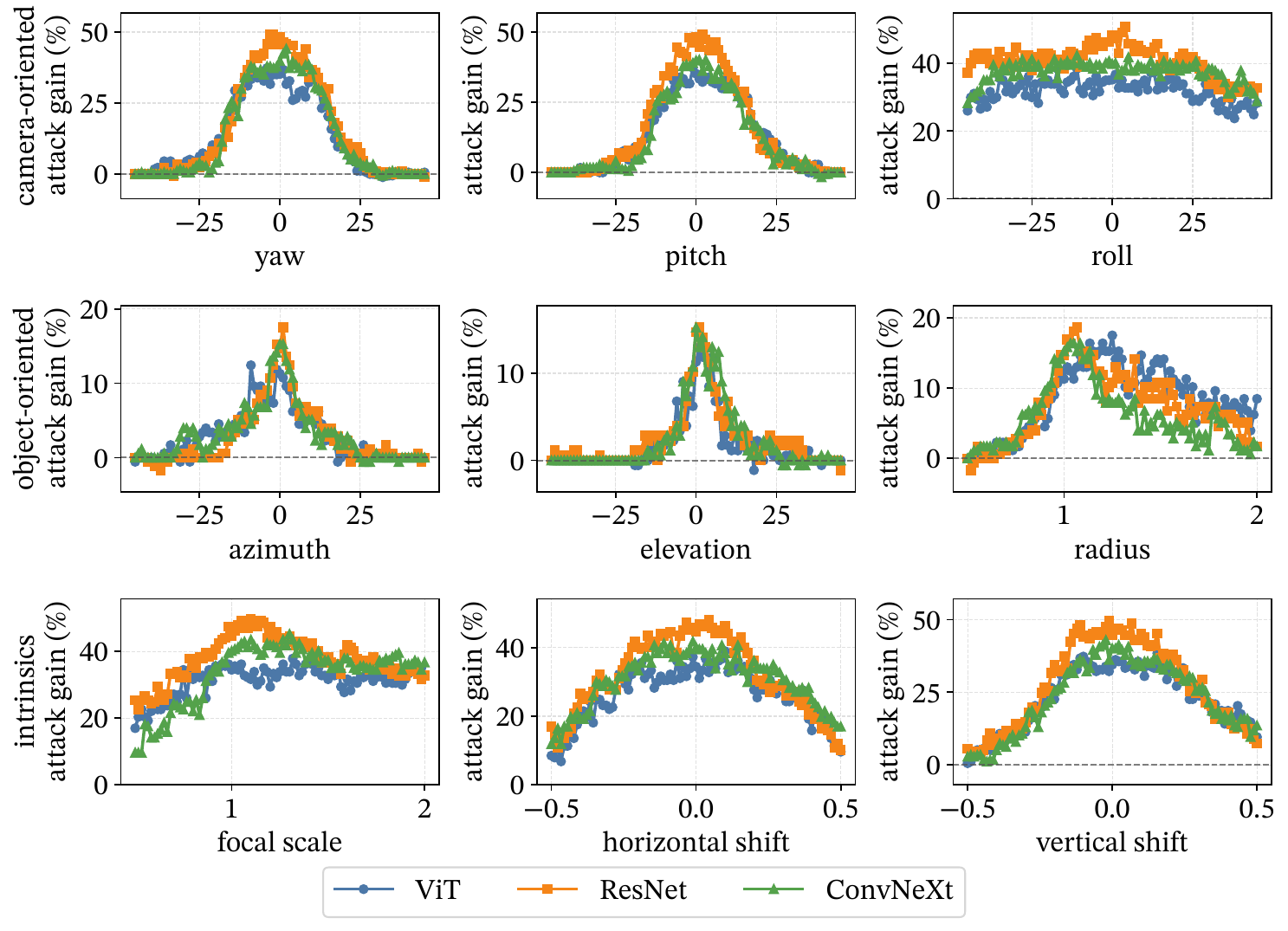}
\caption{Untargeted attack gain across observation dimensions on CO3D.
Attack gain remains concentrated around favorable viewing and distance conditions, showing that untargeted attacks are also scene-dependent.}
\label{fig:co3d_untargeted_ag}
\vspace{-1em}
\end{figure}
\begin{figure}[!tb]
\centering
\includegraphics[width=\linewidth]{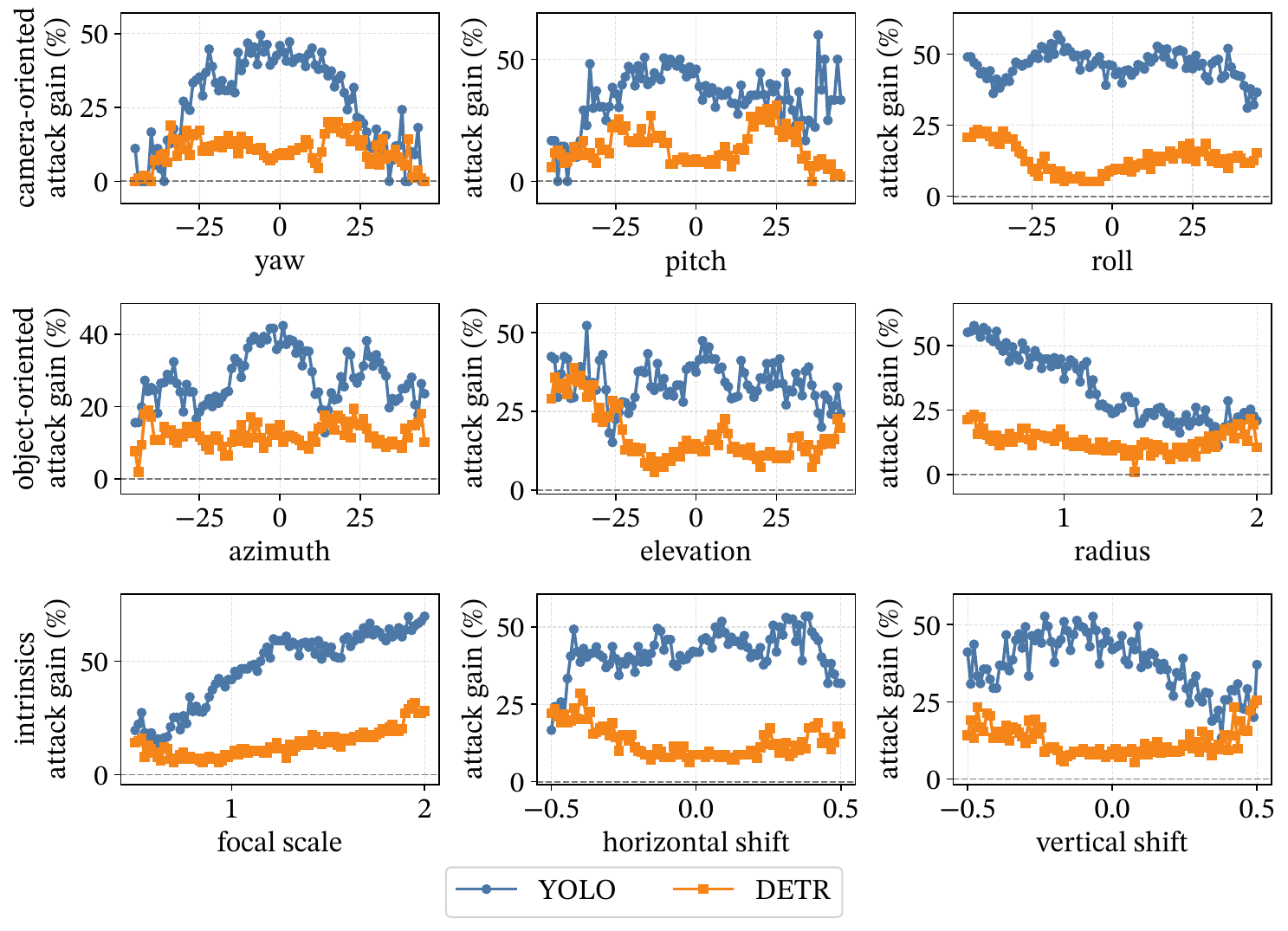}
\caption{Disappearance attack gain across observation dimensions on CO3D.
Both detectors exhibit scene-dependent robustness, but their sensitivity patterns differ under the same rendered view changes.}
\label{fig:co3d_disappear_ag}
\vspace{-1em}
\end{figure}

Figures~\ref{fig:co3d_targeted_asr},~\ref{fig:co3d_untargeted_ag}, and~\ref{fig:co3d_disappear_ag} show scene-conditioned robustness on CO3D across targeted classification, untargeted classification, and disappearance attacks. These figures provide the full one-dimensional sweeps over camera-oriented changes (yaw, pitch, roll), object-oriented changes (azimuth, elevation, radius), and intrinsic changes (focal scale, horizontal shift, vertical shift). The targeted classification plots report ASR, while the untargeted classification and disappearance plots report attack gain to separate adversarial failures from benign failures under difficult views.

\vspace{.3em}
\noindent\textbf{Finding 2.} \textit{Anchor-view success is local and often overestimates deployed robustness.}
For targeted classification, Figure~\ref{fig:co3d_targeted_asr} shows that ASR is highest near the anchor-centered condition and drops as the camera or object-relative viewpoint moves away from the optimized pose. The drop is especially clear for yaw, pitch, azimuth, elevation, radius, focal scale, and image-plane shifts, where changes alter projected patch shape, visible area, scale, or location. This demonstrates that a patch that succeeds in the anchor image does not necessarily remain effective after deployment.
The untargeted and disappearance results in Figures~\ref{fig:co3d_untargeted_ag} and~\ref{fig:co3d_disappear_ag} show broader robustness than targeted attacks, but their attack gain still decreases under larger viewpoint, radius, focal-scale, and image-shift changes. For untargeted classification, attack gain forms concentrated peaks around anchor-centered angular and radius conditions for ViT, ResNet, and ConvNeXt, while performance drops toward large offsets. For disappearance attacks, YOLOv8 and RT-DETR show different absolute gains and sensitivity patterns, but both reveal that the deployed patch does not maintain uniform effectiveness across scene-induced views.

\vspace{.3em}
\noindent\textbf{Finding 3.} \textit{Scene robustness is controlled unevenly by observation dimensions.}
Yaw, pitch, azimuth, elevation, radius, focal scale, and image-plane shifts change change the projected patch shape, visible area, image location, or local context, and therefore produce clear drops in ASR or AG. In contrast, roll has a weaker effect because it largely preserves the projected patch footprint and keeps the patch attached to the same object region. Focal scale shows another distinct pattern: decreasing focal scale zooms out and reduces the patch's pixel support, while increasing focal scale enlarges the projected footprint and improves ASR until the effect saturates or context is cropped. These results show that scene robustness depends on how each observation dimension changes the physical visibility and effective footprint of the deployed patch.

\subsubsection{Waymo Evaluation Results}
\label{sec:eval_waymo}
\begin{figure}[!tb]
\centering
\includegraphics[width=\linewidth]{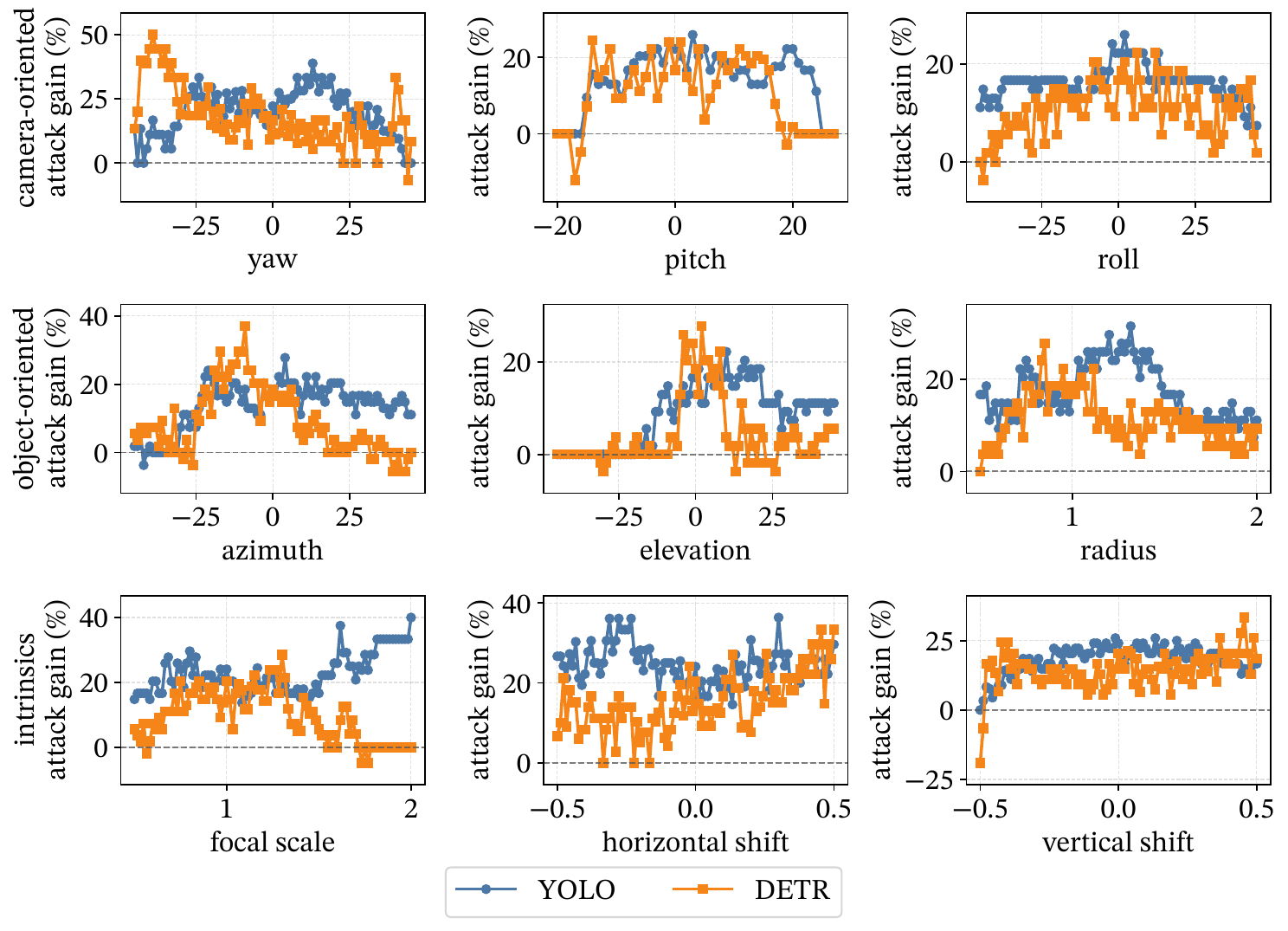}
\caption{Disappearance attack gain across observation dimensions on Waymo.
Driving scenes produce directional and asymmetric operational envelopes because roadside signs are observed along vehicle-induced trajectories.}
\label{fig:waymo_disappear_ag}
\vspace{-1.5em}
\end{figure}

Figure~\ref{fig:waymo_disappear_ag} reports the disappearance attack performance on Waymo. Compared with CO3D, the attack gain is more uneven across view changes. The patch can still increase disappearance rates under favorable camera and scale conditions, but the gain drops more quickly as the target undergoes larger viewpoint, distance, and image-plane changes. This reflects the additional difficulty of driving scenes, where the target is smaller, more frequently viewed from oblique angles, and embedded in more complex backgrounds.

\vspace{.3em}
\noindent\textbf{Finding 4.}
\textit{Driving scenes produce directional, asymmetric operational envelopes.}
Unlike object-centered CO3D captures, Waymo observations are induced by vehicle trajectories, lane geometry, camera mounting, and roadside object placement. In our experiments, the attacked objects are mainly roadside traffic signs, and the selected anchor view is not a frontal canonical view. Therefore, perturbations around the anchor condition are inherently directional: moving the camera in one direction may make the traffic sign more oblique, reduce the visible patch area, or push the target toward the image boundary. This asymmetry is amplified by the driving setting because the vehicle observes the sign along a one-sided trajectory rather than from a uniformly sampled viewing sphere. As a result, several curves in Figure~\ref{fig:waymo_disappear_ag} are not symmetric around the anchor condition. This indicates that the deployed attack envelope in driving scenes is directional rather than symmetric.

\vspace{.3em}
\noindent\textbf{Finding 5.}
\textit{Operational envelopes are model-dependent, even under the same scene trajectory.}
The detector comparison further shows that YOLOv8 and RT-DETR exhibit different sensitivity patterns under the same rendered view changes. A plausible explanation is that the two detectors rely on different localization and feature aggregation mechanisms. YOLOv8 uses dense local predictions and can be more sensitive to local texture disruption near the sign region, while RT-DETR uses query-based global reasoning and may rely more on contextual and object-level cues. Therefore, the same projected patch footprint can affect the two detectors differently as distance, scale, and viewpoint change.

This result also explains why single-view evaluation is unreliable for scene-robustness measurement. Two detectors may exhibit similar attack outcomes at the anchor view, but their vulnerability regions can diverge once distance, scale, and viewpoint change. Thus, anchor-view success alone cannot determine whether a deployed patch has a broad, model-independent operational envelope.

\subsubsection{Impact of Patch Size and Shape}
\label{sec:eval_patch_size_shape}
\begin{figure}[!tb]
\centering
\includegraphics[width=0.9\linewidth]{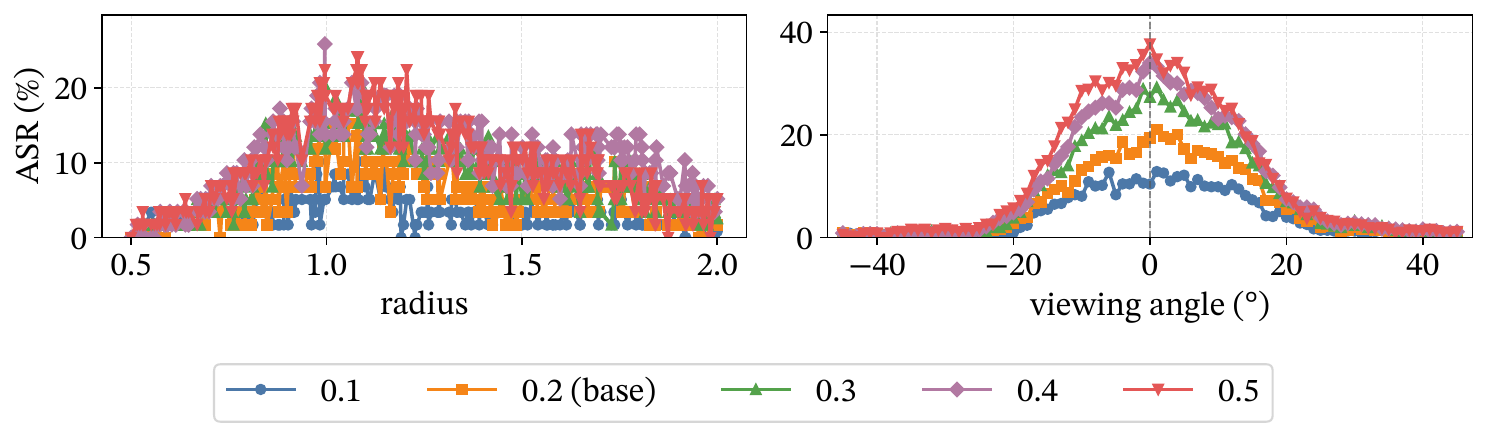}
\caption{Impact of adversarial patch size.}
\label{fig:patch_size}
\vspace{-1em}
\end{figure}

\begin{figure}[!tb]
\centering
\includegraphics[width=\linewidth]{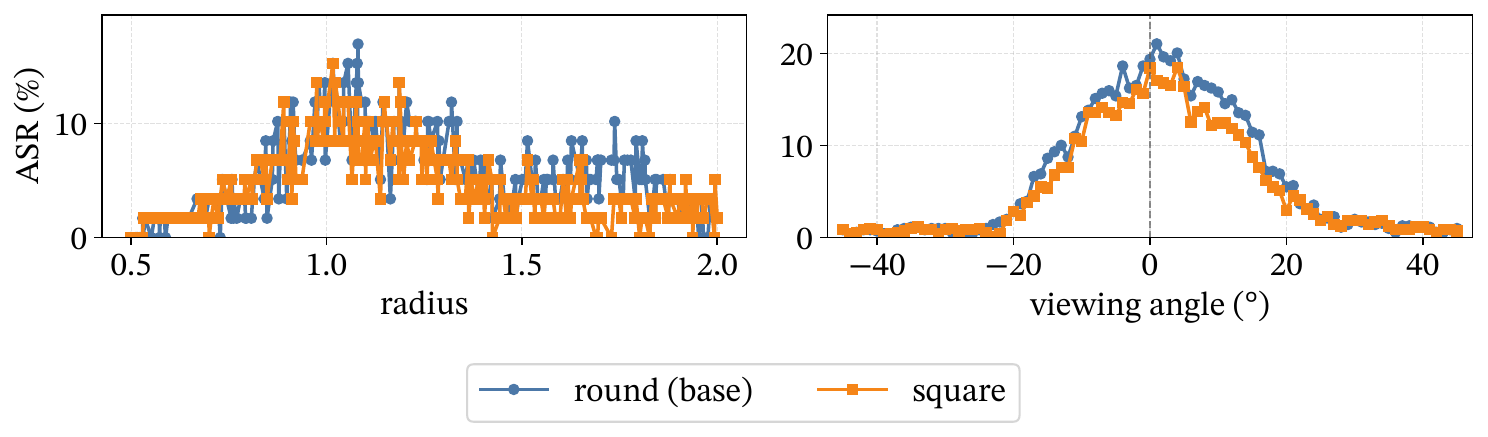}
\caption{Impact of adversarial patch shape.}
\label{fig:patch_shape}
\end{figure}

Figure~\ref{fig:patch_size} shows that larger patches generally achieve higher ASR because they preserve a larger projected adversarial footprint. However, all sizes still degrade outside the favorable radius range and under oblique viewing. Figure~\ref{fig:patch_shape} shows a similar pattern for patch shape: round and square patches have comparable envelopes, with only modest differences in peak ASR. Thus, simple patch-geometry changes affect the strength of the signal but do not remove the underlying distance and viewing-angle failure modes.

\vspace{.3em}
\noindent\textbf{Finding 6.}
\textit{Patch footprint increases attack strength but does not remove geometric failure modes.}
Increasing patch size generally increases ASR by preserving a larger projected adversarial footprint. However, Figure~\ref{fig:patch_size} shows that all evaluated sizes still degrade outside favorable radius and viewing-angle ranges. Similarly, Figure~\ref{fig:patch_shape} shows that round and square patches produce similar envelope shapes. Thus, simple patch geometry changes affect signal strength, but the deployed envelope remains constrained by visibility, projection, and viewing geometry.

\subsection{Effectiveness of Physical Attack Techniques}
\label{sec:eval:attack-components}

We further use \proposed to study whether common physical attack techniques improve scene robustness. Each patch variant is optimized once from the anchor view, embedded into the reconstructed scene using the same APSE procedure, and evaluated over the same controlled view space. Thus, the comparison measures whether an adversarial patch technique improves the deployed patch's operational envelope, rather than only improving anchor-view success.
These are attack-design comparisons, not APSE component ablations: the evaluation framework is kept fixed. All variants use the same targeted classification attack on the CO3D dataset.

\subsubsection{EOT magnitude.}
\begin{figure}[!tb]
\centering
\includegraphics[width=0.9\linewidth]{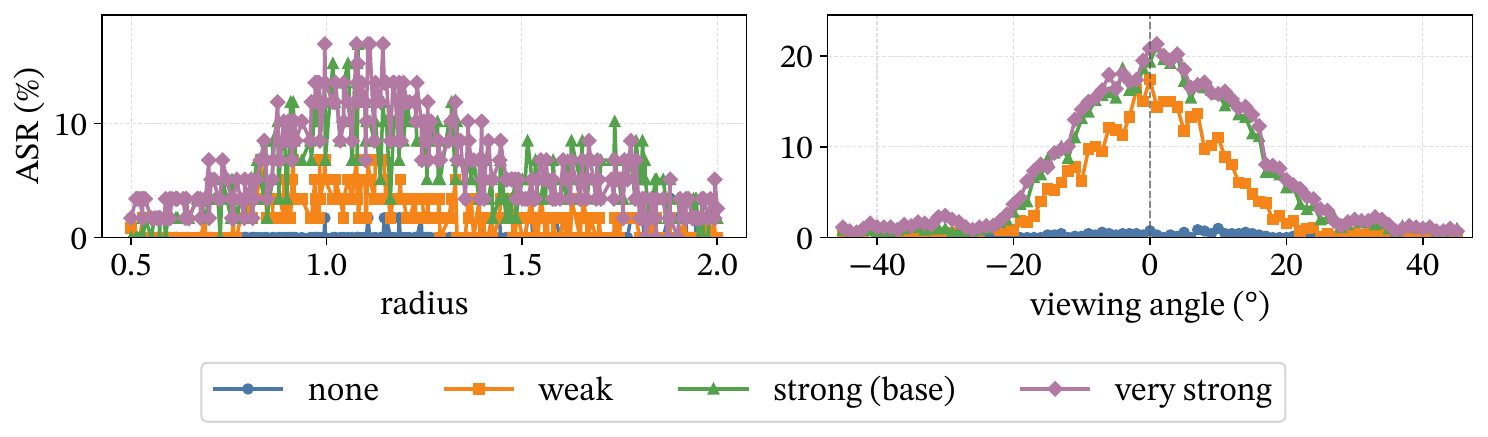}
\caption{Effectiveness of EOT magnitude.}
\label{fig:ab_eot_magnitude}
\vspace{-1em}
\end{figure}
Expectation over Transformation (EOT) optimizes the patch under sampled transformations so that it remains effective beyond the anchor view.
We vary the magnitude of EOT transformations. Detailed settings are provided in Appendix~\ref{sec:app:eot_settings}. As shown in Figure~\ref{fig:ab_eot_magnitude}, without EOT, the deployed patch has almost no scene robustness. Weak EOT improves ASR slightly but remains insufficient across both radius and viewing angle. Strong EOT substantially expands the operational envelope, producing higher ASR around the favorable radius range and a wider effective viewing-angle region. Very strong EOT gives comparable or slightly higher ASR near the anchor-centered region, but its gain over strong EOT is not consistent. This suggests a threshold effect: EOT must cover sufficiently large transformations to match scene-induced variation, but excessively increasing the range does not guarantee proportional improvement.

\subsubsection{EOT components.}
\begin{figure}[!tb]
\centering
\includegraphics[width=0.9\linewidth]{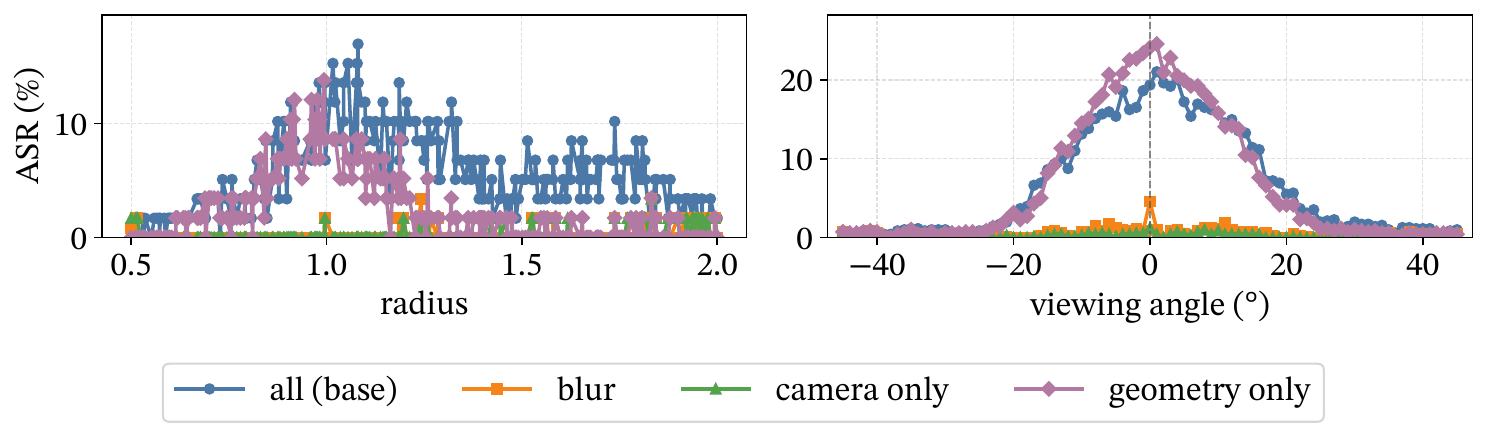}
\vspace{-1em}
\caption{Effectiveness of EOT components.}
\label{fig:ab_eot_components}
\vspace{-1em}
\end{figure}

We compare three EOT component variants: geometry, blur, and camera perturbations. The geometry component applies crop, perspective, rotation, translation, and scale transforms; the blur component applies blur, noise, and JPEG compression; and the camera-only setting applies camera downscaling and quantization.
Figure~\ref{fig:ab_eot_components} shows that geometry is the dominant EOT component. The geometry-only variant achieves the strongest or near-strongest ASR around the anchor-centered viewing direction and largely matches full EOT in the viewing-angle sweep. In contrast, blur-only and camera-only perturbations remain close to zero across most radius and viewing-angle conditions, showing that generic image degradation alone does not provide meaningful scene coverage. Along the radius dimension, full EOT gives broader support than geometry alone, suggesting that combining perturbations can improve distance stability. Overall, the main gain comes from geometry-aware transformations that expose the patch to pose, scale, and projection changes during optimization.

\subsubsection{Non-printability score (NPS) regularization.}
\begin{figure}[!t]
\centering
\includegraphics[width=0.9\linewidth]{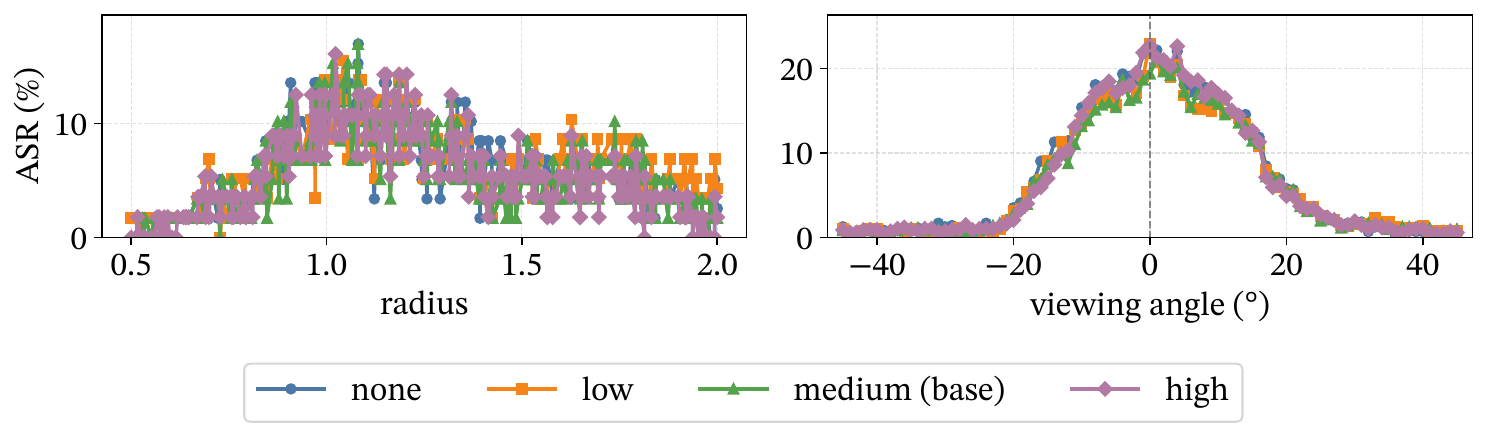}
\caption{Effectiveness of non-printability score (NPS) regularization.}
\label{fig:ab_nps}
\vspace{-1em}
\end{figure}
The NPS encourages the optimized patch to use printable colors. We set the weight for no NPS to $0.0$, low NPS to $0.02$, the baseline to $0.1$, and high NPS to $0.5$. 
As shown in Figure~\ref{fig:ab_nps}, the curves under different NPS weights are close across radius and viewing angle. High NPS occasionally yields slightly higher ASR near the favorable radius range or around the anchor-centered viewing direction, but the improvement is not systematic. NPS also does not noticeably shrink the operational envelope, suggesting that printable-color constraints do not necessarily weaken scene-level attack performance in our setting. However, NPS does not expand the envelope by itself. It mainly serves as a physical deployability constraint rather than a primary robustness mechanism.

\subsubsection{Total variation (TV) regularization.}
\begin{figure}[!t]
\centering
\includegraphics[width=0.9\linewidth]{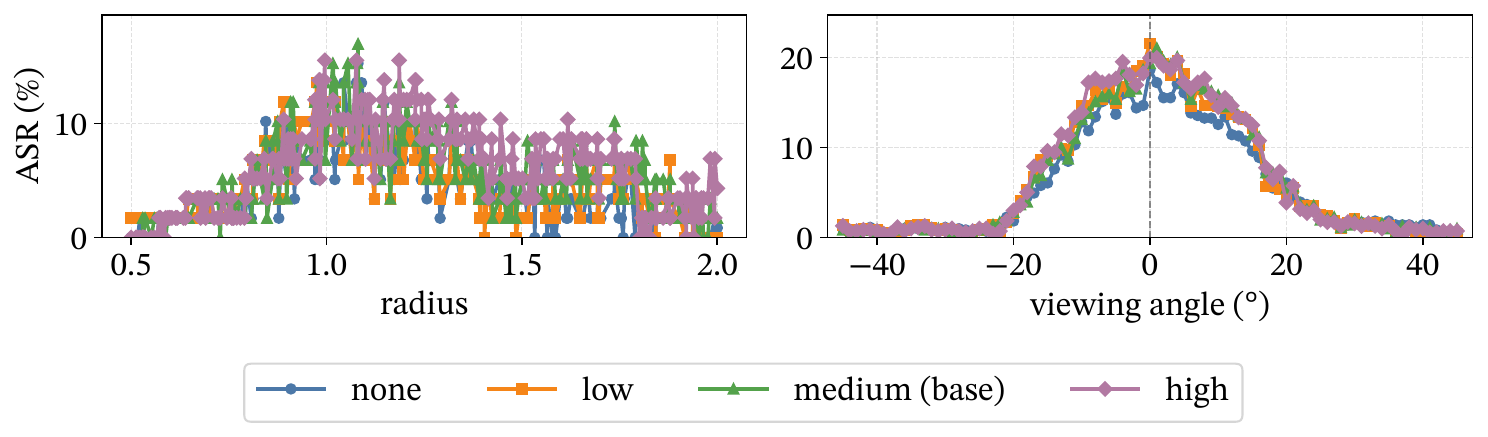}
\caption{Effectiveness of total variation (TV) regularization.}
\label{fig:ab_tv}
\vspace{-1em}
\end{figure}

TV regularization encourages spatial smoothness in the optimized patch. We set the weight for no TV to $0.0$, low TV to $0.5$, the baseline to $2.5$, and high TV to $5.0$. As shown in Figure~\ref{fig:ab_tv}, TV has a modest but more visible effect than NPS in some regions. Higher TV weights tend to improve ASR around the favorable radius range and maintain slightly higher ASR as the camera moves away from the anchor distance. The viewing-angle curves also show small gains near the anchor-centered region, but the overall envelope shape remains similar across TV settings. Thus, spatial smoothness can improve local stability, but it does not remove the main geometric failure modes caused by distance and oblique viewing.

\vspace{.3em}
\noindent\textbf{Finding 7.}
\textit{Geometry-aware EOT expands the envelope, with diminishing returns beyond a sufficient range.}
Among the physical attack techniques we evaluate, EOT is the only one that substantially expands the scene-level region where the deployed patch remains effective. In particular, geometry-aware EOT is essential because it directly models the pose, scale, and projection changes that occur after deployment. In contrast, NPS and TV regularization do not fundamentally change where the patch succeeds or fails in the scene. NPS mainly constrains printability, while TV encourages spatial smoothness and provides mild local stability in some regions. However, neither overcomes the main geometric sources of scene-level failure. This result reinforces the need for operational-envelope evaluation: a technique that improves physical plausibility or anchor-view optimization does not necessarily improve deployed robustness across realistic viewing conditions.

\subsubsection{Impact of $\ell_2$ Regularization}
\label{sec:eval_l2}

\begin{figure}[!tb]
\centering
\includegraphics[width=0.9\linewidth]{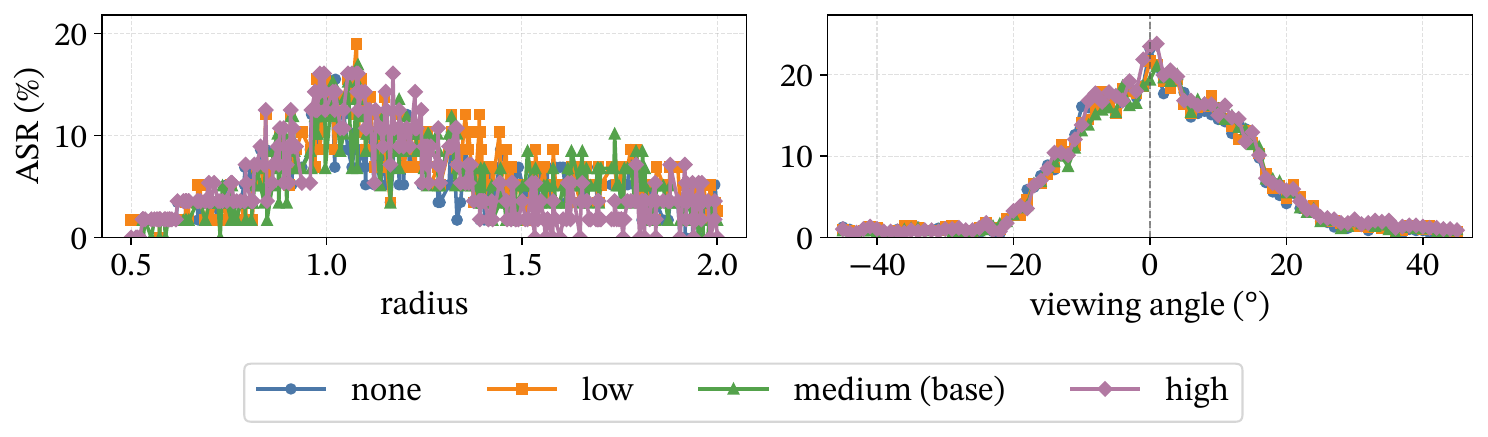}
\caption{Impact of $\ell_2$ regularization.}
\label{fig:ab_l2}
\vspace{-1em}
\end{figure}

$\ell_2$ regularization constrains the pixel-level magnitude of the optimized patch and can make the pattern less visually aggressive. However, strong magnitude constraints may also suppress adversarial features needed for attack success. This tradeoff helps explain why recent physical patch attacks often emphasize deployment-oriented objectives, such as EOT, printability constraints, total variation, and naturalistic priors, rather than treating $\ell_2$ as the main physical-robustness mechanism~\cite{eykholt2018robust,thys2019fooling,hu2021naturalistic,zhu2023tpatch}.

Our scene-conditioned evaluation clarifies this tradeoff. Figure~\ref{fig:ab_l2} shows that adding $\ell_2$ regularization does not substantially degrade deployed attack performance within the operational envelope.
The curves for no, low, medium, and high $\ell_2$ regularization largely overlap, indicating that the operational envelope is not determined by the pixel-level magnitude constraint. Although $\ell_2$ changes the optimized patch appearance and may slightly affect anchor-view performance, its effect is much weaker than scene-induced factors such as projected footprint, visibility, and viewing geometry.

\vspace{.3em}
\noindent\textbf{Finding 8.}
\textit{$\ell_2$ regularization is not a scene-level robustness bottleneck for deployed patches.}
Recent physical attacks often prioritize deployment-oriented robustness terms, such as EOT, NPS, and TV, over explicit $\ell_2$ regularization. While this design trend improves physical reliability, it may leave the patch magnitude less constrained and make the optimized patch visually less stealthy. Our results show that adding $\ell_2$ does not meaningfully shrink the operational envelope in our evaluated setting. This suggests that $\ell_2$ can be used to constrain patch magnitude without substantially sacrificing scene-level attack robustness.

\section{Discussion and Limitations}
\label{sec:discussion_limitations}

\subsection{Implications for Evaluation and Attack Design}
\proposed suggests that physical patch risk should be evaluated as a scene-conditioned property rather than a single-image outcome. For evaluators, the operational envelope identifies where a fixed deployed patch remains effective and where image-centric evaluation may overestimate or underestimate risk. For attack design, our results separate deployment-oriented techniques into different roles: geometry-aware EOT expands the envelope, NPS and TV mainly affect printability or smoothness, and $\ell_2$ regularization can constrain patch magnitude without shrinking the evaluated envelope. Thus, physical robustness should be optimized against the deployment variations that actually dominate the target scene.

\subsection{Limitations}

\noindent\textbf{Attack scope.}
\proposed evaluates fixed, printable, surface-attached 2D patches. It does not evaluate adversarial 3D objects, shape modifications, volumetric perturbations, time-varying displays. This is intentional: our goal is to measure how an existing adversarial patch behaves after deployment, not to synthesize a stronger scene-specific 3D attack. Three-dimensional attacks require different assumptions about fabrication, material properties, geometry changes, and scene-specific optimization, and therefore require a separate evaluation protocol.

\vspace{.5em}\noindent\textbf{Reconstruction and pose quality.}
\proposed relies on the fidelity of the reconstructed 3DGS scene and the accuracy of camera pose estimation. Poorly reconstructed targets due to sparse data can affect both visual fidelity and attack behavior. Our matched physical validation partially calibrates this issue, but rendered evaluation remains an approximation of physical deployment.

\vspace{.5em}\noindent\textbf{Physical and system factors beyond geometry.}
Our evaluation focuses on dominant geometric and imaging variables, including viewpoint, distance, projected scale, camera pose, and camera intrinsics. Real deployments may also involve lighting changes, shadows, weather, motion blur, print artifacts, surface reflectance, temporal dynamics, and sensor-specific processing. These factors are complementary to the scene-geometry effects studied here and can be incorporated in future extensions. In addition, our experiments evaluate RGB classifiers and detectors, not full autonomous-driving stacks with LiDAR, radar, or sensor fusion.

\section{Conclusion}
\label{sec:conclusion}

This paper presents \proposed, a scene-grounded framework for evaluating the robustness of deployed adversarial patches. Rather than reporting a single-view success rate, \proposed measures the patch's operational envelope: the region of scene-induced observation conditions under which the attack remains effective. To enable this measurement, APSE lifts a single anchor-view attack into a reconstructed 3DGS scene while preserving patch appearance, clean-scene content, and target-surface attachment.

Our evaluation shows that \proposed preserves physical attack behavior under matched physical and rendered observations and reveals scene-dependent failure modes missed by image-centric testing. Across object-centered and driving scenes, anchor-view success often does not imply broad deployment robustness. Attack effectiveness varies with distance, viewpoint, projected footprint, and scene geometry. We further find that geometry-aware EOT expands the operational envelope more effectively than common regularizers. Notably, $\ell_2$ regularization can constrain patch magnitude without substantially shrinking the evaluated envelope, suggesting that visual magnitude and scene-level robustness are not always in direct conflict. Overall, we conclude that physical patch risk is better understood as an envelope, not a point estimate. By making this envelope measurable, \proposed provides a more faithful basis for evaluating physical adversarial risk in real scenes.

\newpage
\bibliographystyle{ACM-Reference-Format}
\bibliography{bib/3dgs.bib,bib/adv.bib,bib/ai,bib/new}

\newpage
\appendix

\section{Related Work}
\label{sec:related}

\subsection{Physical Adversarial Attacks}

Physical adversarial attacks aim to generate perturbations that remain effective after real-world capture. Early studies showed that adversarial examples can survive printing and camera capture~\cite{kurakin2018adversarial}, and adversarial patches further demonstrated that localized, printable patterns can induce targeted failures when placed in a scene~\cite{brown2017adversarial}. Subsequent work extended these attacks to object detection, including DPatch~\cite{liu2018dpatch}, robust physical perturbations for traffic signs~\cite{eykholt2018robust}, and ShapeShifter for attacking detectors under physical transformations~\cite{chen2018shapeshifter}, as well as person-detector attacks using physical patches or wearable patterns~\cite{thys2019fooling,xu2020advtshirt}.

A central challenge in physical attacks is robustness to environmental variation, including viewpoint, lighting, distance, printing artifacts, and sensing noise. To address this, prior work commonly uses Expectation over Transformation (EOT)~\cite{athalye2018synthesizing}, which models physical variability during optimization by sampling transformations such as rotation, scale, translation, and illumination. Physical constraints and regularizers, including object masking, total variation, and the Non-Printability Score (NPS) proposed in RP2~\cite{eykholt2018robust}, further ensure that perturbations remain fabricable and stable under printing.

More recent work extends physical attacks beyond rigid objects to non-rigid and context-dependent settings. Adversarial clothing and camouflage attacks explicitly model pose variation, cloth deformation, and texture realism to attack person detectors in the physical world~\cite{hu2022adversarial,hu2023physically,guesmi2024dap}. Other studies explore practical deployment constraints, including context-aware patches~\cite{zhang2023capatch}, temporal or trajectory-aware designs~\cite{zhu2023tpatch}, and distance-aware attacks such as FDA~\cite{cheng2024fda}, which evaluates attack effectiveness across viewing distances. In parallel, work on stealthiness and naturalness introduces adversarial camouflage and naturalistic patch designs that blend into the scene~\cite{hu2021naturalistic}. Beyond planar patches, physical adversarial artifacts have also been extended to accessories and 3D objects, including adversarial eyeglasses and meshes~\cite{sharif2016accessorize,xiao2019meshadv}.

Despite these advances in attack design, evaluation remains limited. Most studies report results under a small number of sampled conditions, such as discrete viewpoints or distances, rather than characterizing how attack effectiveness varies across the continuous observation conditions induced by a realistic scene. This motivates a more systematic, scene-grounded evaluation framework.

\subsection{Adversarial Evaluation in Real and Simulated Scenes}

A growing body of work focuses on evaluating adversarial attacks beyond small-scale demonstrations. Dataset-driven approaches collect adversarial examples in real environments. APRICOT~\cite{braunegg2020apricot} provides over 1,000 annotated images of printed adversarial patches captured in natural scenes, while REAP~\cite{hingun2023reap} constructs a large-scale benchmark on real driving images with geometry- and lighting-aware patch insertion. PAN~\cite{li2023pan} complements these efforts by evaluating the perceptual naturalness of adversarial attacks using human studies. These benchmarks provide important evidence that physical patches can affect real perception systems, complementing attack-generation studies on traffic signs, object detectors, and person detectors~\cite{eykholt2018robust,chen2018shapeshifter,hu2022adversarial,hu2023physically}.
Despite their realism, these datasets remain inherently image-centric. Evaluation is performed on fixed captured images or independently transformed instances, limiting the ability to systematically vary viewpoint, camera trajectory, or scene configuration for the same deployed patch within a single scene.

Simulator-based approaches offer stronger control over evaluation conditions. AttackScenes~\cite{huang2020universal} provides a benchmark with multiple viewpoints and lighting conditions in virtual scenes. In autonomous driving, CARLA-based frameworks enable configurable evaluation across camera poses, weather, and traffic conditions~\cite{nesti2022ss,nesti2024carlagear,lan2024carlaa3}. These systems support repeatable sweeps over environmental variables but rely on simulator-native scenes, where geometry, textures, and assets are manually constructed and may not correspond to a specific real deployment site.
\textbf{In contrast}, our approach reconstructs real-world scenes from captured images and enables controlled evaluation within those scenes, bridging the gap between realism and controllability.

Our operational-envelope view is also related to scenario-based testing in autonomous driving, where system behavior is evaluated over parameterized operating conditions rather than isolated examples~\cite{neurohr2020fundamental,cai2022survey,nalic2020scenario}. Unlike these works, our condition space is attack-centric: the outcome is whether a fixed deployed patch remains effective under each observation condition.

\subsection{3D Gaussian Splatting and Scene Editing}

Scene reconstruction and neural rendering provide a natural substrate for controllable evaluation in real scenes. Classical structure-from-motion and multiview stereo estimate camera poses and scene geometry from image collections~\cite{schonberger2016structure,schonberger2016pixelwise}. Neural rendering methods, including LLFF, NeRF, and efficient explicit or factorized radiance-field representations, enable novel-view synthesis from calibrated captures~\cite{mildenhall2019local,mildenhall2021nerf,fridovich2022plenoxels,chen2022tensorf,muller2022instant,barron2022mip}. Recent advances in 3D Gaussian Splatting (3DGS)~\cite{kerbl20233d,Huang2DGS2024,wu2024recent} further enable efficient reconstruction and high-quality rendering from multi-view images.

However, directly applying 3DGS to adversarial evaluation is non-trivial. Standard 3DGS prioritizes photorealistic rendering and does not guarantee preservation of the fine-grained perturbations that determine adversarial effectiveness. Existing 3DGS editing and generation methods support object insertion, texture transfer, instruction-guided editing, and efficient 3D content creation~\cite{chen2024gaussianeditor,wang2024gaussianeditor,tang2023dreamgaussian,xiao2025localizedgs,galerne2025sgsst,cao20263dot}. These methods focus on semantic plausibility or visual quality, often relying on appearance-based objectives or generative priors, which can alter the subtle pixel-level structures critical for adversarial attacks.

As a result, existing 3DGS-based editing techniques are not suitable for evaluating adversarial patches, where preserving attack-critical perturbations across viewpoints is essential.

\subsection{3DGS-based Adversarial Attacks}
Recent work explores the use of 3DGS for generating adversarial attacks. PGA~\cite{lou2025pga} leverages 3DGS to optimize multi-view robust adversarial camouflage on reconstructed objects. Similarly, 3DGAA~\cite{zhang20253dgaa}, BiTAA~\cite{zhang2025bitaa}, and GSAttack~\cite{meng2025gaussian} use 3DGS-style scene representations to generate or optimize multi-view adversarial objects. These approaches demonstrate that 3DGS can effectively support adversarial perturbations that generalize across viewpoints.

Our work differs fundamentally in both objective and setting. Rather than generating a new 3D adversarial object or optimizing a stronger multi-view attack, we evaluate a fixed adversarial patch produced by an external 2D attack pipeline. \proposed uses 3DGS as a scene-grounded evaluation substrate to measure how the deployed patch's effectiveness varies with viewpoint, distance, projected scale, and scene context. This evaluation objective is not addressed by prior 3DGS-based attack-generation methods.

\section{\proposed Visualizations.}
This appendix provides qualitative examples of the rendered scenes used in our validation and evaluation. Unless otherwise noted, each triplet shows the clean anchor image, the clean 3DGS render, and the adversarial 3DGS render.

\subsection{Physical validation examples}
\begin{figure}[!h]
\centering
\begin{subfigure}[t]{0.32\linewidth}
\centering
\includegraphics[width=\linewidth]{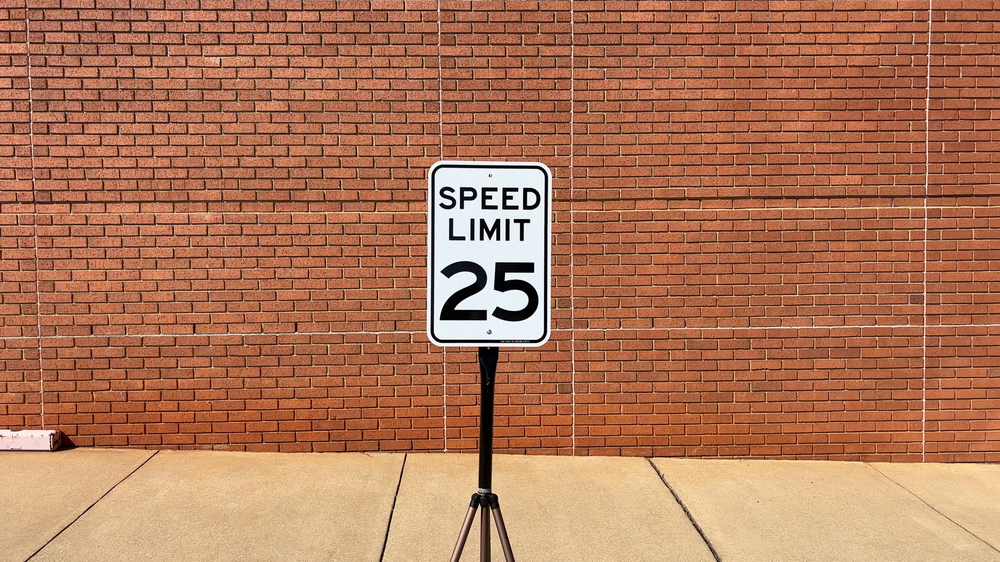}
\end{subfigure}
\begin{subfigure}[t]{0.32\linewidth}
\centering
\includegraphics[width=\linewidth]{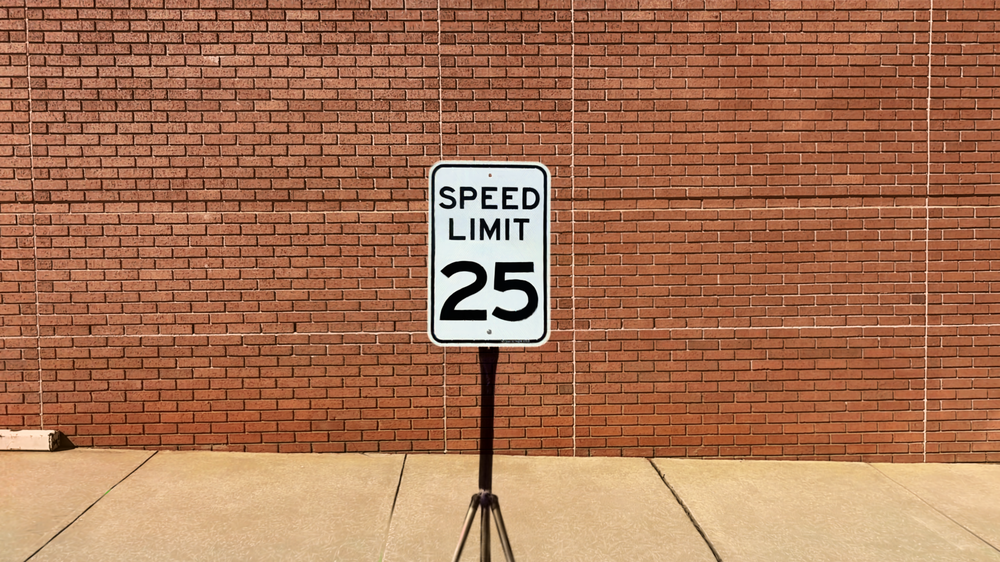}
\end{subfigure}
\begin{subfigure}[t]{0.32\linewidth}
\centering
\includegraphics[width=\linewidth]{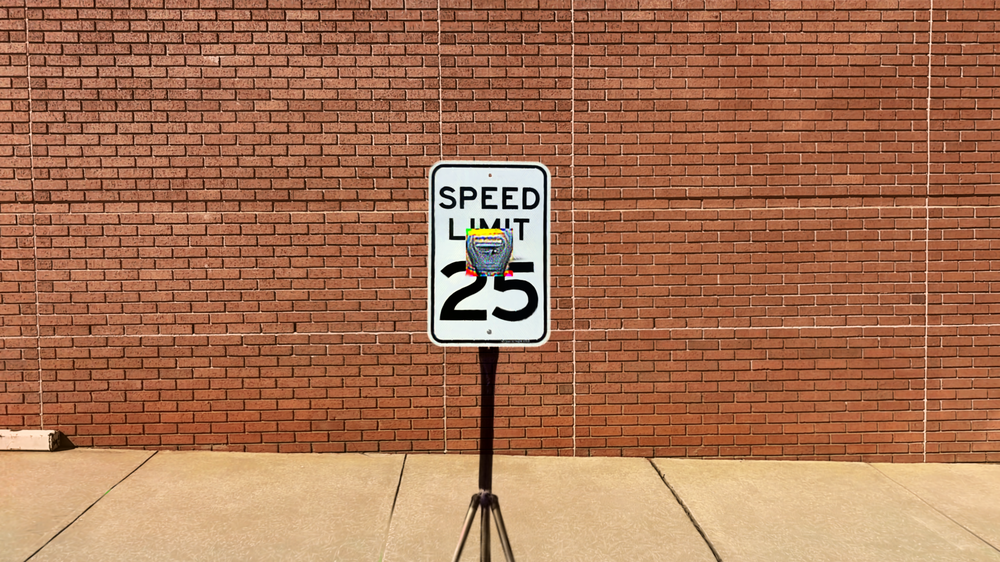}
\end{subfigure}
\caption{Physical validation example: speed limit sign.}
\label{fig:phys_speed}
\end{figure}

\begin{figure}[!h]
\centering
\begin{subfigure}[t]{0.32\linewidth}
\centering
\includegraphics[width=\linewidth]{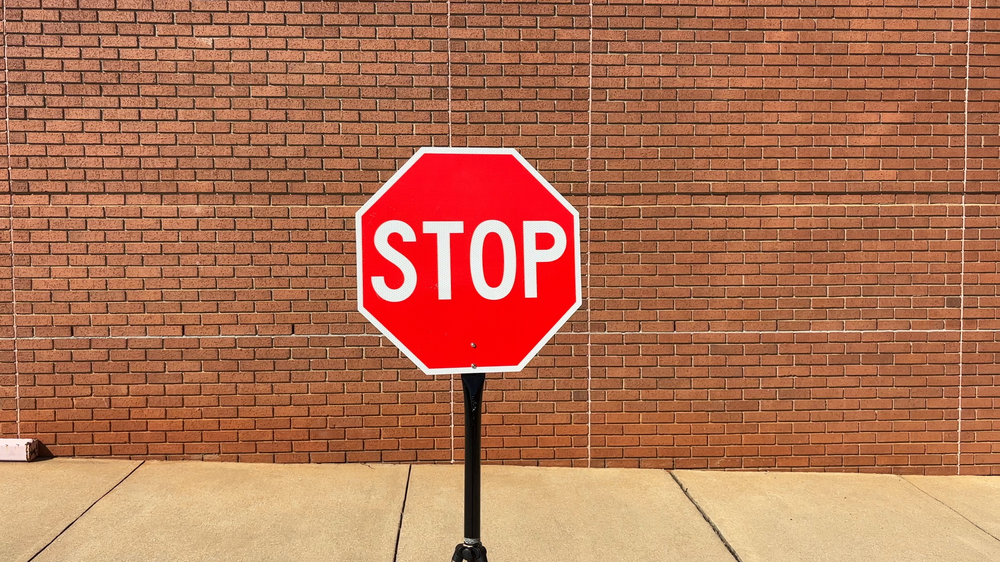}
\end{subfigure}
\begin{subfigure}[t]{0.32\linewidth}
\centering
\includegraphics[width=\linewidth]{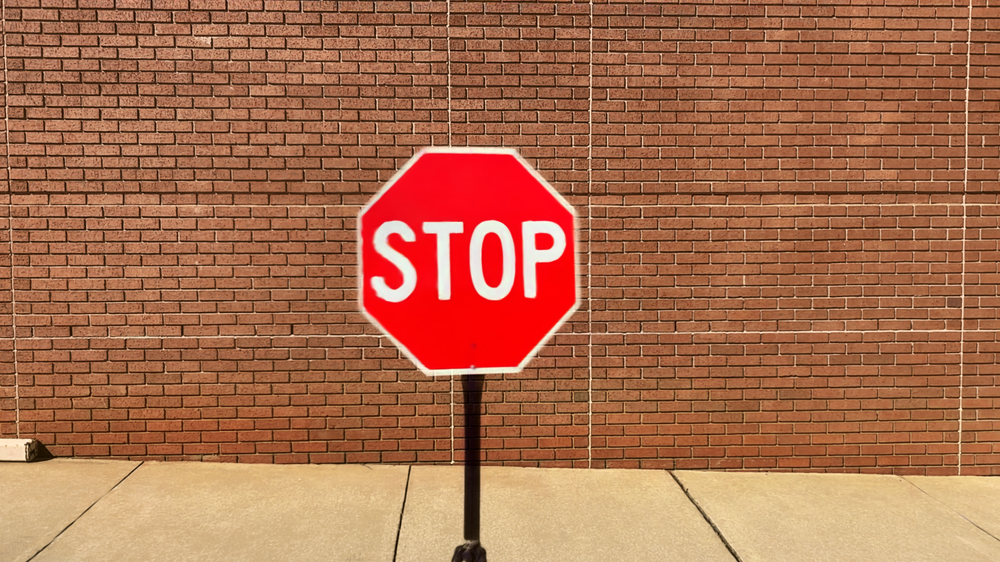}
\end{subfigure}
\begin{subfigure}[t]{0.32\linewidth}
\centering
\includegraphics[width=\linewidth]{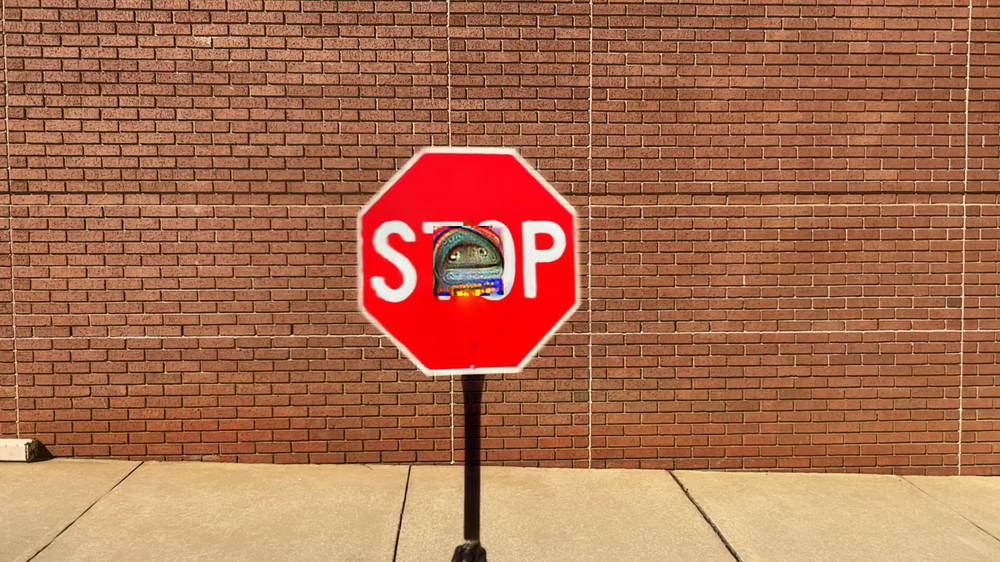}
\end{subfigure}
\caption{Physical validation example: stop sign.}
\label{fig:phys_stop}
\end{figure}

\begin{figure}[!h]
\centering
\begin{subfigure}[t]{0.32\linewidth}
\centering
\includegraphics[width=\linewidth]{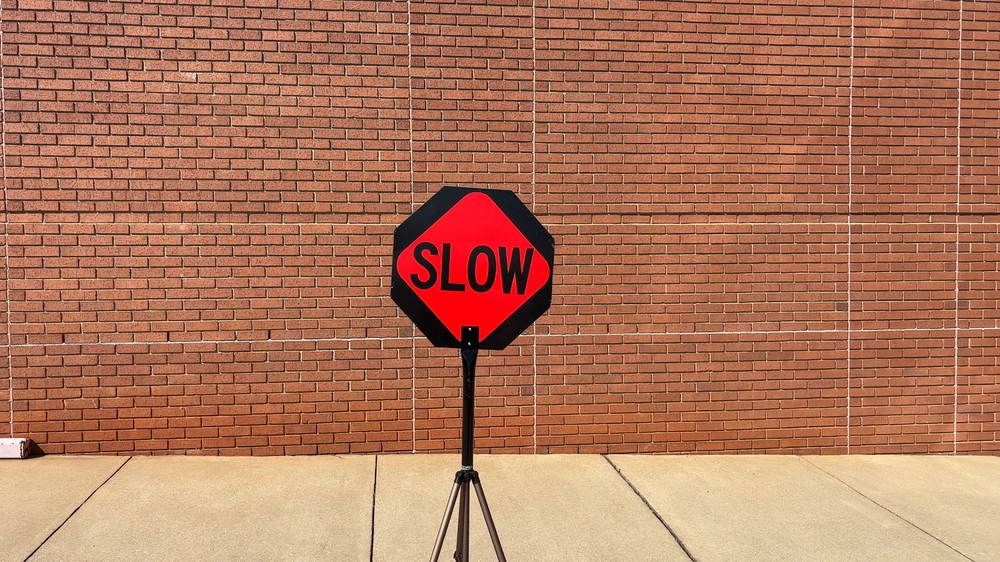}
\end{subfigure}
\begin{subfigure}[t]{0.32\linewidth}
\centering
\includegraphics[width=\linewidth]{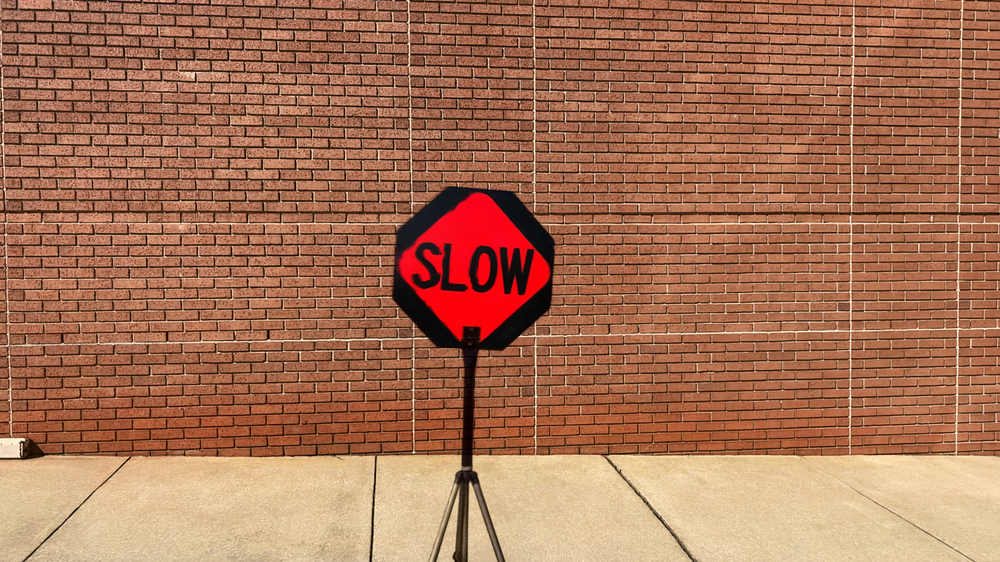}
\end{subfigure}
\begin{subfigure}[t]{0.32\linewidth}
\centering
\includegraphics[width=\linewidth]{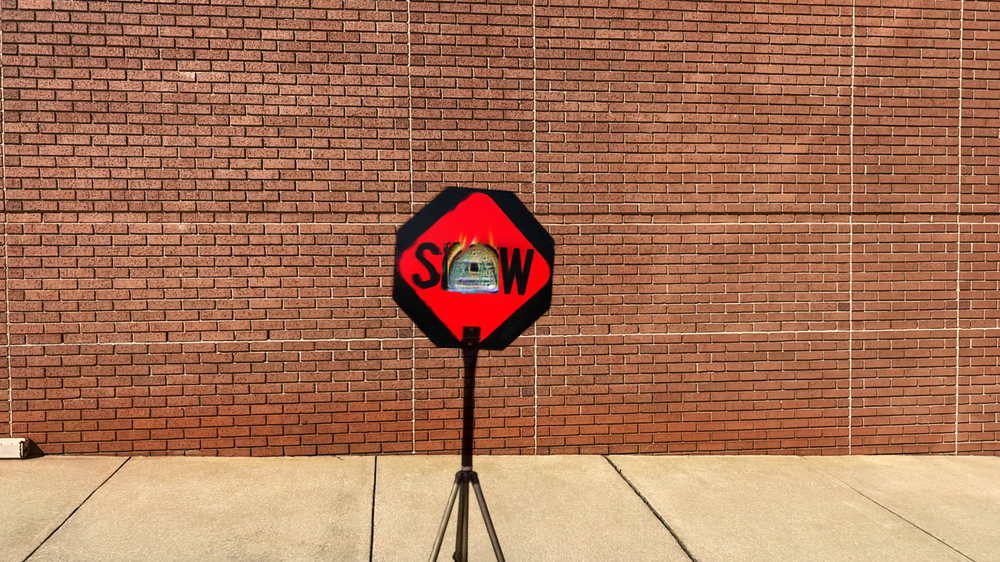}
\end{subfigure}
\caption{Physical validation example: slow sign.}
\label{fig:phys_slow}
\end{figure}

\begin{figure}[!h]
\centering
\begin{subfigure}[t]{0.32\linewidth}
\centering
\includegraphics[width=\linewidth]{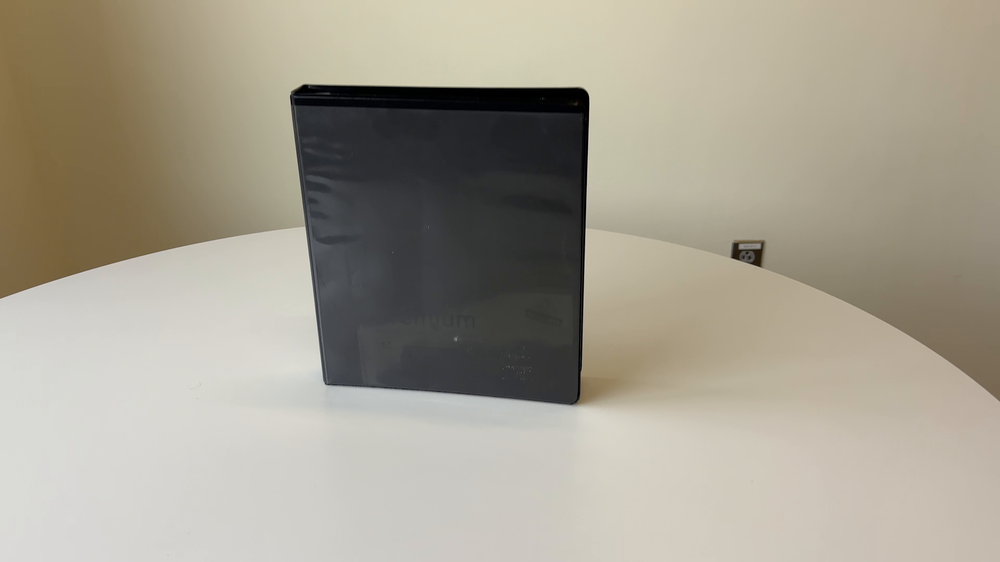}
\end{subfigure}
\begin{subfigure}[t]{0.32\linewidth}
\centering
\includegraphics[width=\linewidth]{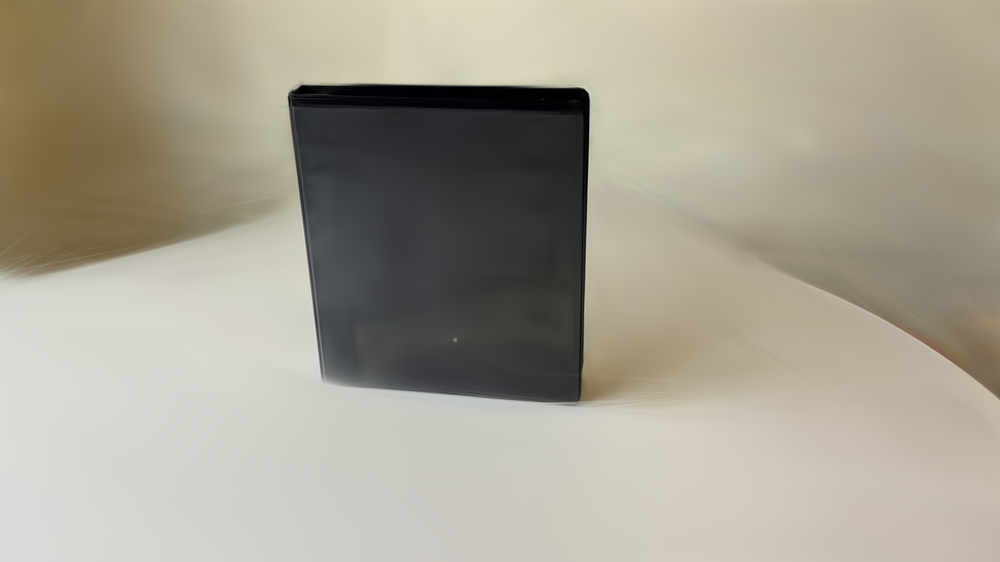}
\end{subfigure}
\begin{subfigure}[t]{0.32\linewidth}
\centering
\includegraphics[width=\linewidth]{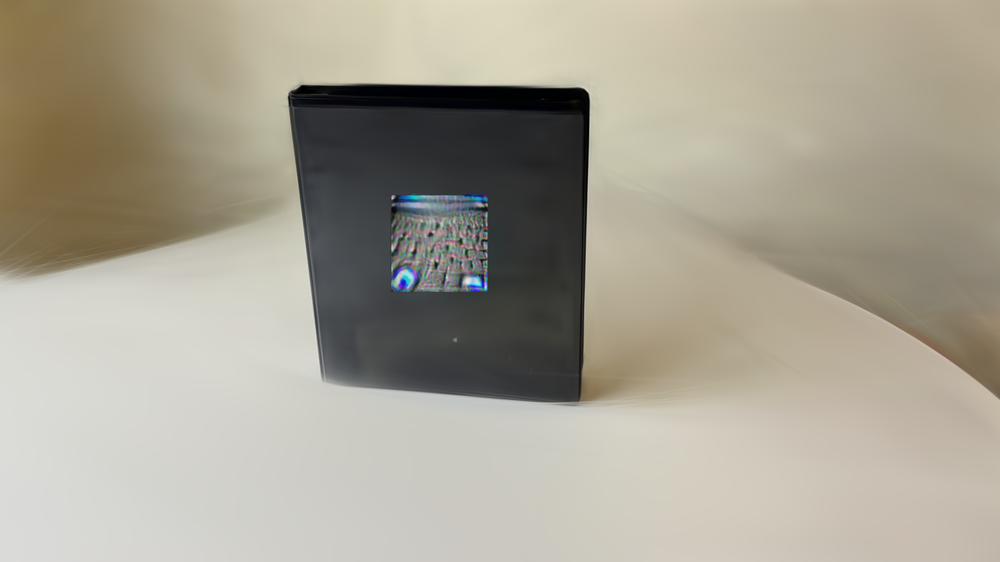}
\end{subfigure}
\caption{Physical validation example: binder.}
\label{fig:phys_binder}
\end{figure}

\begin{figure}[!h]
\centering
\begin{subfigure}[t]{0.32\linewidth}
\centering
\includegraphics[width=\linewidth]{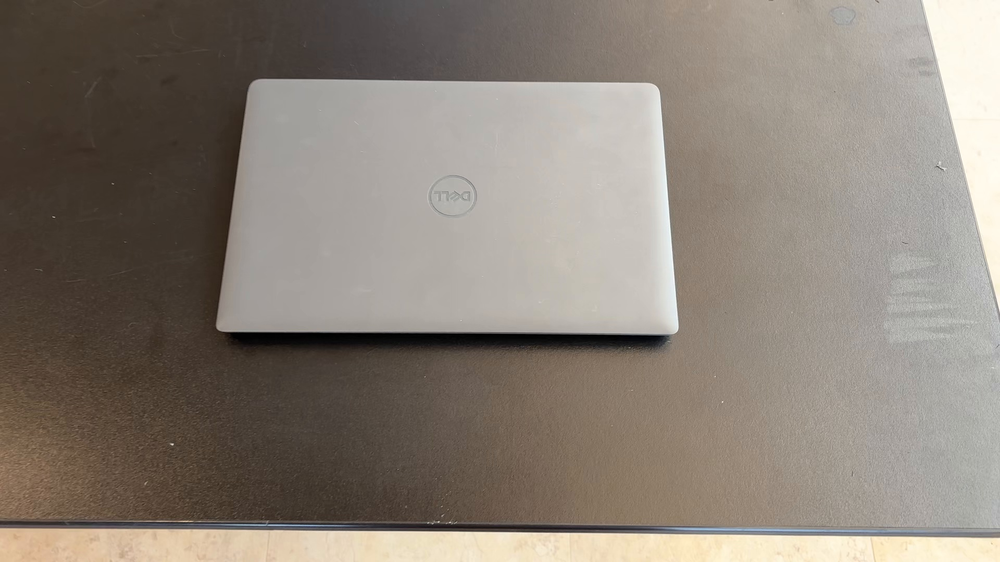}
\end{subfigure}
\begin{subfigure}[t]{0.32\linewidth}
\centering
\includegraphics[width=\linewidth]{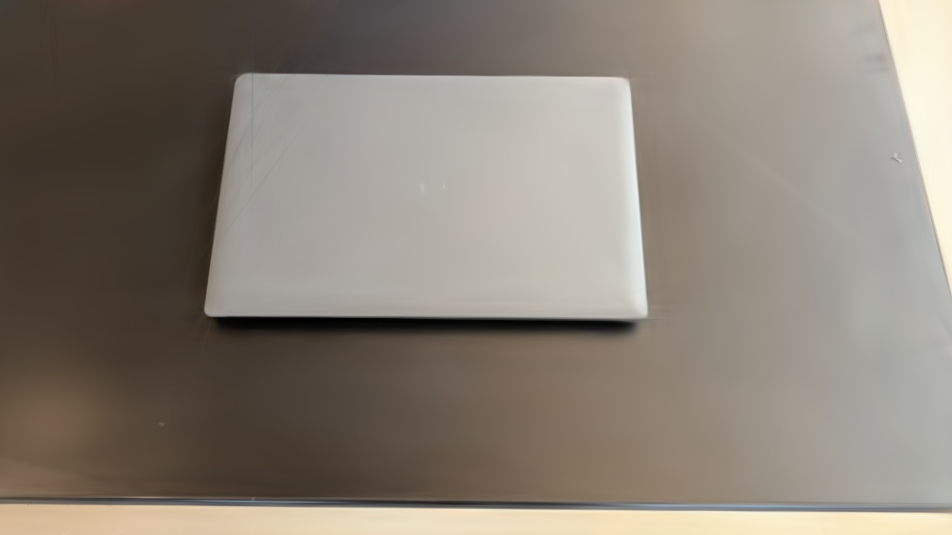}
\end{subfigure}
\begin{subfigure}[t]{0.32\linewidth}
\centering
\includegraphics[width=\linewidth]{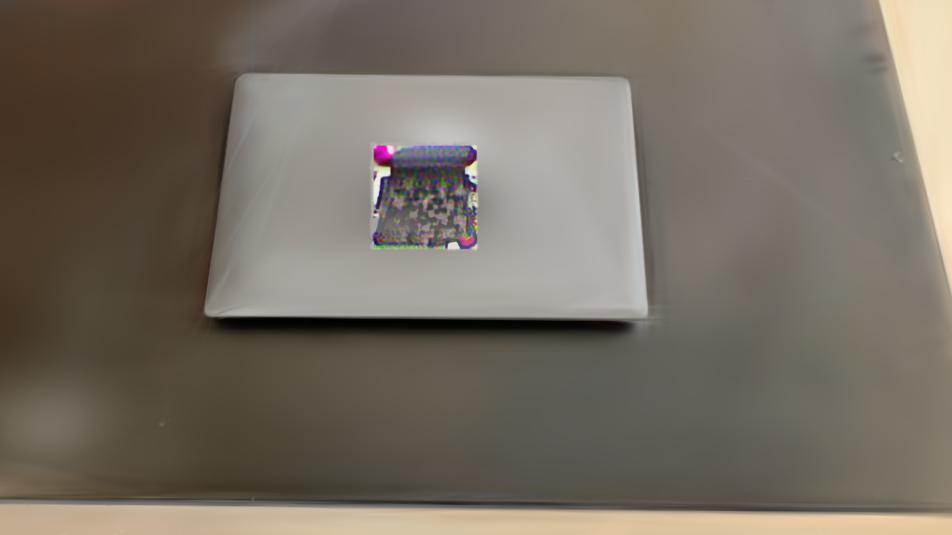}
\end{subfigure}
\caption{Physical validation example: laptop.}
\label{fig:phys_laptop}
\end{figure}

\begin{figure}[!h]
\centering
\begin{subfigure}[t]{0.15\linewidth}
\centering
\includegraphics[width=\linewidth]{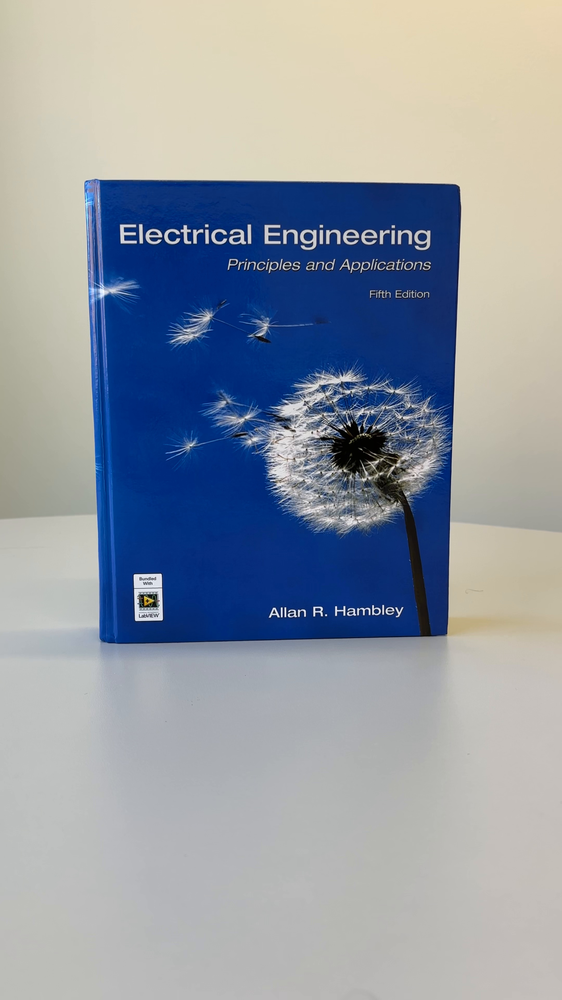}
\end{subfigure}
\begin{subfigure}[t]{0.15\linewidth}
\centering
\includegraphics[width=\linewidth]{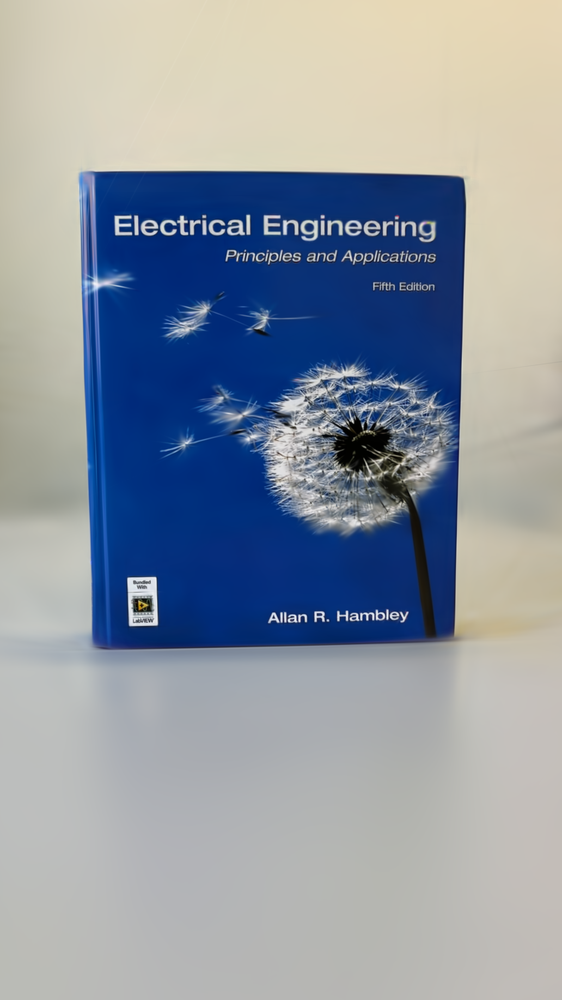}
\end{subfigure}
\begin{subfigure}[t]{0.15\linewidth}
\centering
\includegraphics[width=\linewidth]{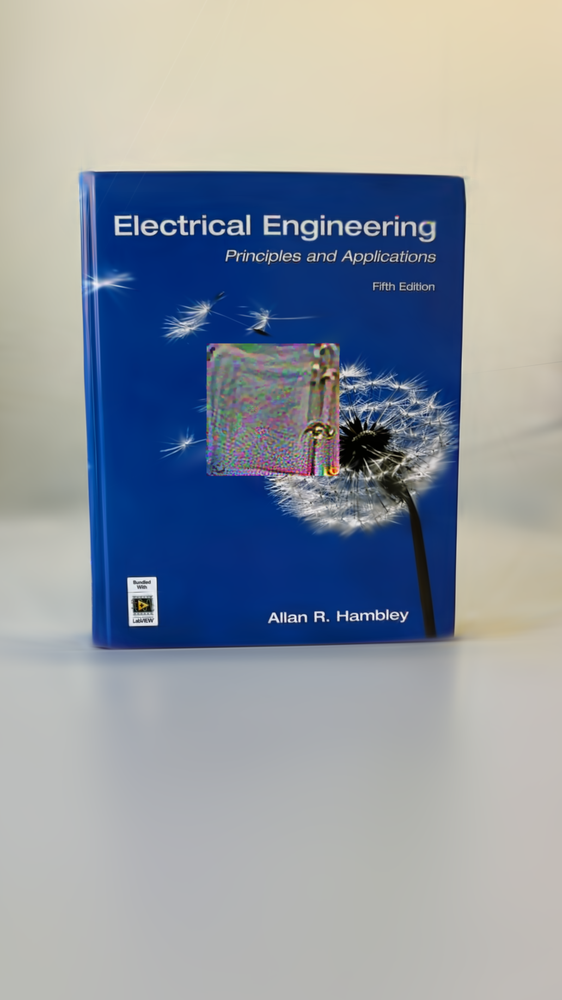}
\end{subfigure}
\begin{subfigure}[t]{0.15\linewidth}
\centering
\includegraphics[width=\linewidth]{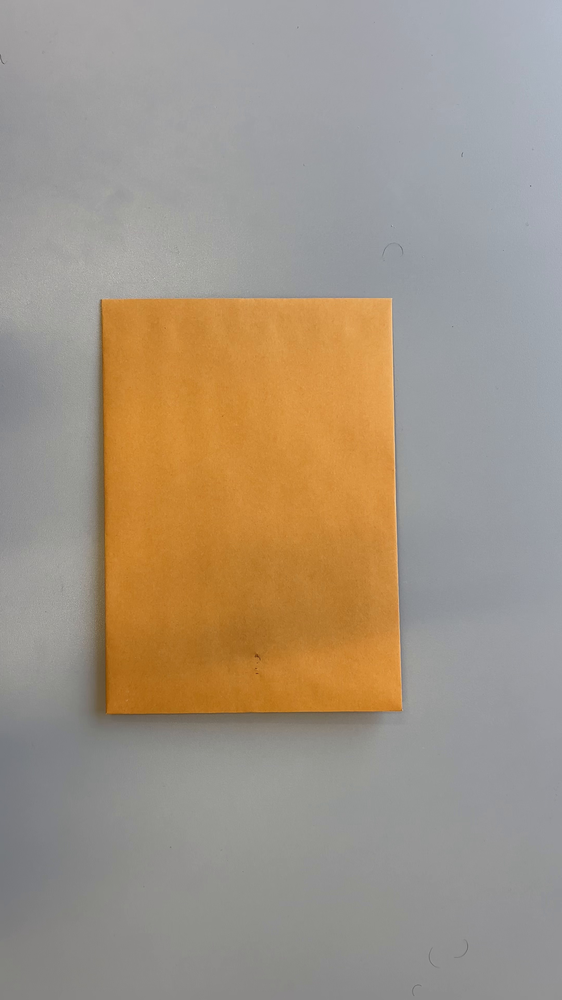}
\end{subfigure}
\begin{subfigure}[t]{0.15\linewidth}
\centering
\includegraphics[width=\linewidth]{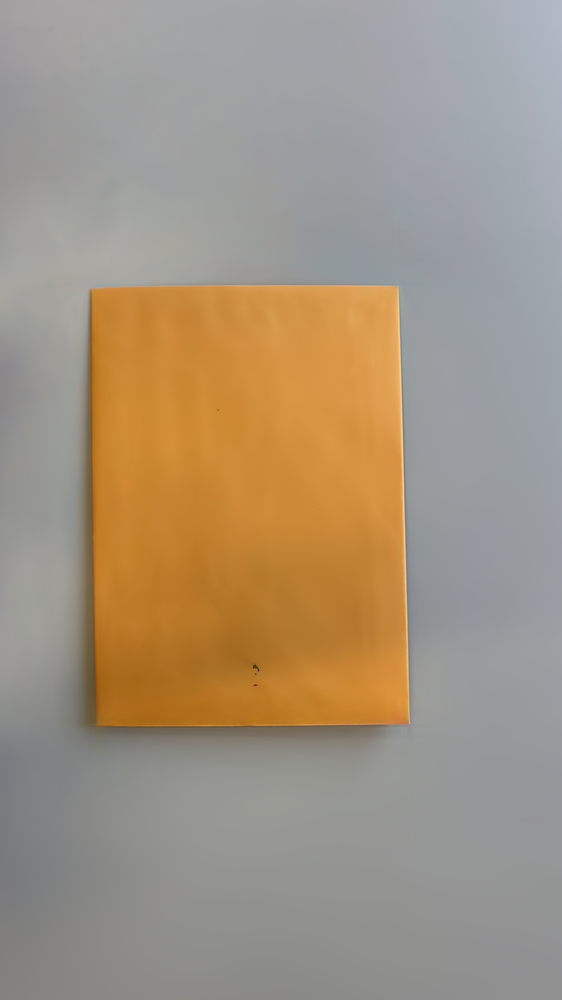}
\end{subfigure}
\begin{subfigure}[t]{0.15\linewidth}
\centering
\includegraphics[width=\linewidth]{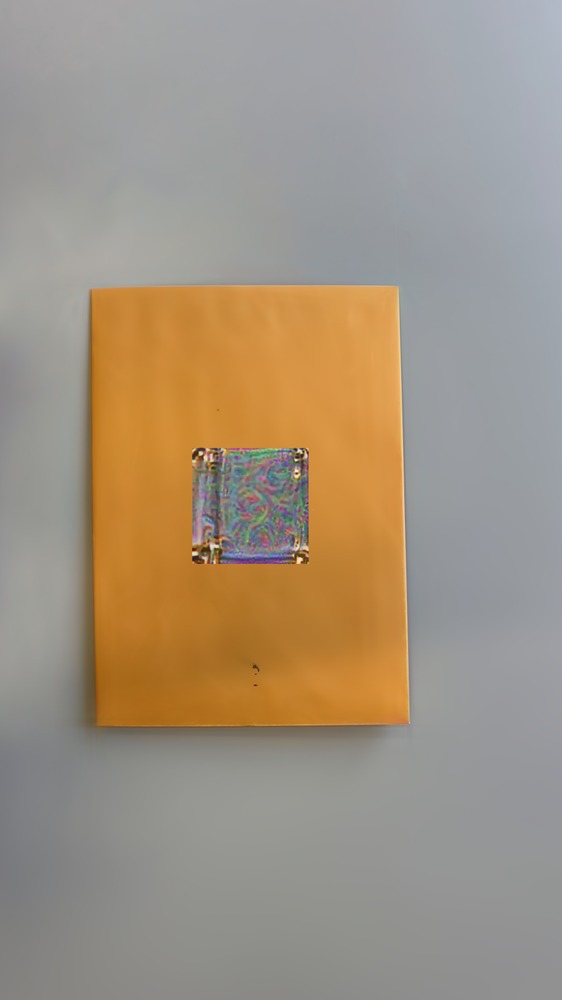}
\end{subfigure}
\caption{Physical validation examples: book and envelope.}
\label{fig:phys_book_envelope}
\end{figure}

Figures~\ref{fig:phys_speed}--\ref{fig:phys_book_envelope} show representative physical validation scenes. These examples illustrate that APSE preserves the target object and surrounding scene while inserting a localized adversarial patch at the anchor view.

\subsection{CO3D Evaluation Examples}
\label{app:co3d_visual_examples}

\begin{figure}[!h]
\centering
\begin{subfigure}[t]{0.32\linewidth}
\centering
\includegraphics[width=\linewidth]{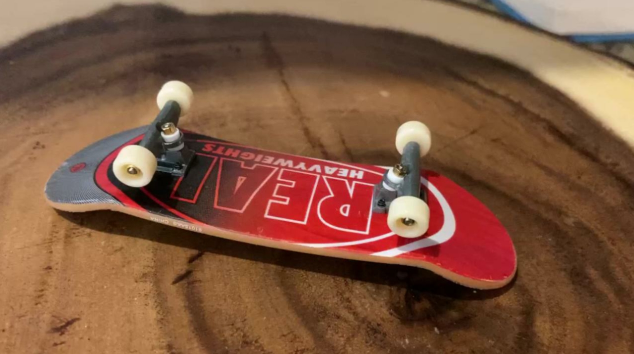}
\end{subfigure}
\begin{subfigure}[t]{0.32\linewidth}
\centering
\includegraphics[width=\linewidth]{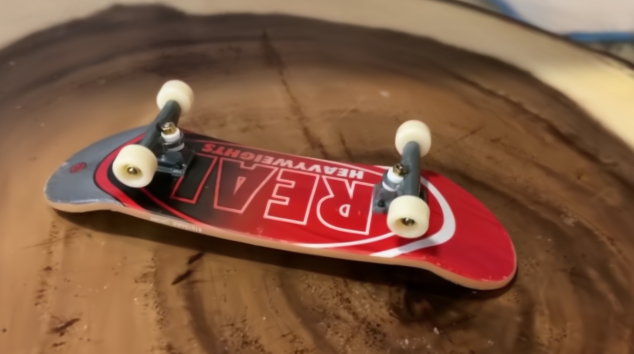}
\end{subfigure}
\begin{subfigure}[t]{0.32\linewidth}
\centering
\includegraphics[width=\linewidth]{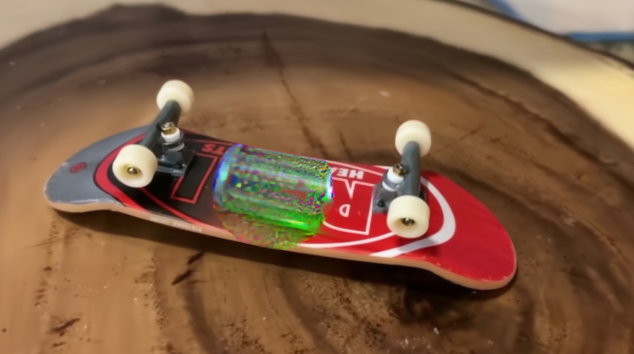}
\end{subfigure}
\caption{CO3D evaluation example: skateboard.}
\label{fig:co3d_skateboard}
\end{figure}

\begin{figure}[!h]
\centering
\begin{subfigure}[t]{0.15\linewidth}
\centering
\includegraphics[width=\linewidth]{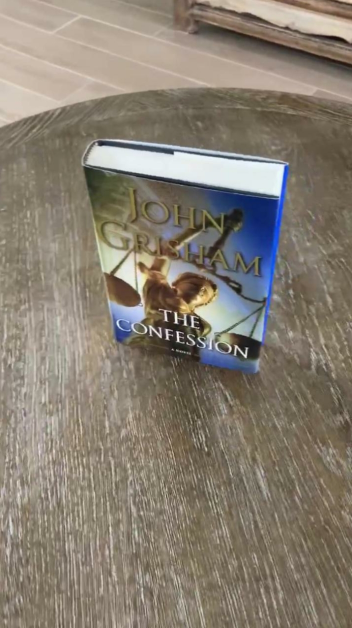}
\end{subfigure}
\begin{subfigure}[t]{0.15\linewidth}
\centering
\includegraphics[width=\linewidth]{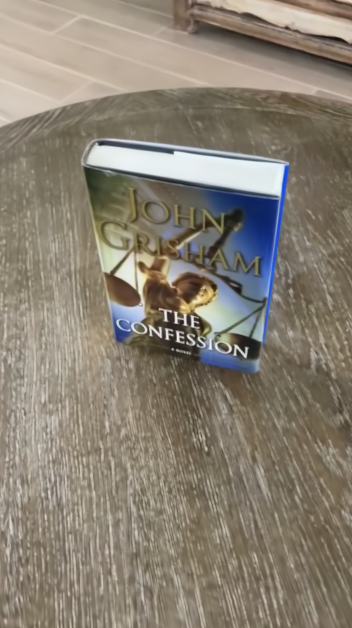}
\end{subfigure}
\begin{subfigure}[t]{0.15\linewidth}
\centering
\includegraphics[width=\linewidth]{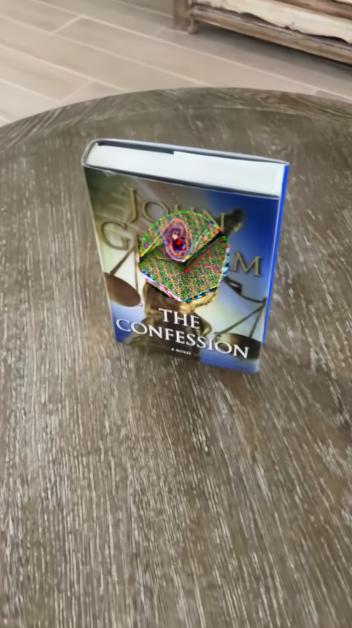}
\end{subfigure}
\caption{CO3D evaluation example: book.}
\label{fig:co3d_book}
\end{figure}

\begin{figure}[!h]
\centering
\begin{subfigure}[t]{0.32\linewidth}
\centering
\includegraphics[width=\linewidth]{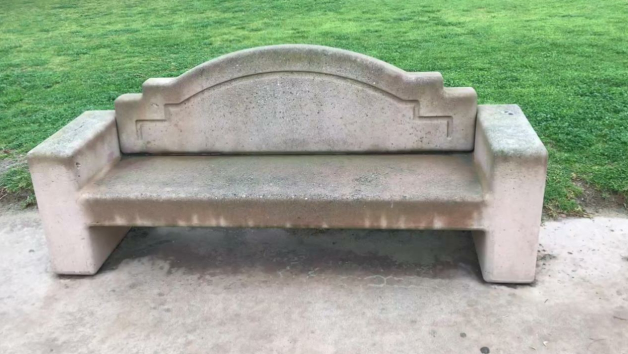}
\end{subfigure}
\begin{subfigure}[t]{0.32\linewidth}
\centering
\includegraphics[width=\linewidth]{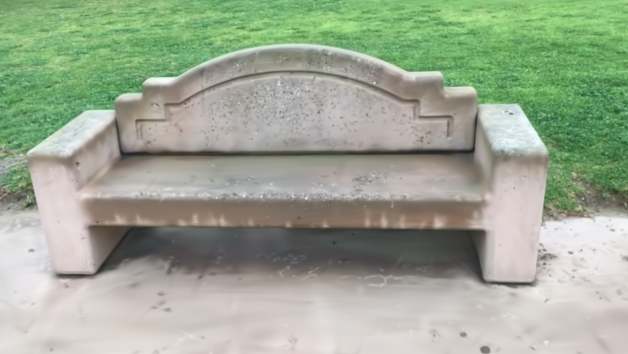}
\end{subfigure}
\begin{subfigure}[t]{0.32\linewidth}
\centering
\includegraphics[width=\linewidth]{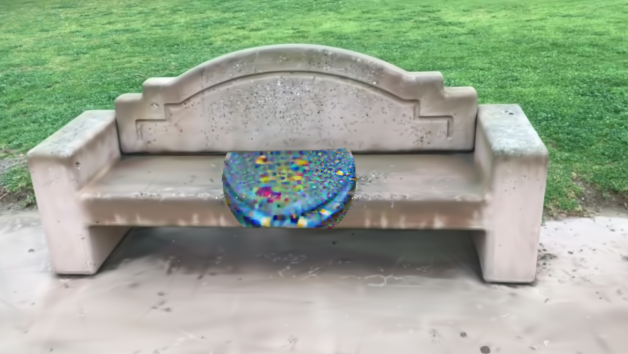}
\end{subfigure}
\caption{CO3D evaluation example: bench.}
\label{fig:co3d_bench}
\end{figure}

\begin{figure}[!h]
\centering
\begin{subfigure}[t]{0.32\linewidth}
\centering
\includegraphics[width=\linewidth]{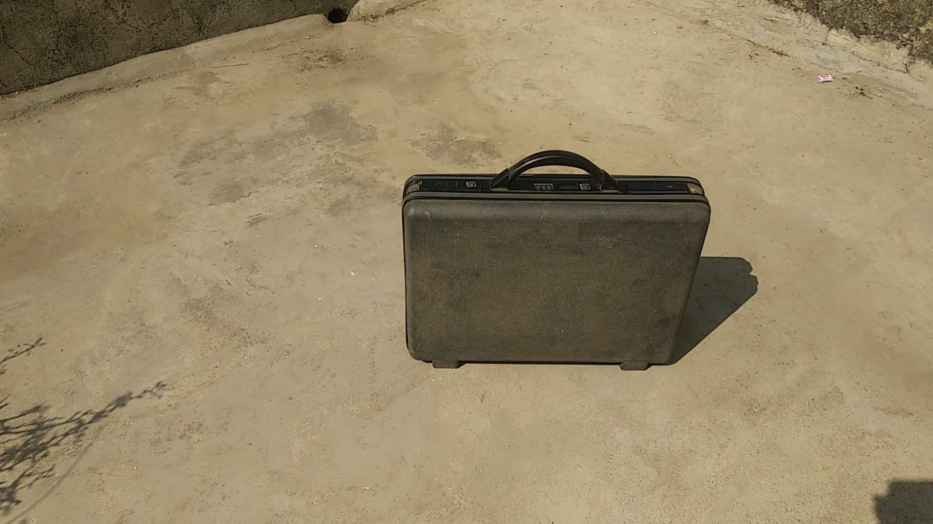}
\end{subfigure}
\begin{subfigure}[t]{0.32\linewidth}
\centering
\includegraphics[width=\linewidth]{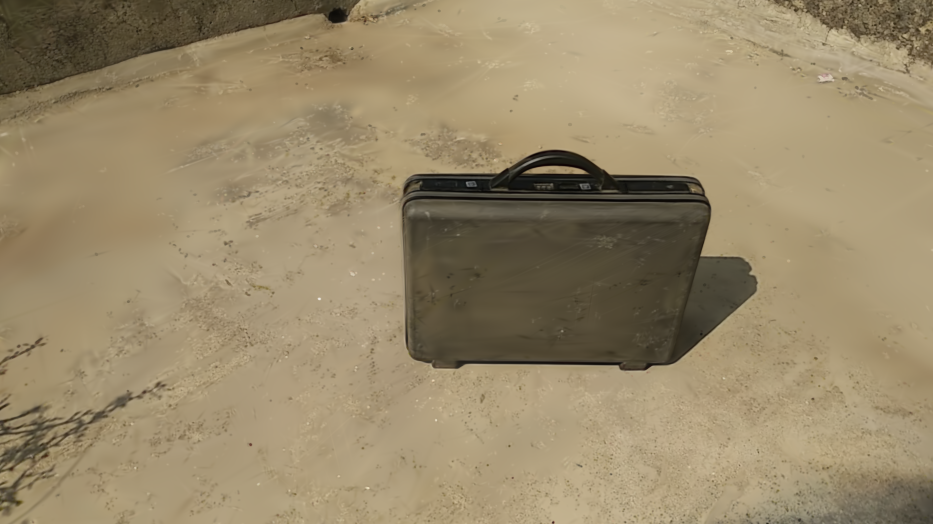}
\end{subfigure}
\begin{subfigure}[t]{0.32\linewidth}
\centering
\includegraphics[width=\linewidth]{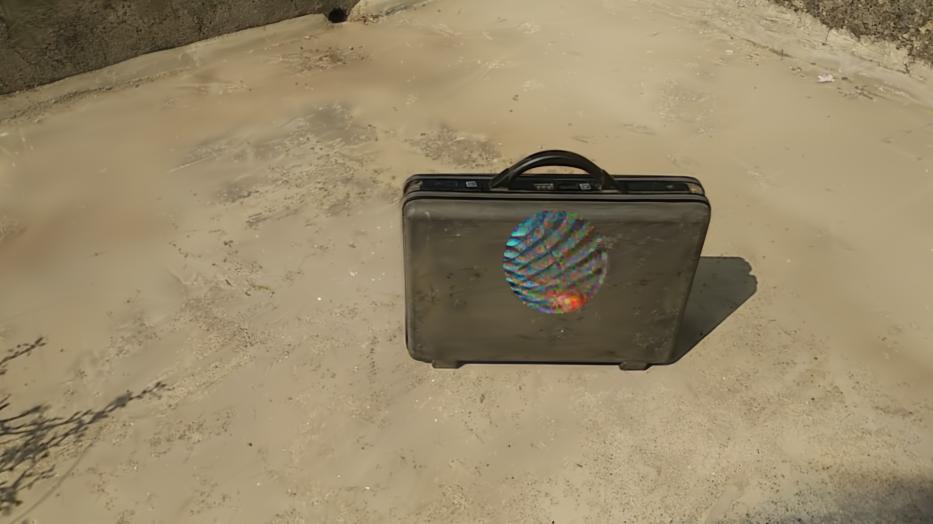}
\end{subfigure}
\caption{CO3D evaluation example: suitcase.}
\label{fig:co3d_suitcase}
\end{figure}

Figures~\ref{fig:co3d_skateboard}--\ref{fig:co3d_suitcase} show representative CO3D scenes used in the object-centered evaluation. The examples cover different object categories and demonstrate the same clean-render and adversarial-render comparison used for scene-level measurement.

\subsection{Waymo Evaluation Examples}
\label{app:waymo_visual_examples}

\begin{figure}[!h]
\centering
\begin{subfigure}[t]{0.32\linewidth}
\centering
\includegraphics[width=\linewidth]{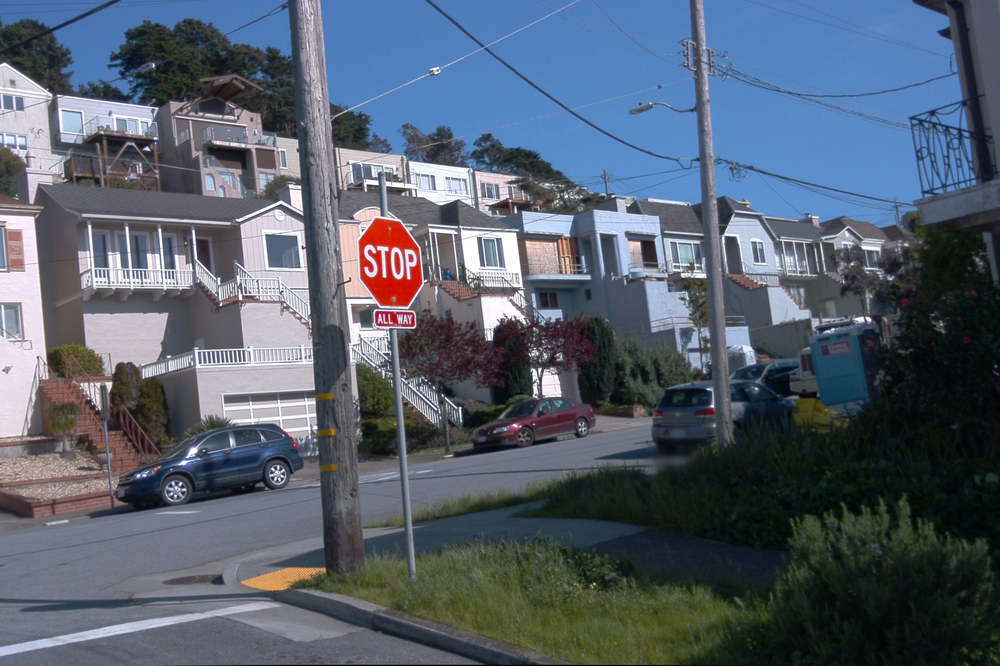}
\end{subfigure}
\begin{subfigure}[t]{0.32\linewidth}
\centering
\includegraphics[width=\linewidth]{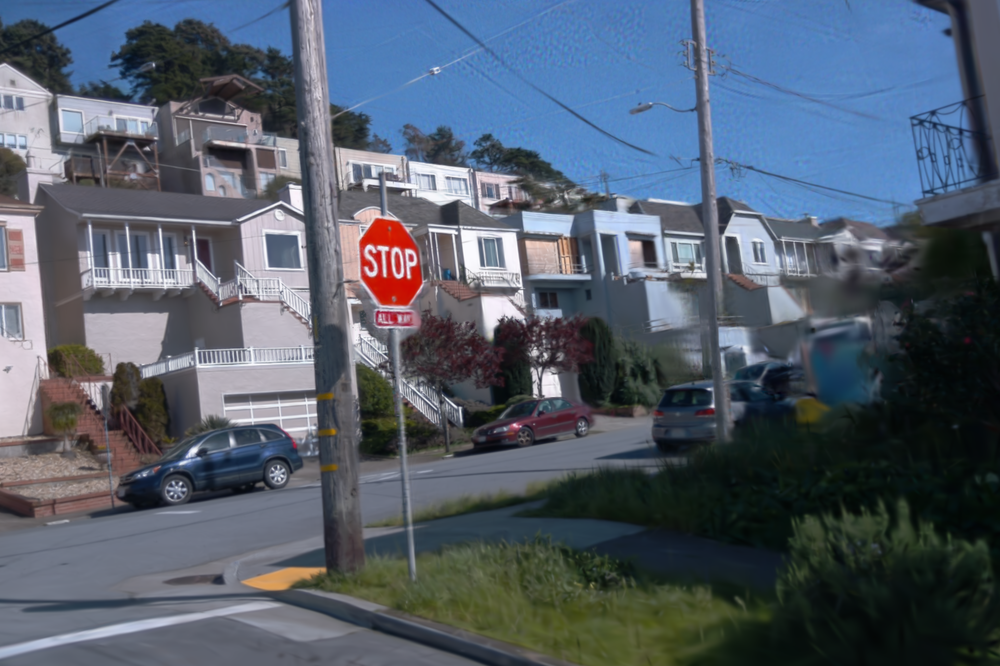}
\end{subfigure}
\begin{subfigure}[t]{0.32\linewidth}
\centering
\includegraphics[width=\linewidth]{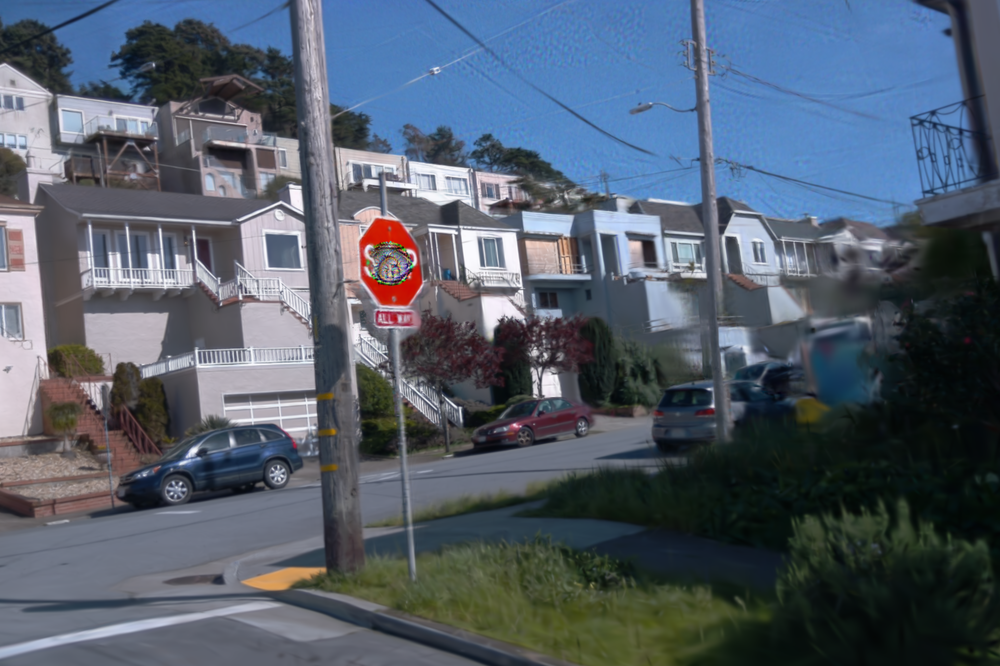}
\end{subfigure}
\caption{Waymo evaluation example.}
\label{fig:waymo_027}
\end{figure}

\begin{figure}[!h]
\centering
\begin{subfigure}[t]{0.32\linewidth}
\centering
\includegraphics[width=\linewidth]{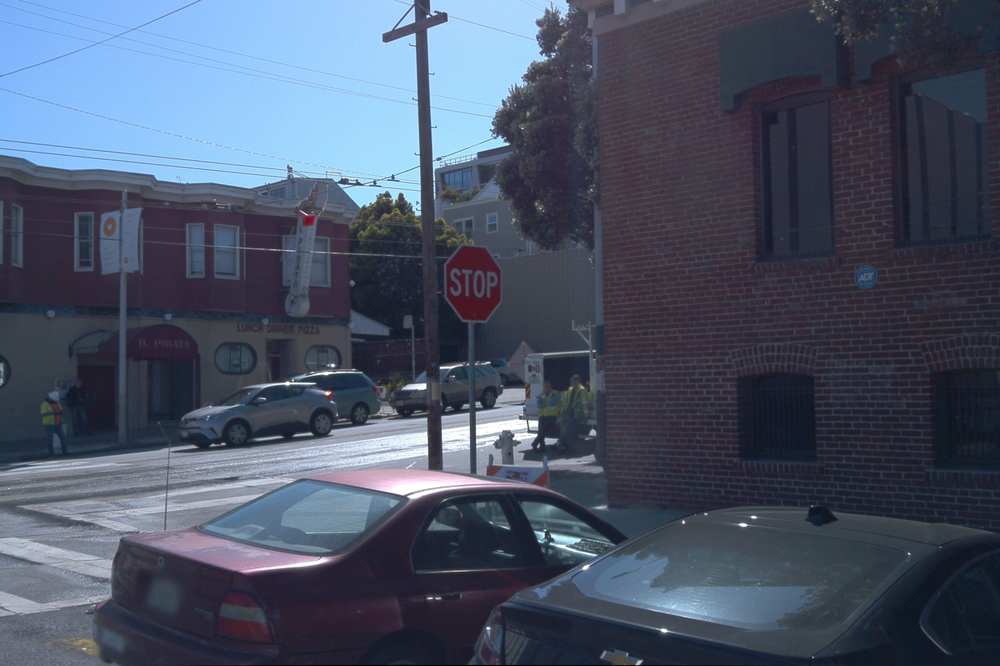}
\end{subfigure}
\begin{subfigure}[t]{0.32\linewidth}
\centering
\includegraphics[width=\linewidth]{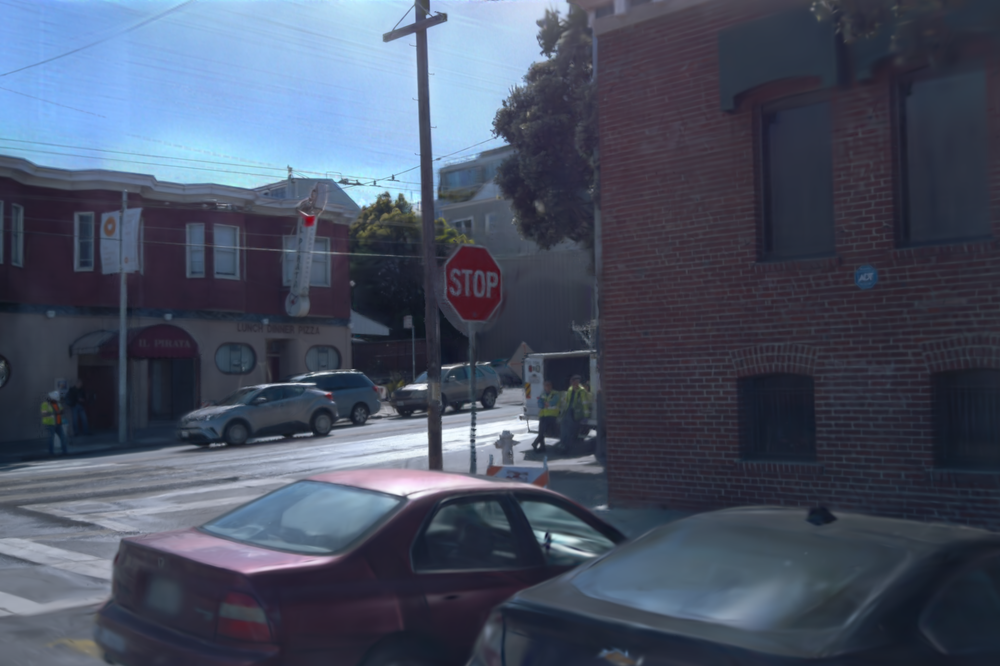}
\end{subfigure}
\begin{subfigure}[t]{0.32\linewidth}
\centering
\includegraphics[width=\linewidth]{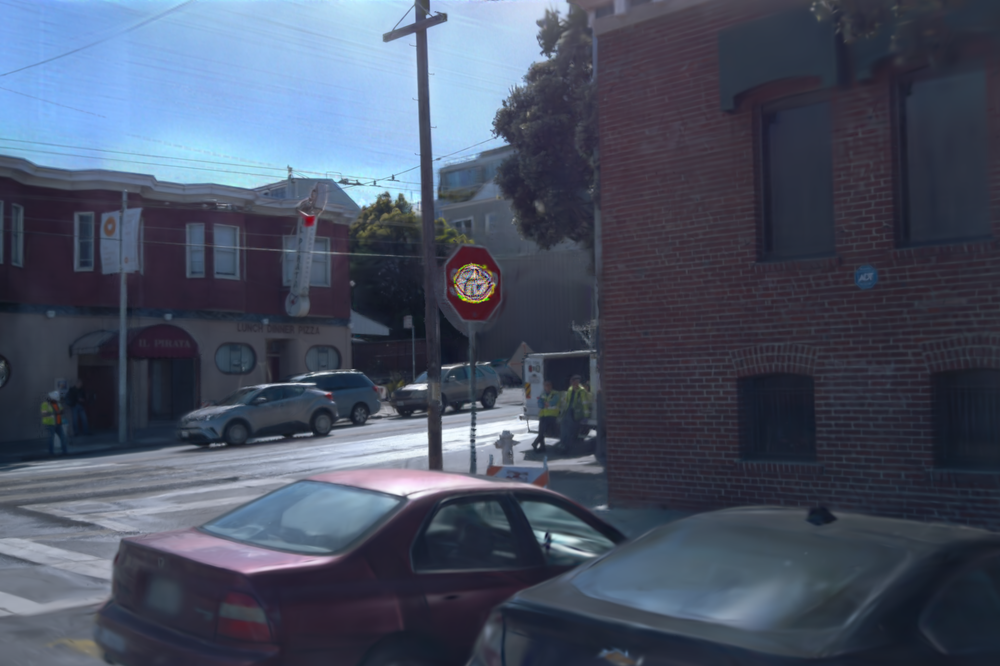}
\end{subfigure}
\caption{Waymo evaluation example.}
\label{fig:waymo_282}
\end{figure}

\begin{figure}[!h]
\centering
\begin{subfigure}[t]{0.32\linewidth}
\centering
\includegraphics[width=\linewidth]{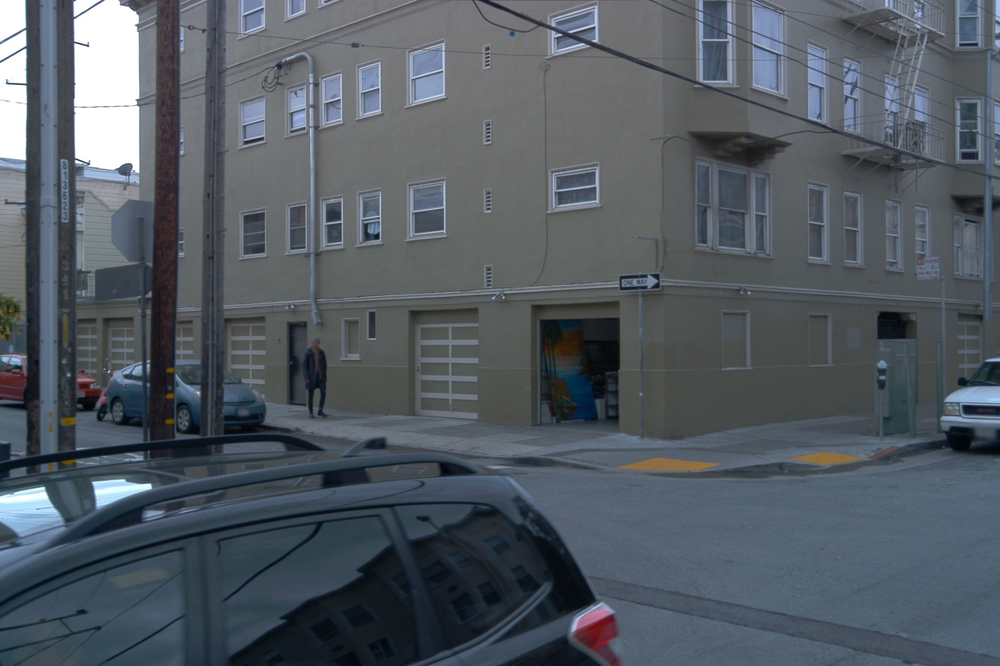}
\end{subfigure}
\begin{subfigure}[t]{0.32\linewidth}
\centering
\includegraphics[width=\linewidth]{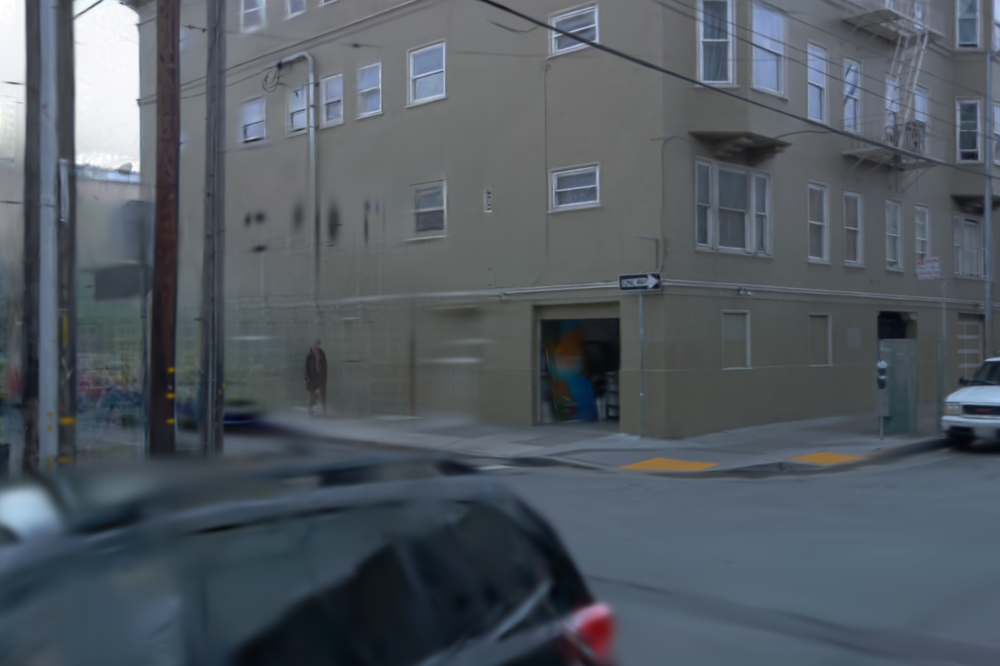}
\end{subfigure}
\begin{subfigure}[t]{0.32\linewidth}
\centering
\includegraphics[width=\linewidth]{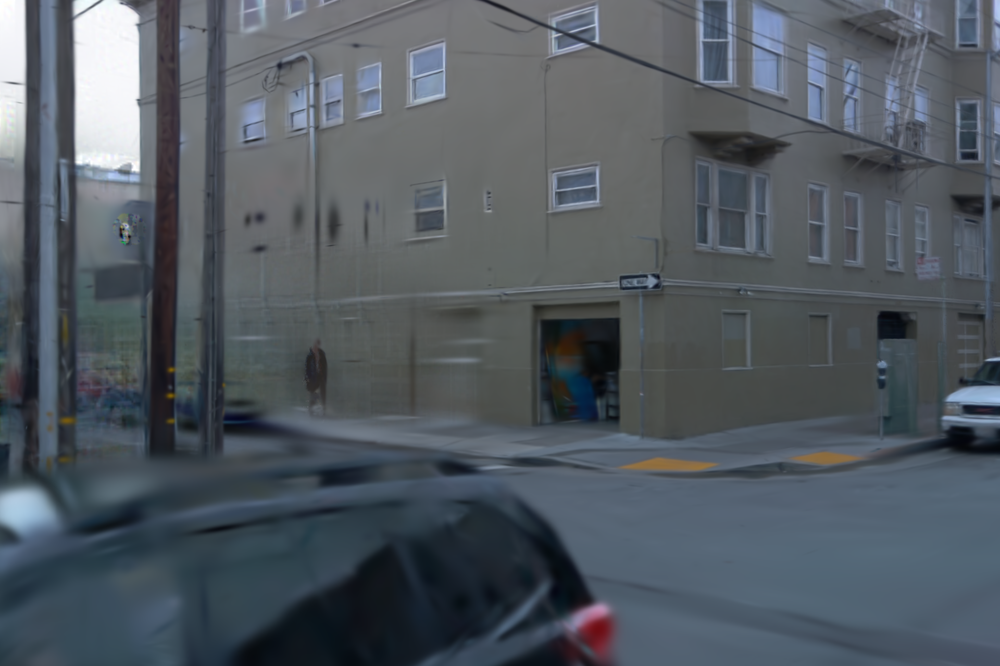}
\end{subfigure}
\caption{Waymo evaluation example.}
\label{fig:waymo_785}
\end{figure}

\begin{figure}[!h]
\centering
\begin{subfigure}[t]{0.32\linewidth}
\centering
\includegraphics[width=\linewidth]{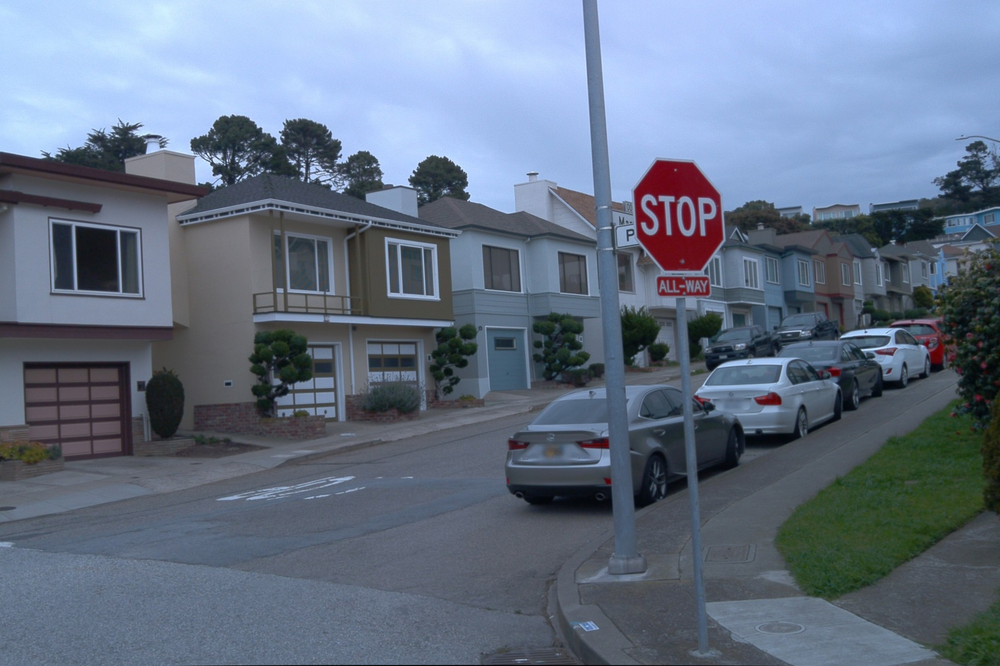}
\end{subfigure}
\begin{subfigure}[t]{0.32\linewidth}
\centering
\includegraphics[width=\linewidth]{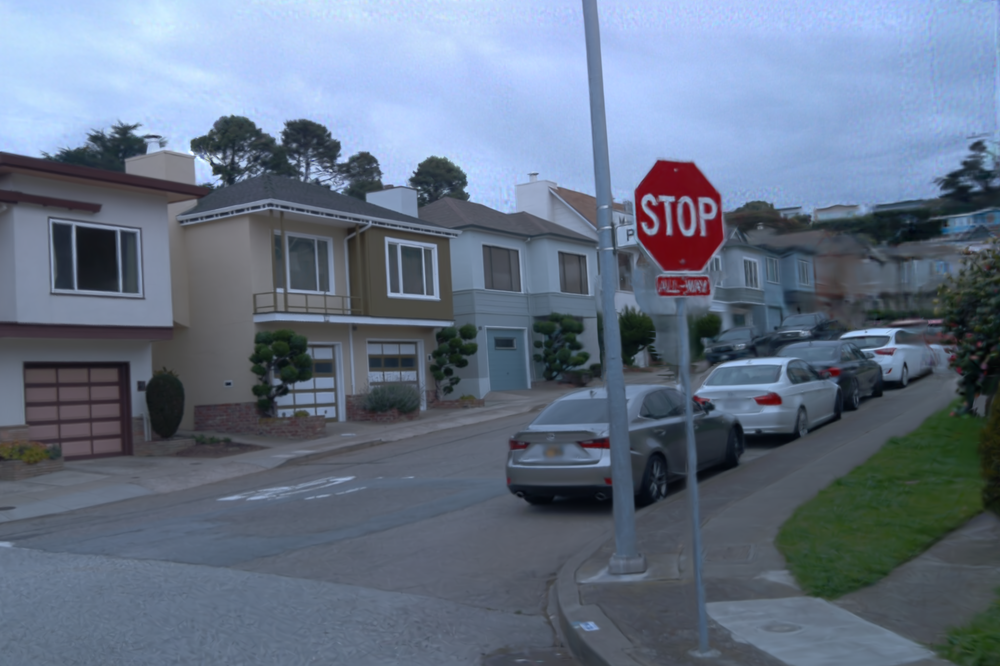}
\end{subfigure}
\begin{subfigure}[t]{0.32\linewidth}
\centering
\includegraphics[width=\linewidth]{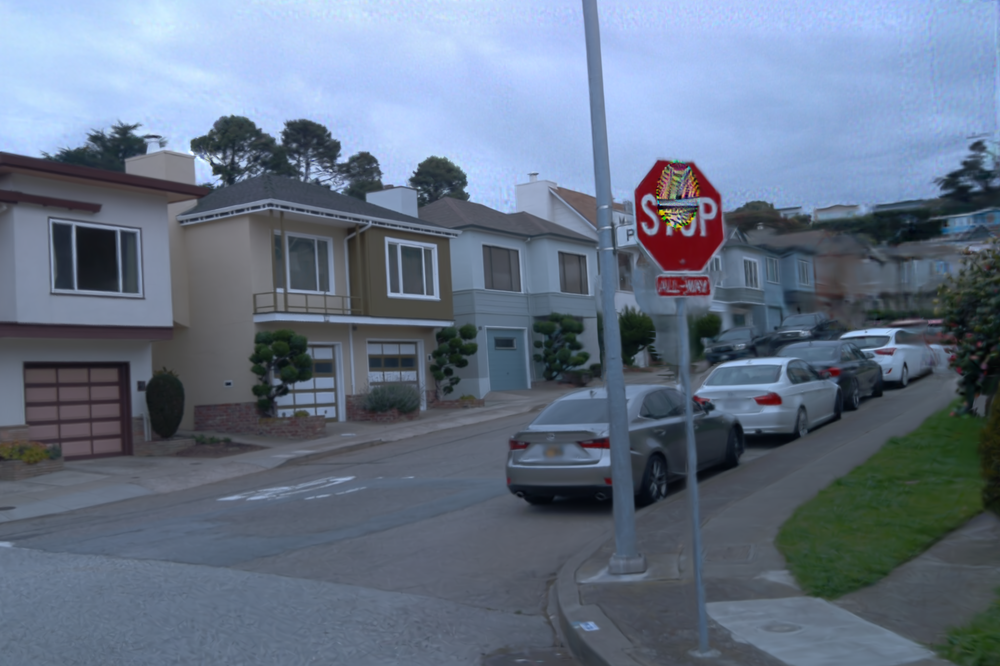}
\end{subfigure}
\caption{Waymo evaluation example.}
\label{fig:waymo_095}
\end{figure}

Figures~\ref{fig:waymo_027}--\ref{fig:waymo_095} show representative Waymo scenes used in the driving evaluation. These examples illustrate how the embedded patch is evaluated on real driving imagery with scene-specific background, target placement, and camera geometry.

\section{Detailed Experimental Settings}
\label{app:settings}

\subsection{Adversarial Patch Optimization}
\label{app:patch_optimization}

We optimize 2D adversarial patches using Expectation over Transformations (EOT) to improve robustness to physical-world variation. At each optimization step, we estimate the objective over a batch of eight transformed observations. The transformation distribution models both photometric and geometric variation. Photometric transformations include multiplicative perturbations to color, brightness, and contrast, each sampled from $[0.75,1.25]$, as well as gamma correction sampled from $[0.8,1.2]$. Geometric transformations include in-plane rotations sampled from $[-20^\circ,20^\circ]$, translations of up to $12$ pixels, and isotropic scaling sampled from $[0.7,1.25]$. 

To further approximate physical imaging effects, we additionally apply random and center-crop perturbations, image-space Gaussian perturbations with standard deviation $0.06$, Gaussian blur, additive noise with magnitude up to $0.1$, JPEG compression with quality sampled between $30$ and $100$, and camera-resolution degradation. The latter is modeled by downscaling the image by a factor between $1.0$ and $0.6$, followed by $8$-bit quantization.

The patch is optimized with a targeted margin loss proposed in Carlini-Wagner attack~\cite{carlini2017towards}. For each attack instance, we use three independent random restarts. Each restart is optimized for at most $5{,}000$ Adam steps with learning rate $0.01$ and momentum parameters $(\beta_1,\beta_2)=(0.9,0.999)$. We employ early stopping with patience $1{,}000$ and a minimum loss improvement threshold of $10^{-4}$. The complete objective consists of the targeted Carlini--Wagner loss together with total variation (TV), non-printability score (NPS), and $\ell_2$ regularization terms. Unless otherwise stated, the corresponding weights are $2.5$ for TV, $0.1$ for NPS, and $1.0$ for $\ell_2$ regularization. The confidence parameter is set to $\kappa=0$. For the NPS term, we use a 30-color printable palette.

\subsection{Attack-Design Settings}
\label{sec:app:ablation_settings}

\subsubsection{EOT Magnitude}
\label{sec:app:eot_settings}

We evaluate the effect of EOT strength by varying the number of transformed observations and the magnitude of the transformation distribution. In the no-EOT setting, optimization is performed only on the anchor view, without stochastic transformations. The weak-EOT setting uses four transformed views per optimization step and applies mild photometric and geometric perturbations: color, brightness, and contrast multipliers are sampled from $[0.95,1.05]$, gamma from $[0.98,1.02]$, crop perturbations are applied with probability $0.25$, perspective noise has magnitude $0.01$, rotations are bounded by $\pm 5^\circ$, translations by $3$ pixels, and scale by $[0.95,1.05]$. This setting also includes mild blur, noise, and JPEG compression, uses downscaling factors in $\{1.0,0.95,0.9\}$, and does not apply camera quantization.

The strong-EOT setting corresponds to our default configuration. It uses eight transformed views per optimization step, with color, brightness, and contrast multipliers sampled from $[0.75,1.25]$, gamma from $[0.8,1.2]$, crop probability $0.5$, perspective noise magnitude $0.06$, rotations bounded by $\pm 20^\circ$, translations up to $12$ pixels, and scale sampled from $[0.7,1.25]$. It further includes blur, additive noise, JPEG compression, downscaling to a factor as low as $0.6$, and $8$-bit quantization.

The very-strong-EOT setting increases both the number and severity of transformations. It uses sixteen transformed views per optimization step, color, brightness, and contrast multipliers sampled from $[0.65,1.35]$, gamma from $[0.7,1.3]$, crop probability $0.75$, perspective noise magnitude $0.09$, rotations bounded by $\pm 30^\circ$, translations up to $18$ pixels, and scale sampled from $[0.6,1.4]$. This setting additionally applies stronger blur, noise, and JPEG compression, uses downscaling factors as low as $0.5$, and includes $8$-bit quantization.

\section{Evaluation Results}
\label{app:evaluation_results}

This appendix reports the full view-wise evaluation curves omitted from the main text for space. Each figure shows attack performance across the complete set of evaluated observation dimensions, complementing the radius and viewing-angle summaries in the main paper.

\subsection{CO3D Results}
\label{app:co3d_results}

\begin{figure}[!h]
\centering
\includegraphics[width=\linewidth]{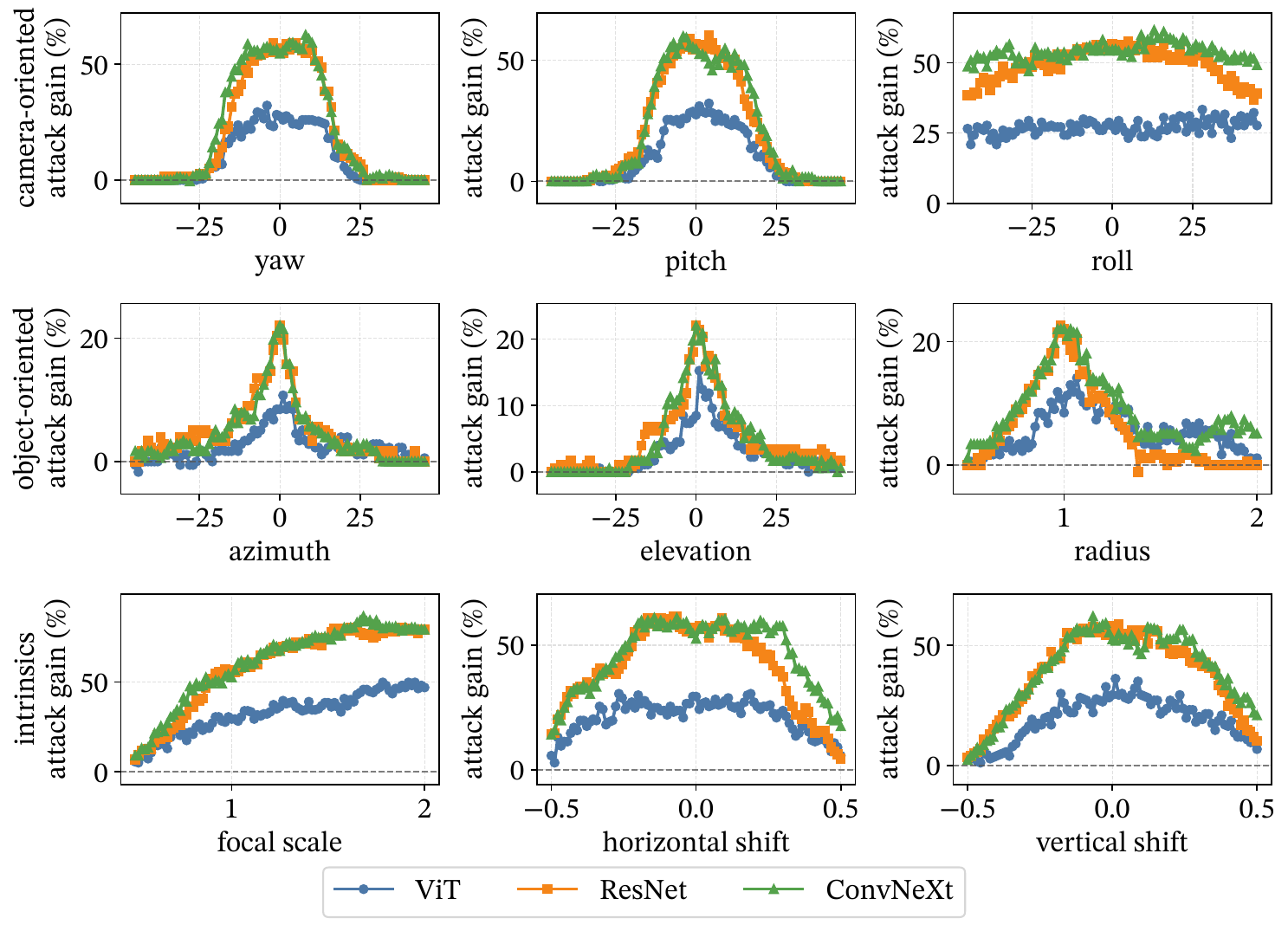}
\caption{Targeted attack performance (AG) across views.}
\label{fig:co3d_targeted_ag}
\end{figure}

\begin{figure}[!h]
\centering
\includegraphics[width=\linewidth]{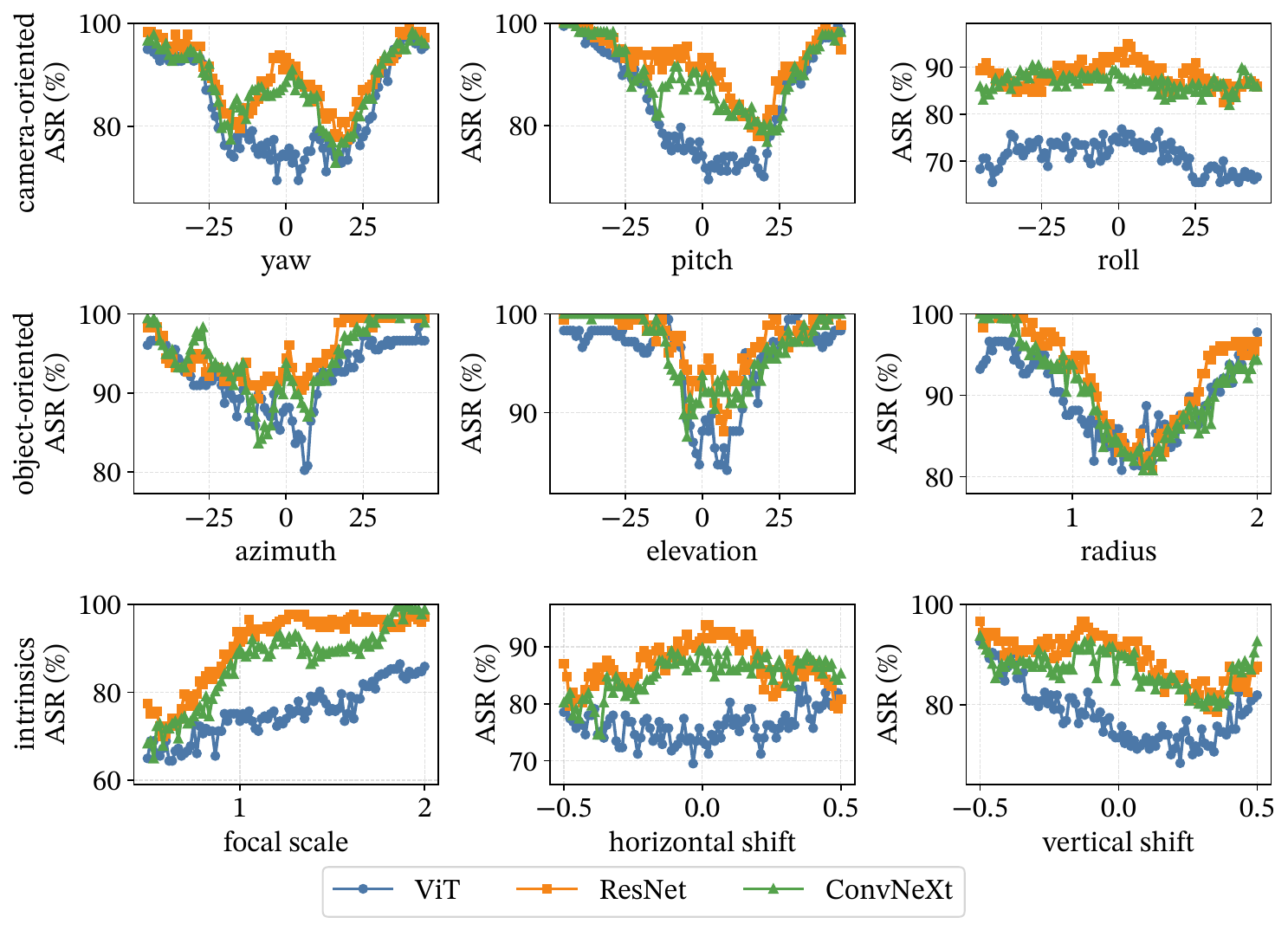}
\caption{Untargeted attack performance (ASR) across views.}
\label{fig:co3d_untargeted_asr}
\end{figure}

\begin{figure}[!h]
\centering
\includegraphics[width=\linewidth]{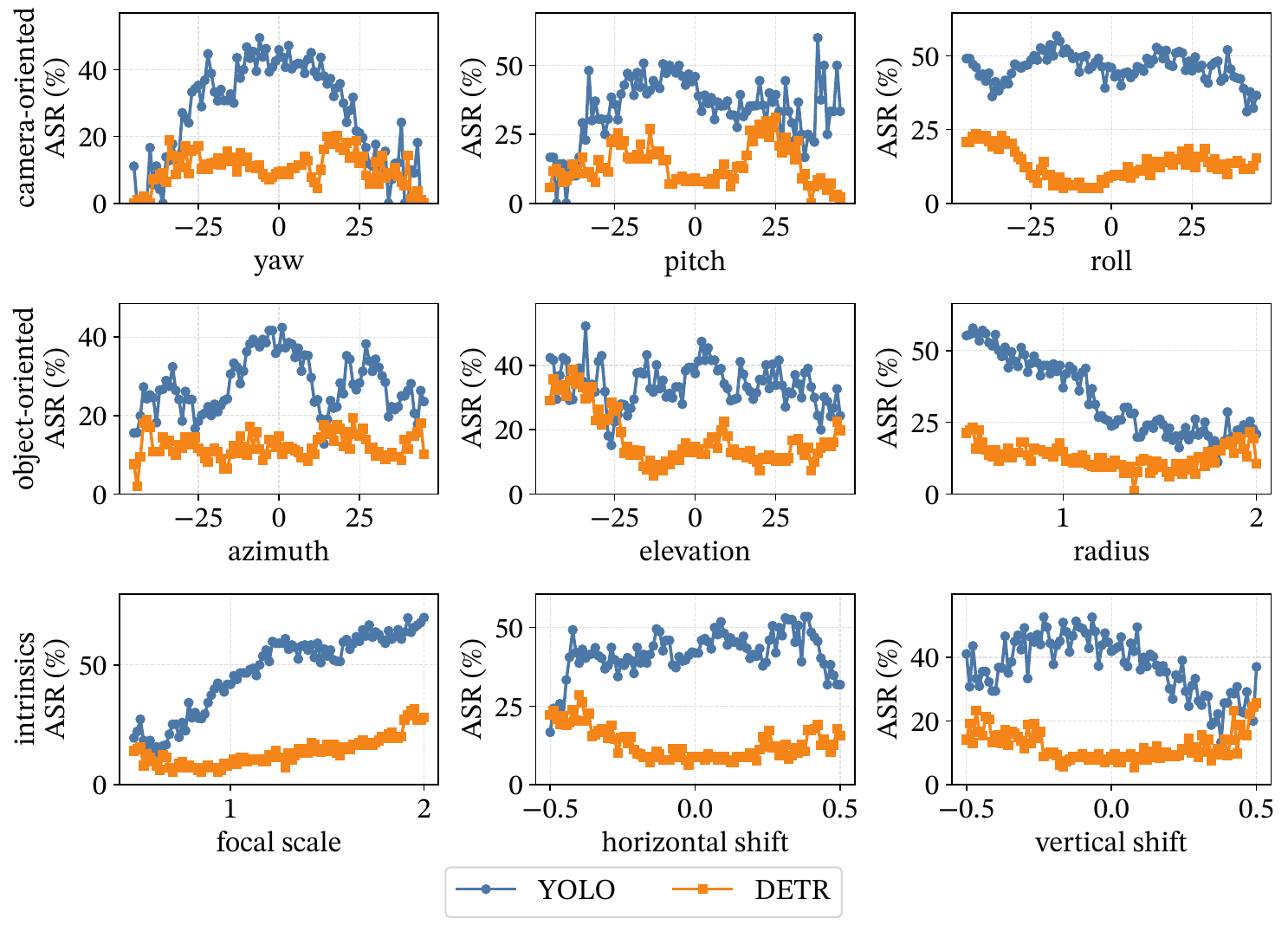}
\vspace{-1em}
\caption{Disappearance attack performance (ASR) across views on CO3D.}
\label{fig:co3d_disappear_asr}
\vspace{-1em}
\end{figure}

Figures~\ref{fig:co3d_targeted_ag}--\ref{fig:co3d_disappear_asr} show the full CO3D evaluation across targeted classification, untargeted classification, and disappearance attacks. These curves provide the complete view-wise behavior behind the main-text finding that attack effectiveness is concentrated near favorable scene conditions and degrades under larger viewpoint, distance, or image-location changes.


\subsection{Waymo Results}
\label{app:waymo_results}

\begin{figure}[!h]
\centering
\includegraphics[width=\linewidth]{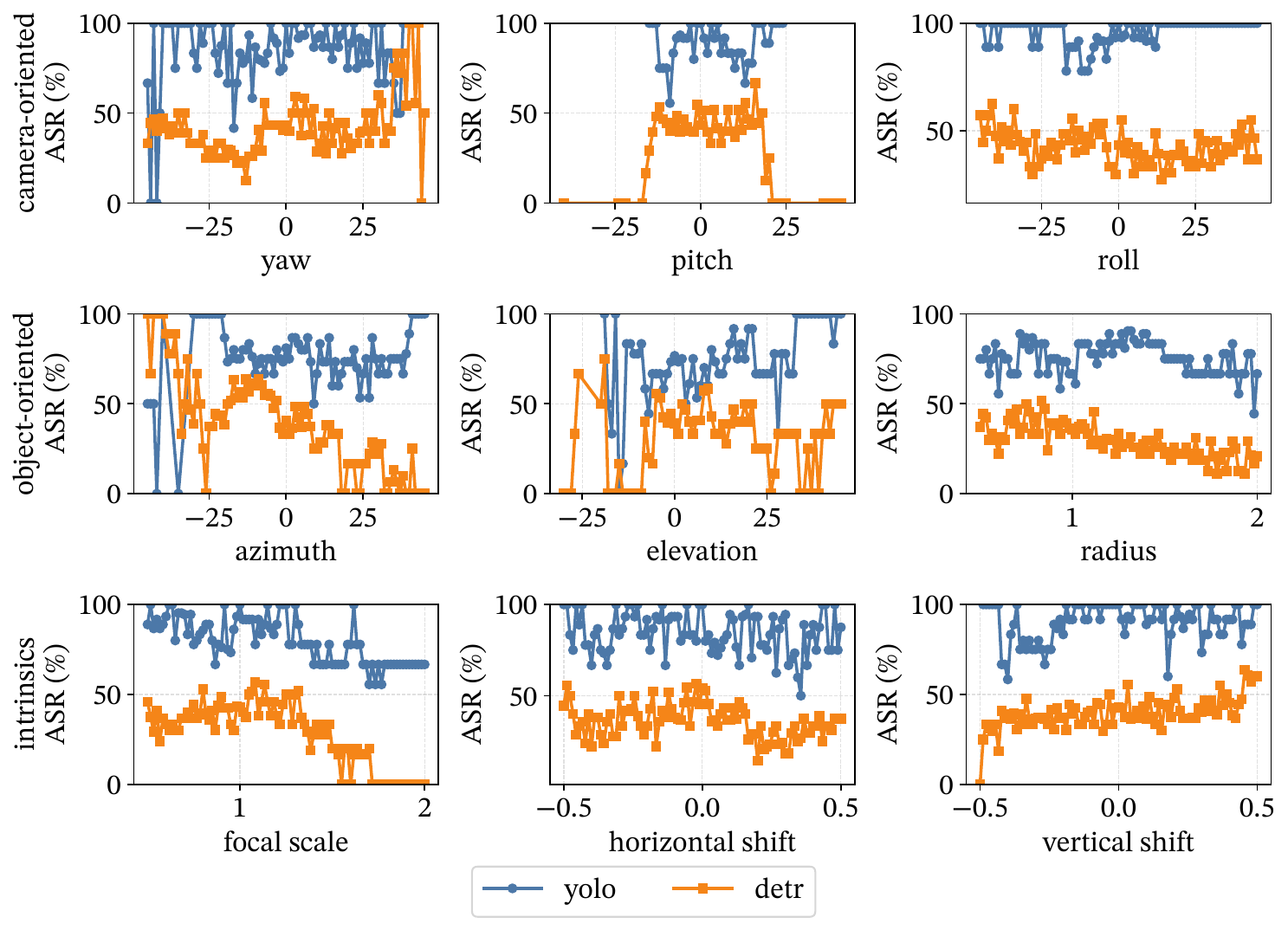}
\vspace{-1em}
\caption{Disappearance attack performance (ASR) across views on Waymo.}
\label{fig:waymo_disappear_asr}
\end{figure}

Figure~\ref{fig:waymo_disappear_asr} reports the full Waymo disappearance-attack curves. Compared with object-centered CO3D scenes, the driving setting shows stronger dependence on trajectory, target placement, and camera geometry, leading to more directional variation across view changes.

\subsection{Patch Geometry Results}
\label{app:patch_geometry_results}

\begin{figure}[!h]
\centering
\includegraphics[width=\linewidth]{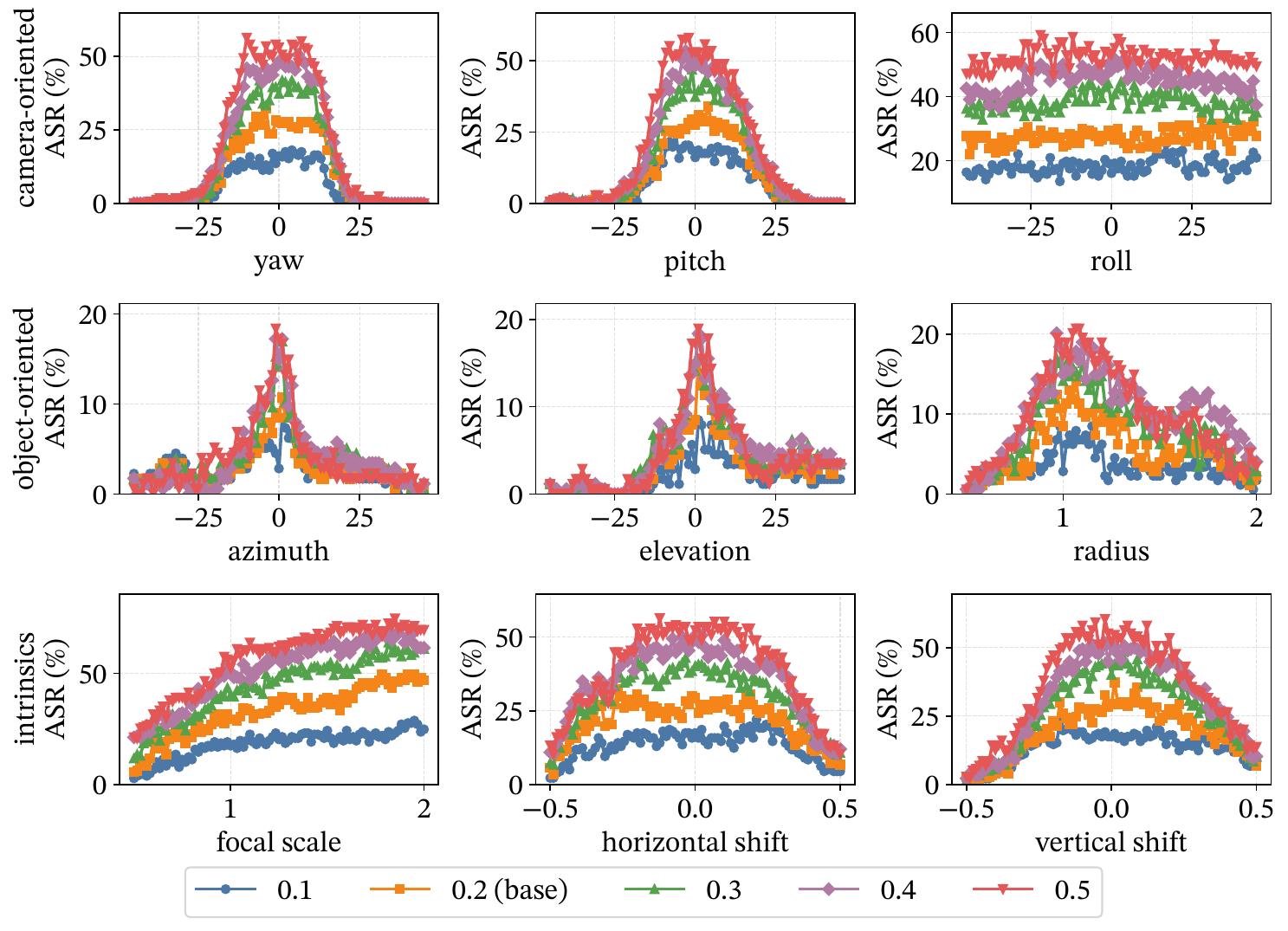}
\caption{Impact of adversarial patch size.}
\label{fig:ablation_patch_size}
\end{figure}

\begin{figure}[!h]
\centering
\includegraphics[width=\linewidth]{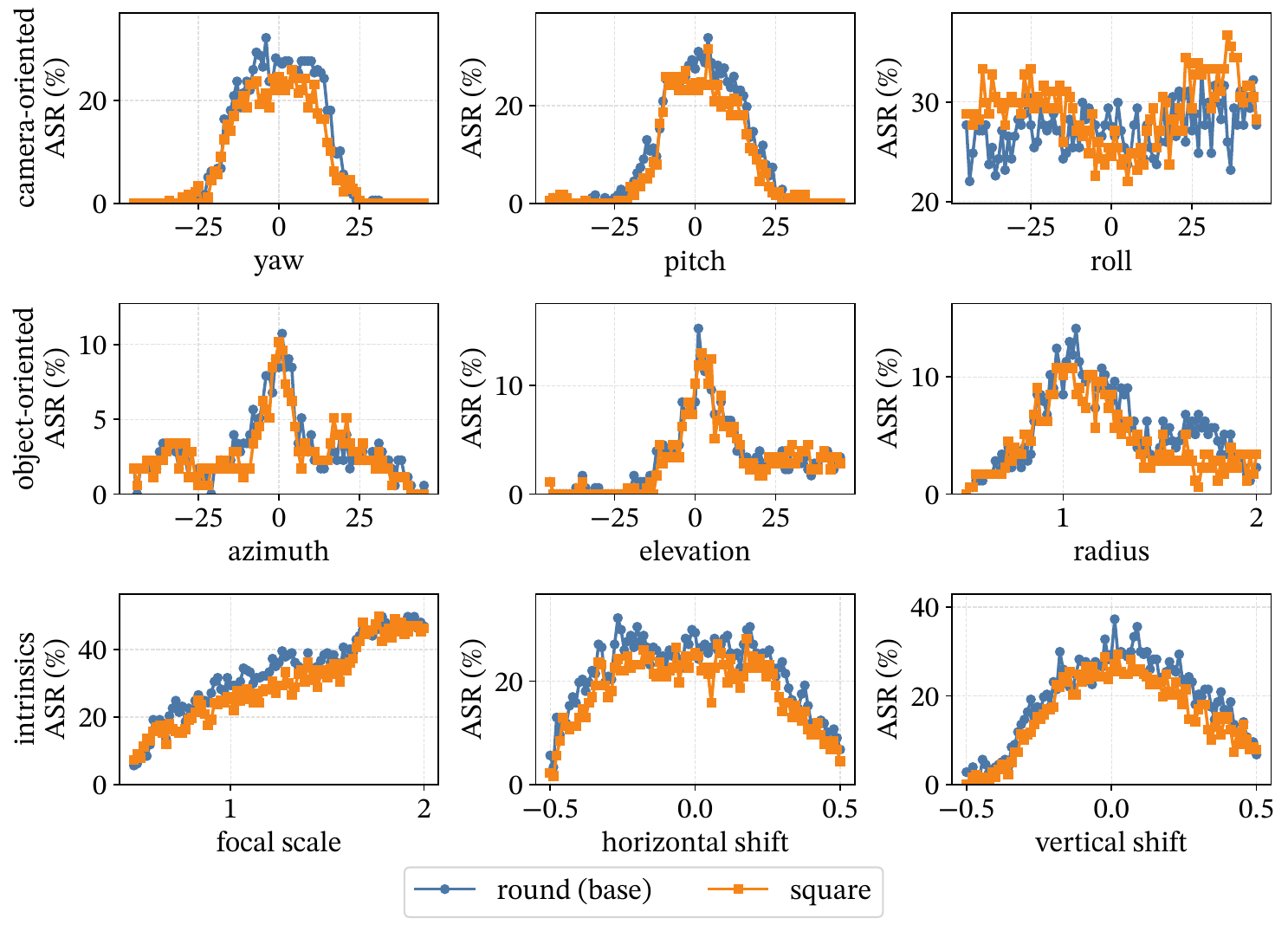}
\caption{Impact of adversarial patch shape.}
\label{fig:ablation_patch_shape}
\end{figure}

Figures~\ref{fig:ablation_patch_size} and~\ref{fig:ablation_patch_shape} provide the full patch-size and patch-shape comparisons. Larger patches generally improve attack strength by increasing the projected adversarial footprint, while changing between round and square shapes has a weaker effect. In both cases, the same distance and oblique-view failure modes remain visible.

\subsection{Attack-Design Results}
\label{app:ablation_results}

\begin{figure}[!h]
\centering
\includegraphics[width=\linewidth]{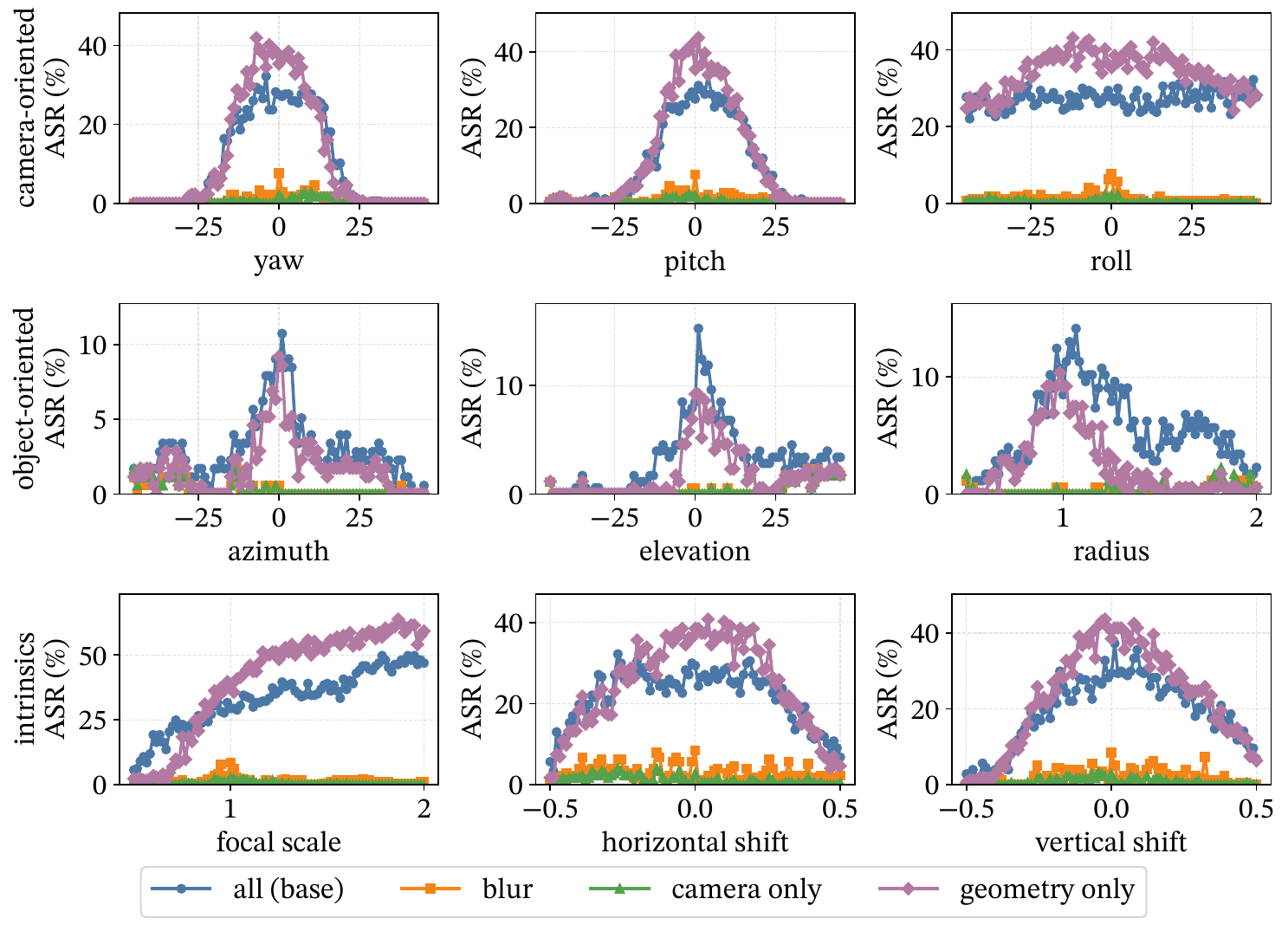}
\caption{Effectiveness of EOT components.}
\label{fig:ablation_eot_component}
\end{figure}

\begin{figure}[!h]
\centering
\includegraphics[width=\linewidth]{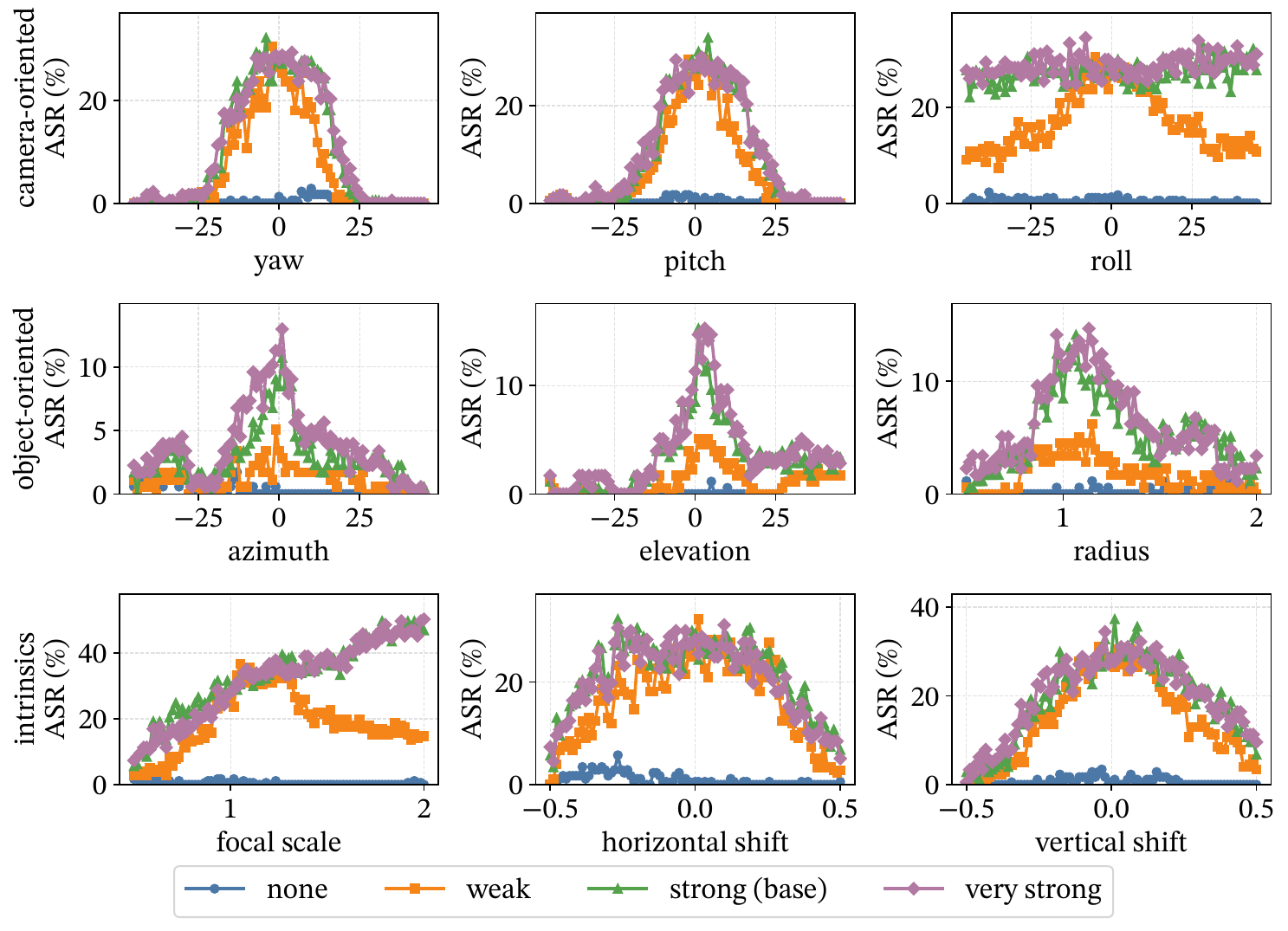}
\caption{Impact of EOT magnitude.}
\label{fig:ablation_eot_magnitude}
\end{figure}

\begin{figure}[!h]
\centering
\includegraphics[width=\linewidth]{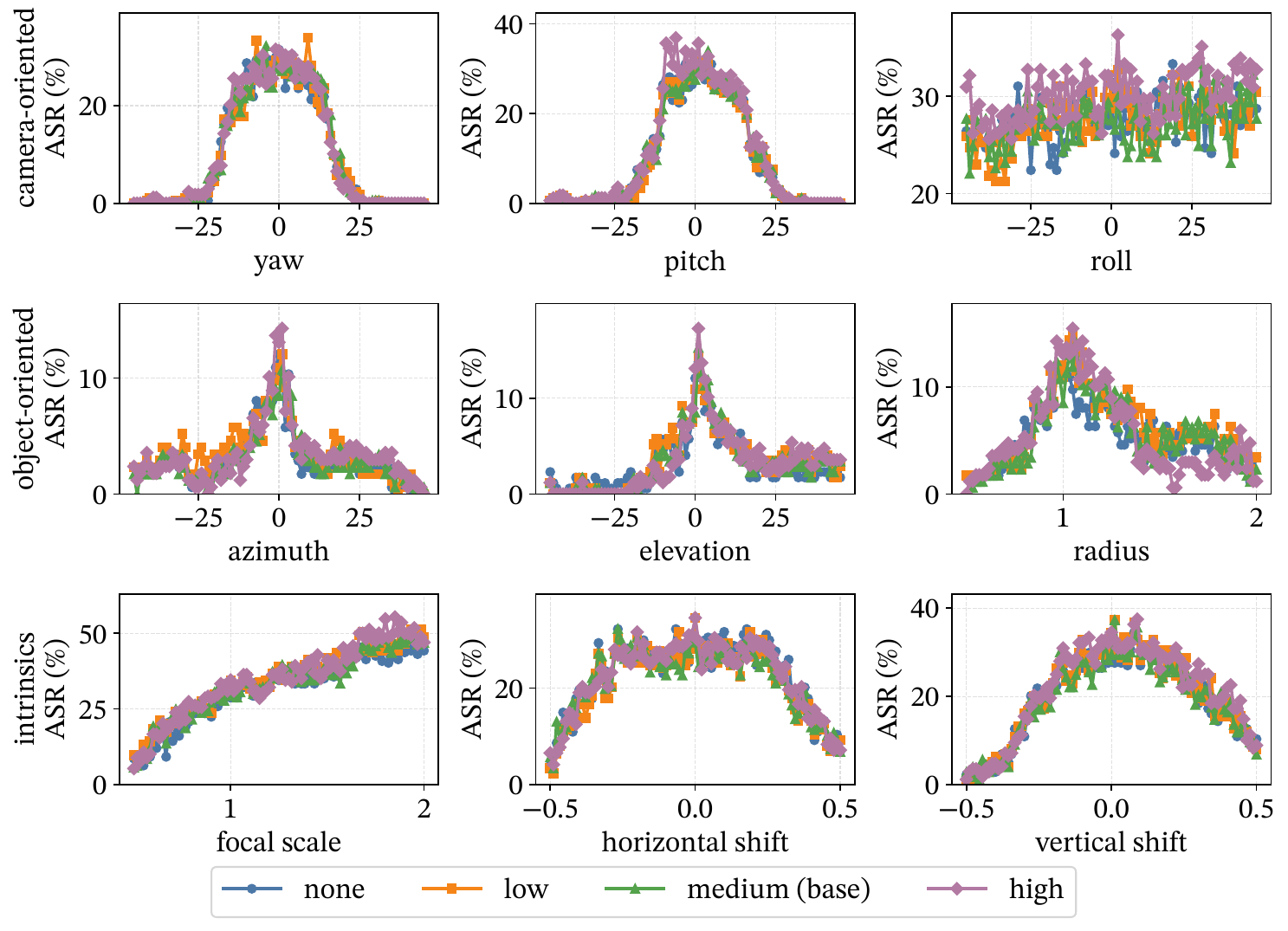}
\caption{Impact of $\ell_2$ loss.}
\label{fig:ablation_l2}
\end{figure}

\begin{figure}[!h]
\centering
\includegraphics[width=\linewidth]{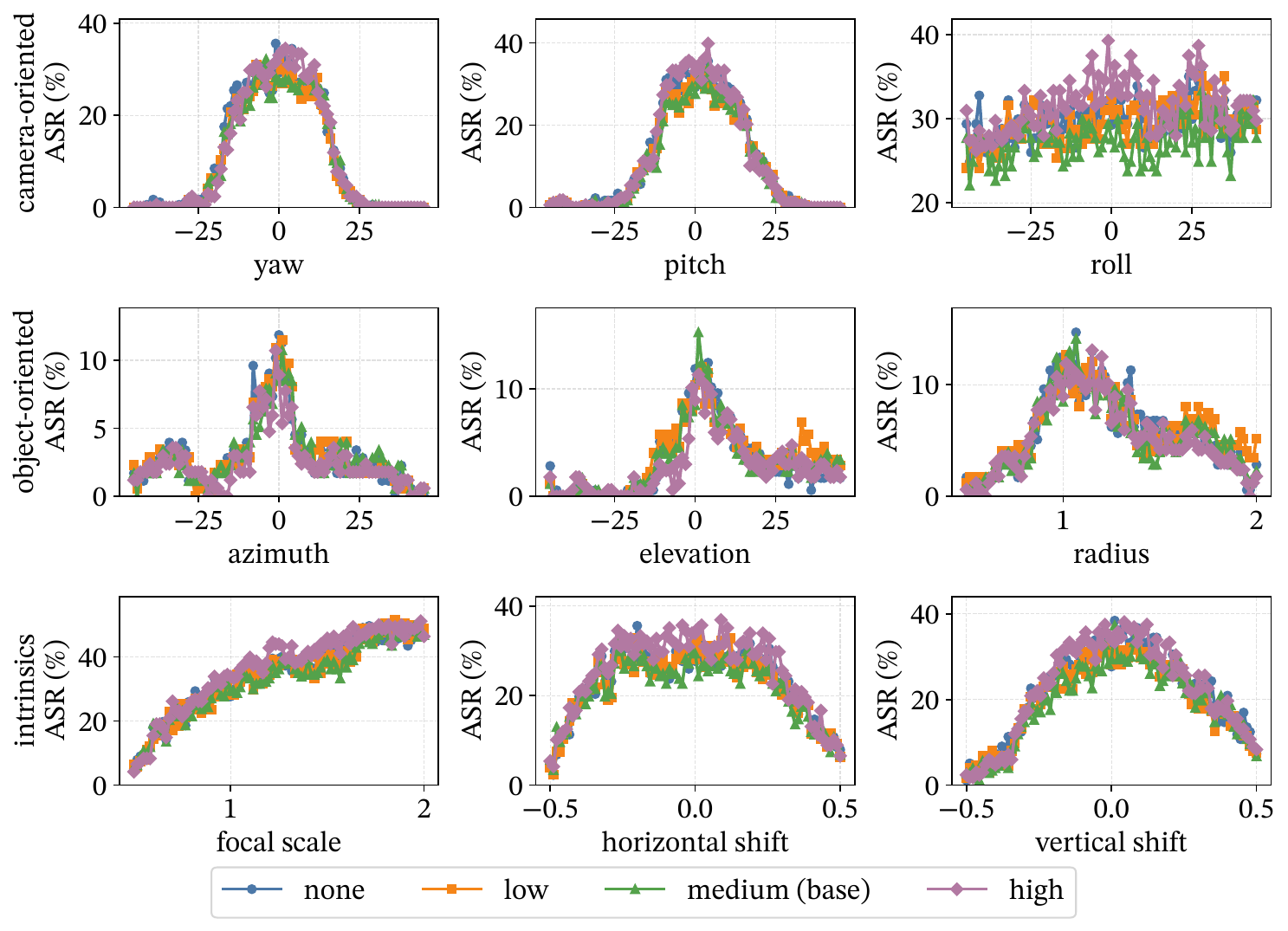}
\caption{Impact of non-printability score (NPS) weights.}
\label{fig:ablation_nps}
\end{figure}

\begin{figure}[!h]
\centering
\includegraphics[width=\linewidth]{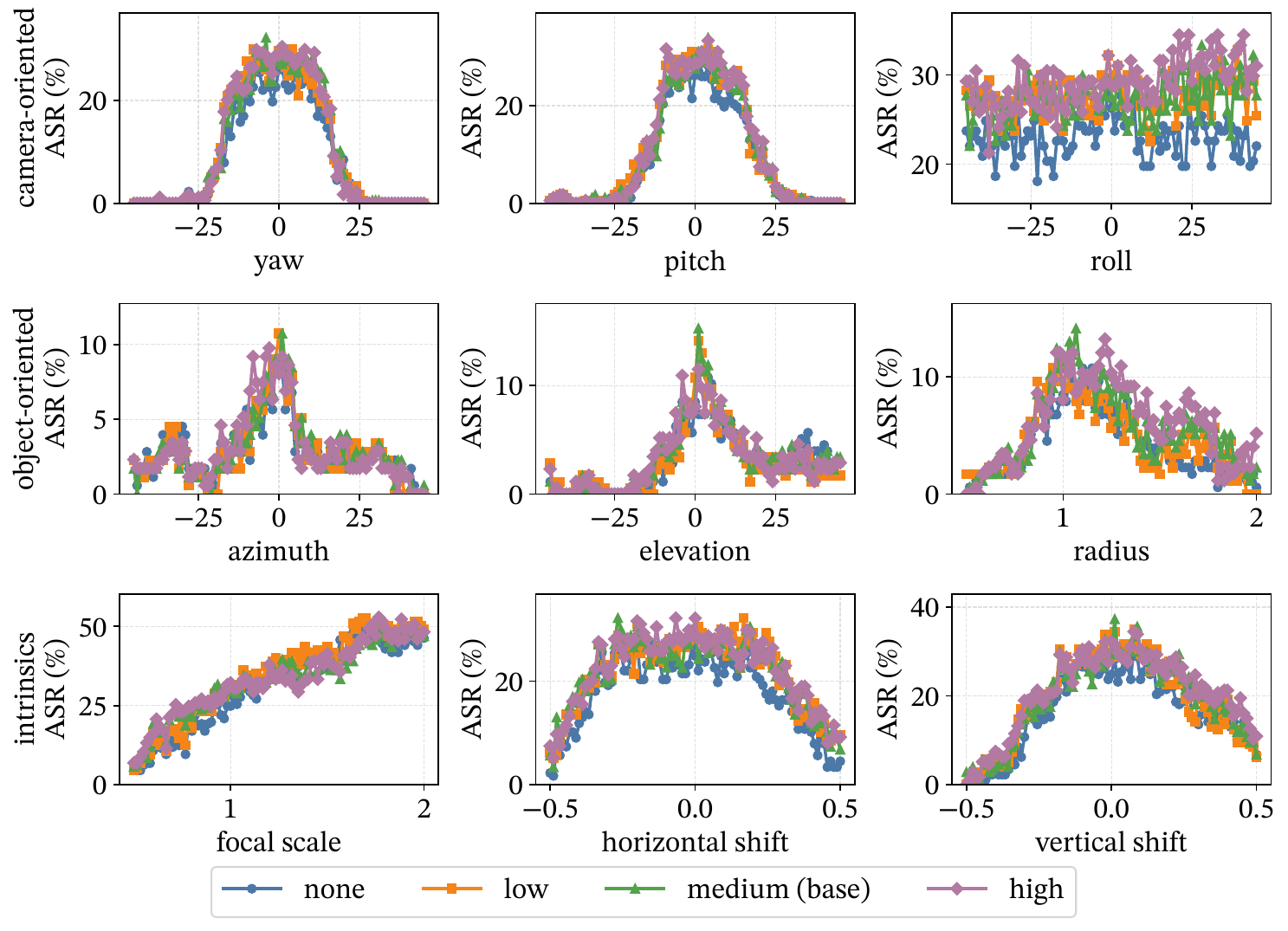}
\caption{Impact of TV loss.}
\label{fig:ablation_tv}
\end{figure}

Figures~\ref{fig:ablation_eot_component},~\ref{fig:ablation_eot_magnitude},~\ref{fig:ablation_l2},~\ref{fig:ablation_nps},~\ref{fig:ablation_tv} report the full attack-design comparisons. The results support the main-text conclusion that geometry-aware EOT is the primary mechanism for expanding scene-level robustness, while $\ell_2$, NPS, and TV mainly affect magnitude, printability, or smoothness rather than removing geometric failure modes.

\section{Ablation study of APSE components.}
We conduct an ablation study to evaluate the effectiveness of each component of APSE. We disable one component in APSE at a time, including the optimization of newly spawned patch Gaussians (Section \ref{subsec:hybrid_gaussian}), the optimization of existing (old) target-surface Gaussians (Section \ref{subsec:hybrid_gaussian}), anchor-view non-mask loss (Eq.~\ref{eq:keep} in Section \ref{subsec:anchor_fidelity}), auxiliary-view containment (Section~\ref{subsec:aux_containment}), and surface attachment (Section~\ref{subsec:thin_object}). 

Table~\ref{tab:apse_ablation} shows that each component contributes to a balanced APSE embedding. We report patch-region reconstruction error, non-patch region reconstruction error, and auxiliary-view depth error, where lower values indicate better fidelity or consistency. Disabling the optimization of newly spawned patch Gaussians substantially degrades patch reconstruction, confirming that the original scene Gaussians alone do not provide sufficient local capacity to reproduce the adversarial appearance. Disabling the optimization of existing target-surface Gaussians also weakens patch fidelity and cross-view consistency, indicating that the spawned Gaussians need the structural support of the reconstructed surface. Removing the anchor-view non-mask loss leaves the supervised patch region largely unchanged, but weakens the overall consistency of the embedding. Removing auxiliary-view containment similarly preserves the anchor-view appearance, but reduces stability under novel views, showing that auxiliary constraints are needed to suppress view-specific artifacts. Finally, disabling surface attachment improves the anchor-view fit, but at the cost of weaker geometric consistency, suggesting that unconstrained Gaussians can overfit the anchor view instead of behaving like a surface-attached patch. Overall, the ablation confirms that APSE relies on the joint effect of all components.

\begin{table*}[!tb]
\small
\caption{Ablation study of key components in APSE}
\label{tab:apse_ablation}
\resizebox{\textwidth}{!}{%
\begin{tabular}{@{}lrrrrr@{}}
\toprule
 & patch region $\ell_1$ & patch region $\ell_2$ & non-patch region $\ell_1$ & non-patch region $\ell_2$ & auxiliary-view depth error $\ell_1$  \\ \midrule
APSE with all components & 0.127 & 0.114 & 0.028 & 0.026 & 0.124 \\
w/o new Gaussian (\$\ref{subsec:hybrid_gaussian}) & 0.333 & 0.268 & 0.030 & 0.034 & 0.046 \\
w/o old Gaussian (\$\ref{subsec:hybrid_gaussian})  & 0.180 & 0.162 & 0.028 & 0.026 & 0.140 \\
w/o anchor view non-mask fidelity (\$\ref{subsec:anchor_fidelity}) & 0.128 & 0.115 & 0.001 & 0.011 & 0.140 \\
w/o auxiliary view (\$\ref{subsec:aux_containment}) & 0.128 & 0.115 & 0.028 & 0.027 & 0.127 \\
w/o surface attachment (\$\ref{subsec:thin_object}) & 0.115 & 0.101 & 0.027 & 0.026 & 0.135 \\ \bottomrule
\end{tabular}%
}
\end{table*}

\section{Cross-scene Evaluation}

\begin{figure}[!h]
\centering
\includegraphics[width=\linewidth]{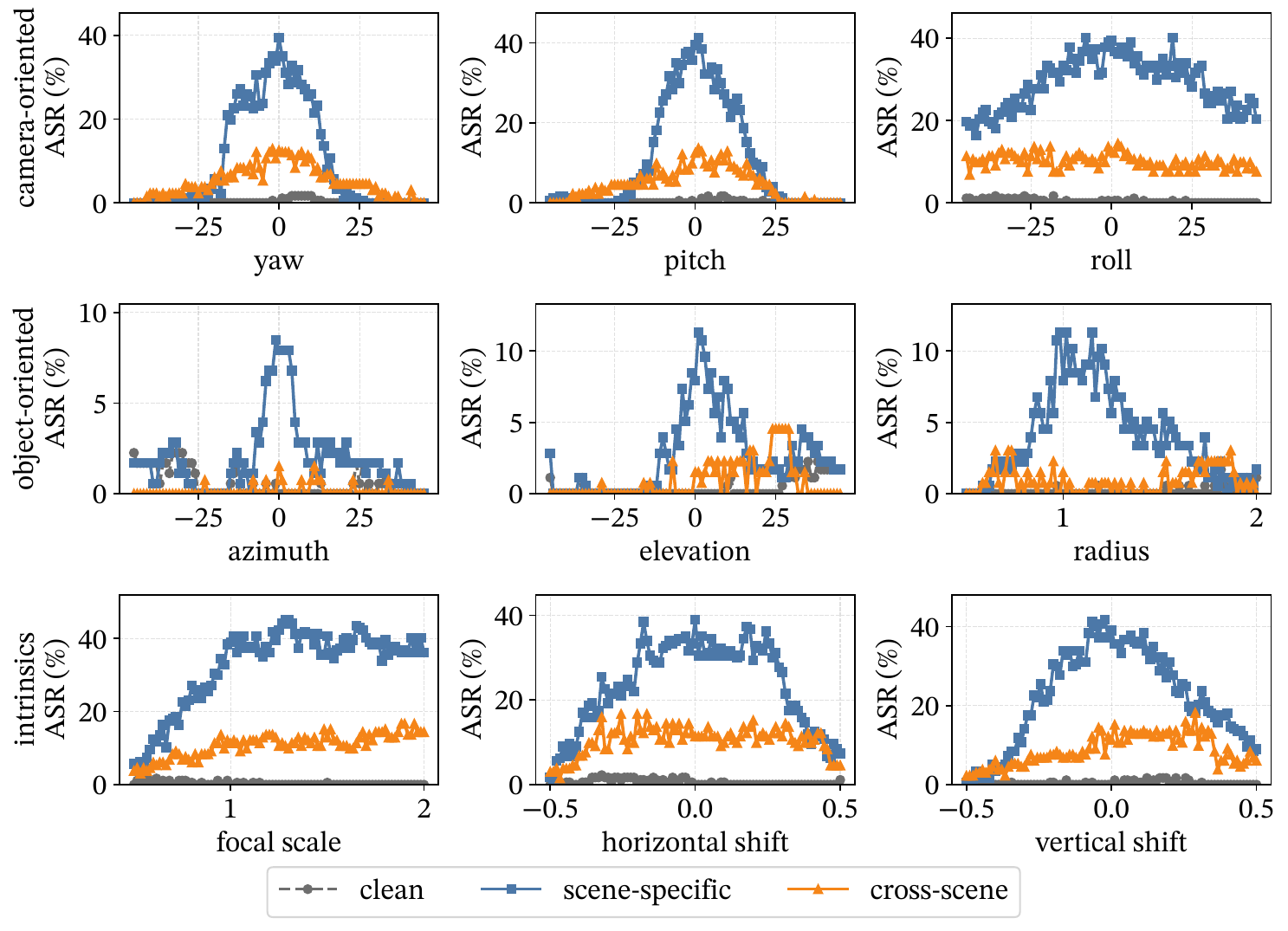}
\caption{Cross-scene evaluation. }
\label{fig:cross_scene_eval}
\end{figure}

We evaluate whether adversarial patches transfer across scenes. In this experiment, a patch is optimized from an anchor view in one source scene and then evaluated in different target scenes that contain the same object category, such as optimizing a patch for a book in one scene and applying it to books in other scenes. This setting tests whether a fixed patch captures object-level adversarial features that generalize across scene context, or whether its effectiveness is tied to the source scene. We perform this evaluation on CO3D under the targeted classification setting.

Figure~\ref{fig:cross_scene_eval} reports the cross-scene results. We compare three settings: clean, which measures the target-label success rate without an adversarial patch; scene-specific, where the patch is optimized and evaluated within the same scene; and cross-scene, where the patch is optimized in one source scene and evaluated in different target scenes. The scene-specific setting achieves substantially higher ASR than the cross-scene setting, while cross-scene performance drops close to the clean baseline. This result shows that targeted patch effectiveness is strongly scene-dependent. In particular, a patch optimized for one reconstructed scene does not generally preserve its operational envelope after transfer to another scene, even when the target object category is unchanged. This supports the need for scene-grounded evaluation rather than evaluating patch robustness as an object-only or image-only property.

\end{document}